\documentclass[review]{elsarticle}

\usepackage{lineno,hyperref}
\usepackage{amsmath}
\usepackage{amssymb}
\usepackage{float}
\usepackage{cancel}
\usepackage{bm}
\usepackage{subcaption}
\usepackage{verbatim}
\usepackage{color}
\modulolinenumbers[5]

\journal{Elsevier}

\newcommand{\sigmat}{\sigma_\mathrm{t}}
\newcommand{\sigmas}{\sigma_\mathrm{s}}

\newcommand{\PN}{P$_N$}
\renewcommand{\vec}[1]{\bm{#1}} 
\newcommand{\vd}{\bm{\cdot}} 
\newcommand{\grad}{\vec{\nabla}} 

\usepackage{graphicx} 
\usepackage{booktabs} 
\usepackage{microtype} 
\usepackage[linesnumbered,boxruled]{algorithm2e}

\begin{document}
\begin{frontmatter}

\title{A high-order / low-order (HOLO) algorithm for preserving conservation in time-dependent low-rank transport calculations}
\author[mymainaddress]{Zhuogang Peng}
\ead{zpeng5@nd.edu}

\author[mymainaddress]{Ryan McClarren\corref{mycorrespondingauthor}}
\cortext[mycorrespondingauthor]{Corresponding author}
\ead{rmcclarren@nd.edu}

\address[mymainaddress]{Department of Aerospace and Mechanical Engineering, University of Notre Dame, Notre Dame, IN, 46545, USA}


\begin{abstract}
Dynamical low-rank (DLR) approximation methods have previously been developed for time-dependent radiation transport problems. One crucial drawback of DLR is that it does not conserve important quantities of the calculation, which limits the applicability of the method. Here we address this conservation issue by solving a low-order equation with closure terms computed via a high-order solution calculated with DLR. We observe that the high-order solution well approximates the closure term, and the low-order solution can be used to correct the conservation bias in the DLR evolution. We also apply the linear discontinuous Galerkin method for the spatial discretization to obtain the asymptotic limit. We then demonstrate with the numerical results that this so-called high-order / low-order (HOLO) algorithm is conservative without sacrificing computational efficiency and accuracy.
\end{abstract}


\end{frontmatter}

\newpage

\section{Introduction}
The radiation transport equation (RTE) describes the movement of particles (e.g., photons or neutrons) through a background medium. Solving the RTE is of great interest across many research areas, namely, nuclear engineering \cite{Azmy2010}, astrophysics \cite{Rybick1985}, and optics \cite{KIM2003}. It is a challenging problem because the RTE has seven independent variables, including one in time, three in position, two in direction, and one in energy, which requires an extravagant computational burden in terms of both memory and operations. Thus, developing a computationally inexpensive, yet accurate, algorithm is a continuing concern. 

There is a long history of methods designed to reduce the complexity and size of radiation transport calculations.  Many of these efforts have focused on the direction, or angular, variables. The diffusion method, and its flux-limited variants \cite{Olson1999,morel2000diffusion,levermore1981flux,Kershaw:1976cf}, represent the direction variables with a two-moment representation based on Fick's law. Nevertheless, a two-moment description of the directional variables is inadequate to describe the behaviour in many problems \cite{Brunner2001}. The expansion order can be increased by using spherical harmonics methods \cite{pomraning2005equations, heizler2010asymptotic}, though these methods have limitations, including negative densities \cite{mcclarren2008solutions}, that need to be addressed with either filters \cite{mcclarren2010simulating, mcclarren2010robust,radice2013new, laboure2016implicit}, or nonlinear closures that increase in the computational complexity \cite{minerbo1978maximum,olbrant2012realizability,zheng2016moment,Levermore:1984tj,laiu2016positive}. The simplified spherical harmonics method \cite{mcclarren2010theoretical}, is an intermediate approximation between the diffusion method and spherical harmonics \cite{pomraning1993asymptotic}, that also has issues with accurately solving many problems \cite{zhang2020convergence}. Recent work on wavelets \cite{buchan2005linear,dargaville2020scalable,soucasse2017goal} and adaptive discrete ordinates methods \cite{stone2008adaptive,jarrell2011discrete,lau2017discrete} have shown that it is possible to reduce complexity by focusing the effort on angular degrees where most necessary.

Alongside the investigations into angular discretizations, there is a record of work addressing the other complexities in transport problems. This includes the so-called second-order forms such as the even-parity, odd-parity \cite{LewisMiller}, self-adjoint \cite{morel1999self}, and least-squares \cite{drumm2011least,Hansen:2015jq} forms.  The second-order forms in many cases require the solution of half the number of equations with the additional benefits that spatial discretizations of second-operators possess \cite{latimer2020spatial, latimer2020geometry}. The second-order forms do have issues in voids \cite{wang2014diffusion,laboure2017globally,zheng2018accurate} that cause either inaccuracy or loss of the self-adjoint character of the equations. Adaptivity in space \cite{ragusa2010two,turcksin2010goal,wang2011standard,hanuvs2016use} and space-angle \cite{kophazi2015space}, as well as selective reduction of degrees of freedom \cite{Sun2019} have all been explored as well.

In this paper, we continue a more recent trend of applying dynamical low-rank (DLR) approximation methods to radiation transport problems.  These methods project the RTE onto a reduced basis in space and angle that evolves dynamically during a calculation.
DLR methods were considered for the RTE in work by the authors \cite{peng2019lowrank}. The idea of this method comes from a conventional paradigm in solving the time-dependent Schrödinger equation for multidimensional dynamical systems, known as Multi-Configuration Time-Dependent Hartree (MCTDH), which gives a rank-1 approximation for a multivariate wave function \cite{Lubich2008}. The DLR approximates the time derivative of the objective matrices or tensors by applying  tangent space projection \cite{Koch2007,Nonnenmacher2008,Koch2010}. It was shown to be robust even with small singular values \cite{Kieri2016} when the time integration is performed using splitting \cite{Lubich2014,Lubich2018}. The application in other kinetic equations can be found in \cite{Einkemmer2018,Einkemmer2019,Einkemmer2020}. The implementation in radiation transport calculations produces high-fidelity results obtained from the low-rank scheme with a fraction of memory usage and computational time. However, this approach does not preserve the total number of particles (i.e., the methods are not guaranteed to be conservative), which may result in a significant error at long times or in steady-state. There is an attempt to recover the conservation by enforcing the conservation law to the low-rank scheme, which results in an optimization problem that needs to be solved \cite{Einkemmer2018qc} that increases the computational cost of the method.


In this paper, we propose a high-order/low-order (HOLO) algorithm \cite{chacon2017multiscale,Bolding2017} to solve this conservation issue. In our algorithm, we take the high-order, low-rank solution to calculate a closure term in the low-order two-moment approximation of the transport equation, in an approach very similar to the quasi-diffusion method \cite{Goldin1964,anistratov2005consistent}. The low-order system conserves particles independent of the closure. This allows us to demonstrate that our HOLO algorithm overcomes the conservation difficulty while preserving the computational efficiency and accuracy. Additionally, we demonstrate that this approach can also preserve the asymptotic diffusion limit \cite{larsen1987asymptotic,larsen1989asymptotic} of the radiation transport equation.

We begin with a brief review of the low-rank method in Section 2. Then we present the low-order system with the closure term to preserve the number of particles. We further introduce the HOLO scheme which couples the low-order system with the low-rank solutions. To guarantee the consistency between the two systems, we develop a moment-based correction method to fix the conservation in the low-rank evolution. After that, we design a numerical scheme for the 2-D RTE with a discontinuous Galerkin discretization in space and a spherical harmonic (P$_N$) expansion in angle. In our results section, we demonstrate the efficacy of our algorithm with numerical results to validate the memory-reduction, conservation and asymptotic preserving properties.

\section{Dynamical low-rank approximation}
We consider a time-dependent radiative transfer equation with one energy group: 
\begin{equation} \label{eq: RadiativeTransfer} 
\begin{split}
    \frac{1}{c} \frac{\partial \psi (\vec{r}, \hat{\Omega}, t)}{\partial t} + \hat{\Omega} \vd \grad \psi (\vec{r}, \hat{\Omega}, t)  +  \sigmat(\vec{r}) \psi (\vec{r}, \hat{\Omega}, t) = 
\frac{1}{4\pi}\sigmas(\vec{r}) \phi(\vec{r}, t) + Q(\vec{r},t).
\end{split}
\end{equation}
The radiation intensity $\psi(\vec{r}, \hat{\Omega}, t)$ [energy/area/steradian/time] is a function of position $\vec{r}$, time $t$, and the unit angle vector $\hat{\Omega}(\mu, \varphi)$, where $\mu$ is the cosine of the polar angle and $\varphi$ is the azimuthal angle. The total and isotropic scattering macroscopic cross-sections with units of inverse length are denoted as $\sigma_t(\vec{r})$ and $\sigma_s(\vec{r})$, respectively, $c$ is the particle speed, and $Q(\vec{r},t)$ is a prescribed source. We set c = 1 in the following derivations for simplicity. The scalar intensity, $\phi(\vec{r}, t)$, is the integral of $\psi(\vec{r}, \hat{\Omega}, t)$ over all angles:
\begin{equation}
\phi(\vec{r},t) = \int_{4 \pi}\psi(\vec{r}, \hat{\Omega}, t) \, d \hat{\Omega}.
\end{equation}
The low-rank method aims to approximate the solution to Eq.~\eqref{eq: RadiativeTransfer} with rank $r$ using the form 
\begin{equation}\label{eq: LR_form}
\psi(\vec{r},\hat{\Omega},t) \approx \sum_{i,j = 1}^r X_i(\vec{r},t)S_{ij}(t)W_j(\hat{\Omega},t);
\end{equation}
where $X_i$ is an orthonormal basis for $\vec{r}$ and $W_j$ is an orthonormal basis for $\hat{\Omega}$ using the inner products
\[\langle f,g\rangle _{\vec{r}} = \int_{D} f  g\, d\vec{r}, \quad \langle f,g\rangle _{\hat{\Omega}} = \int_{4 \pi} f g \, d\hat{\Omega} . \]
The expansion is unique with  orthonormality $\langle X_i, {X}_j \rangle_{\vec{r}} = \langle W_i, {W}_j \rangle_{\hat{\Omega}} = \delta_{ij}$ and gauge conditions $\langle X_i, \dot{X}_j \rangle_{\vec{r}} = 0$ and $\langle W_i, \dot{W}_j \rangle_{\hat{\Omega}}  = 0$. The rank $r$ should be less than the number the degrees of freedom in the either of the bases $X_i$ and $W_j$. We then define the orthogonal projectors to the low-rank ansatz spaces $ \bar{X} = \{X_1, X_2, ..., X_r\}$ and $\bar{W} = \{W_1, W_2, ..., W_r\}$ as  
\begin{equation}\label{Projection1}
P_{\bar{X}} g = \sum_{i=1}^{r}X_i \langle X_i g \rangle_{\vec{r}},
\end{equation}
\begin{equation}\label{Projection2}
P_{\bar{W}} g = \sum_{j=1}^{r}W_j \langle W_j g \rangle_{\hat{\Omega}}.
\end{equation}
The full solution $\psi(\vec{r},\hat{\Omega},t)$ can be projected into the low-rank manifold $\mathcal{M}_r$ by the projector $Pg = P_{\bar{W}} g - P_{\bar{X}} P_{\bar{W}} g + P_{\bar{X}} g$. To make the computation more robust, we split the process into three steps \cite{Lubich2014}, where we solve each of the following three equations over a time step:
\begin{multline}\label{eq: MainEq1}
\partial_t \psi(\vec{r}, \hat{\Omega},t)  = P_{\bar{W}} \bigg( - \hat{\Omega} \cdot \grad \psi(\vec{r}, \hat{\Omega}, t)
- \sigma_t(\vec{r}, t) \psi (\vec{r}, \hat{\Omega}, t) \\
+ \frac{1}{4\pi} \sigmas(\vec{r}, t) \phi(\vec{r}, t) 
+ Q(\vec{r},t) \bigg),
\end{multline}

\begin{multline}\label{eq: MainEq2}
\partial_t \psi(\vec{r}, \hat{\Omega},t) = - P_{\bar{X}} P_{\bar{W}} \bigg( - \hat{\Omega} \cdot \grad \psi(\vec{r}, \hat{\Omega}, t)
- \sigma_t(\vec{r}, t) \psi (\vec{r}, \hat{\Omega}, t) \\
+ \frac{1}{4\pi} \sigmas(\vec{r}, t) \phi(\vec{r}, t) 
+ Q(\vec{r},t) \bigg),
\end{multline}

\begin{multline}\label{eq: MainEq3}
\partial_t \psi(\vec{r}, \hat{\Omega},t) = P_{\bar{X}} \bigg( - \hat{\Omega}\cdot \grad \psi(\vec{r}, \hat{\Omega}, t)
- \sigma_t(\vec{r}, t) \psi (\vec{r}, \hat{\Omega}, t) \\
+ \frac{1}{4\pi} \sigmas(\vec{r}, t) \phi(\vec{r}, t) 
+ Q(\vec{r},t) \bigg).
\end{multline}

We then formulate the projections \eqref{eq: MainEq1} - \eqref{eq: MainEq3} explicitly from time $t_0$ to $t_0+h$ where $h$ is the step size. To simplify the notation we define $K_j(\vec{r},t) = \sum^{r}_{i} X_{i}(\vec{r},t)S_{ij}(t)$
and $L_i = \sum_j^r S_{ij}(t)W_{j}(\hat{\Omega}, t)$. The corresponding projected equations are 
\begin{multline}\label{eq: K_eq}
\partial_t K_j = -\sum_{l=1}^{r} \grad K_l \, \langle \hat{\Omega} W_l W_j\rangle_{\hat{\Omega}} - \sigmat K_j 
+ \frac{1}{4\pi} \sigmas \sum_{l=1}^{r}  K_l \langle W_l\rangle_{\hat{\Omega}} \langle W_j\rangle_{\hat{\Omega}} \\+ Q \langle W_j\rangle_{\hat{\Omega}},
\end{multline}

\begin{multline}\label{eq: S_eq}
\frac{d}{dt} S_{ij} = \sum_{kl}^{r} \langle \grad X_k \, X_i \rangle_{\vec{r}} S_{kl} \langle \hat{\Omega} W_l W_j \rangle_{\hat{\Omega}} 
+\sum_{k}^{r} \langle \sigma_t X_k X_i \rangle_{\vec{r}} S_{kj} \\
- \frac{1}{4\pi} \sum_{kl}^{r} \langle \sigma_s X_k X_i \rangle_{\vec{r}}S_{kl}\langle W_l\rangle_{\hat{\Omega}} \langle W_j\rangle_{\hat{\Omega}} - \langle X_i Q \rangle_{\vec{r}} \langle W_j\rangle_{\hat{\Omega}},
\end{multline}

\begin{multline}\label{eq: L_eq}
\frac{d}{dt} L_{i} = -\hat{\Omega} \sum_{k}^{r} \langle \grad X_k \, X_i \rangle_{\vec{r}} L_k - \sum_{k}^{r} \langle \sigma_t X_k X_i \rangle_{\vec{r}} L_{k} \\
+ \frac{1}{4\pi} \sum_{k}^{r} \langle \sigma_s X_k X_i \rangle_{\vec{r}} \langle L_k \rangle_{\hat{\Omega}} 
+ \langle Q X_i \rangle_{\vec{r}}.
\end{multline}

We summarize the procedures to solve equations \eqref{eq: K_eq} - \eqref{eq: L_eq} in Algorithm \ref{alg: DLRA}. Importantly, in the algorithm only the low-rank components $X$, $S$ and $W$ are stored during the time evolution rather than the full solution $\psi$. This allows the low-rank method to save computer memory. 

\begin{algorithm}[ht]
	\textbf{Given}\\
	-initial time $t_0$\\
	-time step $h$\\
	-desired rank $r$\\
	-initial condition $\psi^{(0)}(\vec{r},\hat{\Omega},t_0)$\\
	-initial approximations $X^{(0)}_i(\vec{r}, t_0)$, $S^{(0)}_{ij}(t_0), W^{(0)}_j(\hat{\Omega}, t_0)$\\
	\textbf{Repeat:}\\
	-$t_1 = t_0 + h$ \\
	-Solve equation \eqref{eq: K_eq} for $K_j^{(1)}(\vec{r}, t_1)$ with initial condition $K_j^{(0)}(\vec{r}, t_0) = \sum^{r}_{i} X_{i}^{(0)}S_{ij}^{(0)}$ and then factor into $X_{i}^{(1)}(\vec{r}, t_1)$ and $S_{ij}^{(1)}(t_1)$ using a QR decomposition; $W_{j}^{(1)} = W_{j}^{(0)}$ is preserved in this step.\\
	-Solve equation \eqref{eq: S_eq} for $S_{ij}^{(2)}(t_1)$ with initial condition $S_{ij}^{(2)}(t_0) = S_{ij}^{(1)}$, $X_{i}^{(2)} = X_{i}^{(1)}$ and $W_{j}^{(2)} = W_{j}^{(1)}$ are preserved in this step. \\
	-Solve equation \eqref{eq: L_eq} for $L_i^{(3)}(\hat{\Omega}, t_1)$ with initial condition $L_i^{(3)}(\hat{\Omega}, t_0) = \sum^{r}_{j} S_{ij}^{(2)}W_{j}^{(2)}$ and then factor into $S_{ij}^{(3)}(t_1)$ and $W_{j}^{(3)}(\hat{\Omega}, t_1)$ using a QR decomposition; $X_{i}^{(3)} = X_{i}^{(2)}$ is preserved in this step.\\
	-$t_0 = t_1$ \\
	-$X_i^{(0)} = X_i^{(3)}$, $S_{ij}^{(0)} = S_{ij}^{(3)}$, $W_j^{(0)} = W_j^{(3)}$
	\caption{Dynamical low-rank approximation}
	\label{alg: DLRA}
\end{algorithm}


\section{HOLO algorithm}
In this section we will construct the low-order system that requires closure terms. We apply the quasi-diffusion method \cite{Goldin1964} which is also known as the variable Eddington factor method \cite{Mihalas1978StellarA}, to Eq.~\eqref{eq: RadiativeTransfer} to yield a two-angular-moments formulation \cite{Brunner2001}: 
\begin{equation} \label{eq: ZerothMoment} 
\frac{1}{c} \frac{\partial \phi (\vec{r}, t)}{\partial t} + \grad \vd J (\vec{r}, t)  +  \sigmat(\vec{r}) \phi (\vec{r}, t) 
= \sigmas(\vec{r}) \phi(\vec{r}, t) + Q(\vec{r},t).
\end{equation}
and
\begin{equation} \label{eq: FirstMoment} 
\frac{1}{c} \frac{\partial J (\vec{r}, t)}{\partial t} + \grad \vd \chi \phi (\vec{r}, t)  +  \sigmat(\vec{r}) J (\vec{r}, t) 
= 0.
\end{equation}
where 
\begin{equation}
    J(\vec{r},t) = \int_{4 \pi} \hat{\Omega} \psi(\vec{r}, \hat{\Omega}, t) \, d \hat{\Omega}
\end{equation}
is the radiative flux (or current density), and
\begin{equation}
    \chi = \frac{\int_{4 \pi} \hat{\Omega} \otimes \hat{\Omega} \psi(\vec{r}, \hat{\Omega}, t) \, d \hat{\Omega}}{\phi(\vec{r}, t)}
\end{equation}
is the Eddington tensor and $\otimes$ denotes the outer product. Note that $\phi$ is the zeroth moment, $J$ is the first moment and $\chi$ is the normalized second moment related to radiation pressure. To close Eqs.~\eqref{eq: ZerothMoment} and \eqref{eq: FirstMoment} we need to evaluate the Eddington tensor $\chi$. An approximation $\chi \approx \frac{1}{3} I$ , used in the diffusion method, is based on the assumption that $\psi$ is a linear function of angle, where $I$ is an identity tensor. Such approximations ignore the relation of Eddington factor to higher order moments, which limits their accuracy. 

We propose a closure term $\gamma = \grad \vd \frac{1}{3} \phi (\vec{r}, t) - \grad \vd \chi \phi (\vec{r}, t)$ added to the right hand side of Eq.~\eqref{eq: FirstMoment} that corrects the $\chi \approx \frac{1}{3} I$ approximation based on the solution to the full RTE.  The corrected equation that we solve is
\begin{equation} \label{eq: FirstMoment2} 
\frac{1}{c} \frac{\partial J (\vec{r}, t)}{\partial t} + \frac{1}{3} \grad \vd  \phi (\vec{r}, t)  +  \sigmat(\vec{r}) J (\vec{r}, t) 
= \gamma(\vec{r},t).
\end{equation} 
As we will show later, $\gamma$ is calculated as the difference between the low-rank equations \eqref{eq: K_eq} - \eqref{eq: L_eq} and the quasi-diffusion equation \eqref{eq: ZerothMoment} - \eqref{eq: FirstMoment}. The main idea of our HOLO algorithm is that the solution to the low-order system ~\eqref{eq: ZerothMoment} and \eqref{eq: FirstMoment2} can be accurate if $\gamma$ is evaluated through the solution to the high-order system (\ref{eq: K_eq} - \ref{eq: L_eq}). Additionally, the solution is conservative in $\phi$ regardless of the value of the closure term because $J$ appears in conservative form in Eq.~\eqref{eq: ZerothMoment}.

\subsection{Angular discretization for low-rank equations in 2D}
We choose the angular bases $W_j(\mu, \varphi, t)$ to be the spherical harmonics expansion truncated at $N$: 
\begin{equation}\label{eq: 2D_Discretization2}
W_j(\mu, \varphi, t) \approx \sum_{l=0}^{N} \, \sum_{k=0}^{l} v_{jlk}(t)Y_{l}^{k}(\mu, \varphi)
\end{equation}
where 
$$Y_l^k(\mu, \varphi) = \sqrt{\frac{2l+1}{4 \pi} \, \frac{(l-k)!}{(l+k)!}} \, P_l^k(\mu) \, e^{i \, k\varphi},$$
where $P_{l} ^k (\mu)$ is  the associate Legendre polynomial and $N$ is the expansion order. Here $0 \leq k \leq l \leq N$ and the negative $k$ are omitted because of the symmetry properties of the spherical harmonics \cite{Brown2005}. The total number of moments is $n = \frac{(N+1)(N+2)}{2}$. 

We use vectors ${V_j}$ to collect all the angular elements $v_{jlk}$ in $W_j$ and $\vec{Y}$ for all the spherical harmonics $Y_l^k$ sorted by the index $l$, e.g., $\vec{Y} = [Y_0^0, \ Y_1^0, \ Y_1^1, \ Y_2^0,\ ... \ Y_N^N]^T$. The integration of $\vec{Y}$ in angular domain is computed as $$\int_{0}^{2\pi} \int_{-1}^{1} \vec{Y} \, d\mu d\varphi = \langle \vec{Y} \rangle = [2\sqrt{\pi}, 0, 0, ... 0]$$ for later use. We then substitute the angular basis $W_j = Y^T V_j$ into the projection Eqs.~\eqref{eq: K_eq} and \eqref{eq: S_eq}: 
\begin{multline}\label{eq: K_eq_2}
\partial_t K_j = -\sum_{l=1}^{r} \partial_z K_l {V_l}^T \vec{A_z} {V_j} -\sum_{l=1}^{r} \partial_x K_l {V_l}^T \vec{A_x} {V_j} - \sigmat K_j \\
+ \frac{1}{4\pi} \sigmas \sum_{l=1}^{r}  K_l {V_l}^T \langle \vec{Y} \rangle \langle \vec{Y} \rangle^T \,{V_j} + Q  \int_{0}^{2\pi} \int_{-1}^{1} \langle \vec{Y} \rangle^T \, V_j, 
\end{multline}

\begin{multline}\label{eq: S_eq_2}
\frac{d}{dt} S_{ij} = \sum_{kl}^{r} \langle \partial_z X_k \, X_i \rangle_{\vec{r}} S_{kl} {V_l}^T \vec{A_z} {V_j} + \sum_{kl}^{r} \langle \partial_x X_k \, X_i \rangle_{\vec{r}} S_{kl} {V_l}^T \vec{A_x} {V_j}\\
+\sum_{k}^{r} \langle \sigma_t X_k X_i \rangle_{\vec{r}} S_{kj} 
- \frac{1}{4\pi} \sum_{kl}^{r} \langle \sigma_s X_k X_i \rangle_{\vec{r}}S_{kl}{V_l}^T \langle \vec{Y} \rangle \langle \vec{Y} \rangle^T{V_j}\\ - \langle X_i Q \rangle_{\vec{r}} \langle \vec{Y} \rangle^T \, V_j,
\end{multline}
Note that $A_x = \int_0^{2\pi} \int_{-1}^{1} \sqrt{1-\mu^2} \cos \varphi \vec{Y} \vec{Y}^T \ d\mu d\varphi$, and $A_z = \int_0^{2\pi} \int_{-1}^{1} \mu \vec{Y} \vec{Y}^T \ d\mu d\varphi$ can be calculated using the recursion property and no quadrature rule is required. 

We multiply $\vec{Y}$ on both sides of Eq.~\eqref{eq: L_eq} and integrate over all angles to remove the angular dependence:   
\begin{multline}\label{eq: L_eq_2}
\frac{d}{dt} R_{i} = -\sum_{k}^{r} \langle \partial_z X_k \, X_i \rangle_{\vec{r}} R_k \vec{A_z} -\sum_{k}^{r} \langle \partial_x X_k \, X_i \rangle_{\vec{r}} R_k \vec{A_x} + \sum_{k}^{r} \langle \sigma_t X_k X_i \rangle_{\vec{r}} R_k \\
- \frac{1}{4\pi} \sum_{k}^{r} \langle \sigma_s X_k X_i \rangle_{\vec{r}}R_k \langle \vec{Y} \rangle \langle \vec{Y} \rangle^T - \langle X_i Q \rangle_{\vec{r}} \langle \vec{Y} \rangle^T
\end{multline}
where $R_{i} = \int_{0}^{2\pi} \int_{-1}^{1} L_i(t, \mu, \varphi) \vec{Y}^T \, d\mu d\varphi$. The low-rank \PN~ moments can be obtained from 
$$\psi_l^k(x, z, t) = \int_{0}^{2\pi} \int_{-1}^{1} \psi(x, z, \mu, \varphi, t) Y_l^k(\mu, \varphi) \, d\mu d\varphi = \sum_{ij}^{r} X_i(x, z, t) S_{ij}(t) v_{jlk}(t).$$

\subsection{Time evolution and consistency}
The quantities in our low-order system are related to the spherical harmonics moments by
\begin{equation}\label{eq: 2D_QD_unknowns}
\begin{split} 
&\phi(x,z,t) =  2\sqrt{\pi}\psi_0^0(x, z, t),  \\
&J(x,z,t) = \begin{bmatrix}J_z(x, z, t) \\ J_x(x, z, t)
\end{bmatrix} = \begin{bmatrix}
2\sqrt{\frac{\pi}{3}}\psi_1^0(x, z, t)\\-2\sqrt{\frac{2\pi}{3}}\psi_1^1(x, z, t)
\end{bmatrix},\\
&\gamma(x, z, t) = \begin{bmatrix}
2\sqrt{\frac{\pi}{3}}\gamma_z(x, z, t)\\-2\sqrt{\frac{2\pi}{3}}\gamma_x(x, z, t)
\end{bmatrix}.
\end{split}
\end{equation}
We substitute \eqref{eq: 2D_QD_unknowns} into the quasi-diffusion approximation \eqref{eq: ZerothMoment} and \eqref{eq: FirstMoment2} to get an equivalent P$_{1}$ system
\begin{equation}\label{eq: P1_approximation}
\begin{split} 
    &\frac{\partial \phi_0^0} {\partial t} + \sqrt{\frac{1}{3}}\frac{\partial \phi_1^0} {\partial z} - \sqrt{\frac{2}{3}}\frac{\partial \phi_1^1} {\partial x}
 = \left(\sigmas - \sigmat\right)\phi_0^0 + \frac{1}{2\sqrt{\pi}}Q.\\
&\frac{\partial \phi_1^0} {\partial t} + \sqrt{\frac{1}{3}}\frac{\partial \phi_0^0} {\partial z} = -\sigmat \phi_0^0 + \gamma_z\\
&\frac{\partial \phi_1^1} {\partial t} - \sqrt{\frac{1}{6}}\frac{\partial \phi_0^0} {\partial x} = -\sigmat\phi_0^0 + \gamma_x\\
\end{split}.
\end{equation}

In this work, we adopt the forward Euler method for the time integration. At a time step from $t_n$ to $t_{n+1} = t_n + \Delta t$, we calculate the closure terms in \eqref{eq: P1_approximation} through equations 
\begin{equation}\label{eq: gamma}
\begin{split} 
    &\gamma_z \big |_{n} = \frac{\phi_1^0 \big|^{HO}_{n+1}  - \phi_1^0 \big|^{HO}_{n} }{\Delta t}  + \sqrt{\frac{1}{3}}\frac{\partial \phi_0^0 \big|^{HO}_{n}} {\partial z}  + \sigmat \phi_0^0 \big|^{HO}_{n}, \\
&\gamma_x \big |_{n} = \frac{\phi_1^1 \big|^{HO}_{n+1}  - \phi_1^1 \big|^{HO}_{n} }{\Delta t}  - \sqrt{\frac{1}{6}}\frac{\partial \phi_0^0 \big|^{HO}_{n}} {\partial x}  + \sigmat \phi_0^0 \big|^{HO}_{n}.
\end{split}
\end{equation}
where the three angular moments are solved from the high-order system (\ref{eq: K_eq_2} - \ref{eq: L_eq_2}). Then we use the definitions for $\gamma_x \big |_{n}$ and $\gamma_z \big |_{n}$  and solve the following low-order system
\begin{equation}\label{eq: LO_system}
\begin{split} 
    &\frac{\phi_0^0 \big|^{LO}_{n+1}  - \phi_0^0 \big|^{LO}_{n} }{\Delta t} + \sqrt{\frac{1}{3}}\frac{\partial \phi_1^0 \big|^{LO}_{n}} {\partial z} - \sqrt{\frac{2}{3}}\frac{\partial \phi_1^1 \big|^{LO}_{n} } {\partial x}
 = \left(\sigmas - \sigmat\right)\phi_0^0 \big|^{LO}_{n} + \frac{1}{2\sqrt{\pi}}Q \big|^{LO}_{n},\\
&\frac{\phi_1^0 \big|^{LO}_{n+1}  - \phi_1^0 \big|^{LO}_{n} }{\Delta t} + \sqrt{\frac{1}{3}}\frac{\partial \phi_0^0 \big|^{LO}_{n}} {\partial z} = -\sigmat \phi_0^0\big|^{LO}_{n} + \gamma_z\big|_{n},\\
&\frac{\phi_1^1 \big|^{LO}_{n+1}  - \phi_1^1 \big|^{LO}_{n} }{\Delta t} - \sqrt{\frac{1}{6}}\frac{\partial \phi_0^0 \big|^{LO}_{n}} {\partial x} = -\sigmat\phi_0^0\big|^{LO}_{n} + \gamma_x\big|_{n}.
\end{split}
\end{equation}
This formulation is globally conservative.

\subsection{Conservation fix}
Because the DLR algorithm is not conservative (as has been previously discussed), the accuracy of the closure term is limited. To overcome this drawback, we use the low-order results to update the corresponding term within the low-rank solution in every time step. We update the angular bases $W_j$, as written in Eq.~\eqref{eq: 2D_Discretization2} to make the high-order and low-order solution have the same first two moments ($\phi_0^0$, $\phi_1^0$ and $\phi_1^1$). This correction can be computed without forming the full solution, preserving the low-rank representation. The formulation of the correction is given by
\begin{equation}\label{eq: conservation_fix}
\sum_{j=1}^r (K_j^\mathrm{new} - K_j)\begin{bmatrix} v_{j00}\\v_{j10}\\v_{j11}\end{bmatrix} = \begin{bmatrix}\phi_0^0 \big |^{LO} \\
    \phi_1^0 \big |^{LO} \\ \phi_1^0 \big |^{LO}\end{bmatrix} - \begin{bmatrix}\phi_0^0 \big |^{HO} \\
    \phi_1^0 \big |^{HO} \\ \phi_1^1 \big |^{HO}\end{bmatrix},
\end{equation} 
where we keep $W_j$ unchanged and update $X_i$ and $S_{ij}$ with the low-order results. This is a linear matrix equation for the $K_j^\mathrm{new}$ with a solution that can be found by least-squares. We can then factorize $K_j^\mathrm{new}$ into $X_i^\mathrm{new}$ and $S_{ij}^\mathrm{new}$ by QR decomposition, which are initial conditions for the next time step. 
The full-time evolution algorithm is given in Algorithm \ref{alg: HOLO}.

\subsection{Reduction of Memory Requirements}

The feature of memory saving in the DLR method is maintained in the HOLO algorithm. We use $m$ to denote the degrees of freedom in the spatial discretization. The memory footprint required to store the solution in each time step is 
\begin{equation} \label{eq: memory}
\mathrm{Memory} [\mathrm{bytes}] = 8\times2\times(mr + r^2 + nr + 3m).
\end{equation}
The factor of 8 assumes 8 bytes per floating-point number, the factor of two comes from the fact that the previous and current time step needs to be stored, and the $3m$ accounts for the storage of $\phi$ and $J$ at the $m$ spatial degrees of freedom. The memory requirement for the full-rank update requires at least 
\begin{equation} \label{eq: memory_fullrank}
\mathrm{Memory} [\mathrm{bytes}] = 8 \times 2 mn.
\end{equation}
bytes. When $r \ll m,n$ the HOLO method can require much less memory because there no quadratic terms combining $m$ and $n$. 

\begin{algorithm}[ht]
	\textbf{Given}\\
	-initial time $t_0$\\
	-time step $h$\\
	-desired rank $r$\\
	-initial condition $\phi_0(x, z, t_0)$, $\phi_1(x, z, t_0)$\\
	-initial approximations $X_i(x, z, t_0)$, $S_{ij}(t_0), W_j(\mu, \varphi, t_0)$\\
	\textbf{Repeat:}\\
	-$t_1 = t_0 + h$ \\
	-Using low-rank approximation to calculate $\psi_l^k(x, z, t_1) = \sum_{ij}^{r} X_i(x, z, t_1) S_{ij}(t_1) v_{jlk}(t_1)$ \\
	-Calculate the closure $\gamma$ with $\psi_l^k(x,z,t_1)$ using Eq.~\eqref{eq: 2D_QD_unknowns}\\		
	-Solve the LO system Eq.~\eqref{eq: LO_system} for $\phi^{LO}(x, z,t_1)$ and $J^{LO}(x, z,t_1)$ \\
	-Update the low-rank bases $X_i(x, z,t_1)$ and $S_{ij}(t_1)$ with the LO $\phi^{LO}(x, z,t_1)$ and $J^{LO}(x, z,t_1)$ using Eq.~\eqref{eq: conservation_fix}\\
	-$t_0 = t_1$
	\caption{HOLO algorithm}
	\label{alg: HOLO}
\end{algorithm}
\DecMargin{.5em}

\section{Spatial discretization}
\subsection{Low-order system}
The linear conservation form of Eq.~\eqref{eq: LO_system} is
\begin{equation}\label{eq: LO_system_conservationform}
\frac{\partial \vec{u}}{\partial_t} + \vec{A_z}^{LO}\frac{\partial \vec{u}}{\partial z}+ \vec{A_x}^{LO}\frac{\partial \vec{u}}{\partial x} = C \vec{u} + S
\end{equation}
where 
\[ \vec{u} = \begin{bmatrix}
\phi\\J_z\\J_x
\end{bmatrix}, \,\,\, \vec{A_x}^{LO} = \begin{bmatrix}
0&0&-\sqrt{\frac{2}{3}}\\0&0&0\\-\sqrt{\frac{1}{3}}&0&0\end{bmatrix}, \,\,\, \vec{A_z}^{LO} = \begin{bmatrix}
0&\sqrt{\frac{1}{3}}&0\\\sqrt{\frac{1}{3}}&0&0\\0&0&0\end{bmatrix},\]
\[ C = \begin{bmatrix}
\sigmas-\sigmat&0&0\\0&-\sigmat&0\\0&0&-\sigmat
\end{bmatrix}, \,\,\, S = \begin{bmatrix}
\frac{1}{2\sqrt{\pi}}Q\\
\gamma_z\\
\gamma_x\end{bmatrix}.\]

We apply the bilinear discontinuous (BLD) Galerkin finite element method to discretize  Eq. \eqref{eq: LO_system_conservationform} on rectangular cells in $XZ$ geometry. The solution vector $\vec{u}$ on cell $k$ is expanded with basis functions $B_{k, i}(x, z)$
\begin{equation}
    \vec{u_k}(x, z, t) = \sum_i^{4} B_{k, i}(x, z)\vec{u}_{k, i}(t),
\end{equation}
where the basis functions are 
\begin{equation} \label{eq: basis_functions}
    \begin{split}
    B_{k, 1}(x, z) = \frac{x_R - x}{\Delta x_k}\frac{z_T - z}{\Delta z_k},\\
    B_{k, 2}(x, z) = \frac{x - x_L}{\Delta x_k}\frac{z_T - z}{\Delta z_k},\\
    B_{k, 3}(x, z) = \frac{x - x_L}{\Delta x_k}\frac{z - z_B}{\Delta z_k},\\
    B_{k, 4}(x, z) = \frac{x_R - x}{\Delta x_k}\frac{z - z_B}{\Delta z_k}.
    \end{split}
\end{equation}
The weak form of \eqref{eq: LO_system_conservationform} is obtained by multiplying with the basis function and integrating over cell
\begin{equation}
\begin{split}
    \frac{d}{dt}\int_{z_B}^{z^T}dz\int_{x_L}^{x^R}dx\, B_{k, i} \vec{u_k} + \vec{A_x} \int_{z_B}^{z^T}dz\int_{x_L}^{x^R}dx \, B_{k, i} \frac{d}{d_x}\vec{u_k}\\ +  \vec{A_z} \int_{z_B}^{z^T}dz\int_{x_L}^{x^R}dx \, B_{k, i} \frac{d}{d_z}\vec{u_k} = \\
    C \int_{z_B}^{z^T}dz   \int_{x_L}^{x^R}dx \, B_{k, i} \vec{u_k} + \int_{z_B}^{z^T}dz\int_{x_L}^{x^R} & dx \, B_{k, i} \vec{S_k}.
\end{split}
\end{equation}
By integrating by parts, the stream terms can be written as 
\begin{equation*}
    \vec{A_x} \int_{z_B}^{z^T}dz\int_{x_L}^{x^R}dx \, B_{k, i} \frac{d}{d_x}\vec{u_k} = \vec{A_x}
    \left(\int_{z_B}^{z^T}dz \, B_{k, i} \vec{u_k} - \int_{z_B}^{z^T}dz\int_{x_L}^{x^R}dx \, \frac{d}{d_x}B_{k, i} \vec{u_k} \right),
\end{equation*}
and
\begin{equation*}
    \vec{A_z} \int_{z_B}^{z^T}dz\int_{x_L}^{x^R}dx \, B_{k, i} \frac{d}{d_z}\vec{u_k} = \vec{A_z}
    \left(\int_{x_L}^{x^R}dx \, B_{k, i} \vec{u_k} - \int_{z_B}^{z^T}dz\int_{x_L}^{x^R}dx \, \frac{d}{d_z}B_{k, i} \vec{u_k} \right).
\end{equation*}
Then we collect the vector $\vec{u}_k = [u_{k,1},  u_{k,2}, u_{k,3}, u_{k,4}]^T$, $\vec{B}_k = [B_{k,1}, B_{k,2}, B_{k,3}, B_{k,4}]^T$ and $\vec{S}_k = [S_{k,1}, S_{k,2}, S_{k,3}, S_{k,4}]^T$ to get a $4$ equation system for cell $k$
\begin{equation*}
    M\frac{d\vec{u}_k}{dt} + \vec{A_x} \left(( L\vec{u}_k)^{x, surf} - L_x\vec{u}_k \right) + \vec{A_z} \left(( L\vec{u}_k)^{z, surf} - L_z\vec{u}_k \right) = CM\vec{u}_k + M\vec{S}_k,
\end{equation*}
where 
\begin{equation*}
    M = \int_{z_B}^{z^T}dz\int_{x_L}^{x^R}dx \, \vec{B}_k \vec{B}_k^T = \frac{\Delta x_k \Delta z_k}{36} \begin{bmatrix} 4&2&1&2\\
    2&4&2&1\\
    1&2&4&2\\
    2&1&2&4
    \end{bmatrix},
\end{equation*}

\begin{equation*}
    L_x = -\int_{z_B}^{z^T}dz\int_{x_L}^{x^R}dx \, \frac{\partial \vec{B}_k}{\partial_x} \vec{B}_k^T = -\frac{\Delta z_k}{12} \begin{bmatrix} -2&-2&-1&-1\\
    2&2&1&1\\
    1&1&2&2\\
    -1&-1&-2&-2
    \end{bmatrix},
\end{equation*}

\begin{equation*}
    L_z = -\int_{z_B}^{z^T}dz\int_{x_L}^{x^R}dx \, \frac{\partial \vec{B}_k}{\partial_z} \vec{B}_k^T = -\frac{\Delta x_k}{12} \begin{bmatrix} -2&-1&-1&-2\\
    -1&-2&-2&-1\\
    1&2&2&1\\
    2&1&1&2
    \end{bmatrix},
\end{equation*}

\begin{equation*}
    (L\vec{u}_k)^{x,\mathrm{surf}} = \int_{z_B}^{z^T}dz \, \vec{B}_k (\vec{B}_k \vec{u})^T = \frac{\Delta z_k}{6} \begin{bmatrix}
    -2u_{k,1}^{x-} - 2u_{k,4}^{x-}\\
    2u_{k,2}^{x+} + u_{k,3}^{x+}\\
    u_{k,2}^{x+} + 2u_{k,3}^{x+}\\
    -u_{k,1}^{x-} - 2u_{k,4}^{x-}
    \end{bmatrix},
\end{equation*}
and
\begin{equation*}
    (L\vec{u}_k)^{z,\mathrm{surf}} = \int_{x_L}^{x^R}dx \, \vec{B}_k (\vec{B}_k \vec{u})^T = \frac{\Delta x_k}{6} \begin{bmatrix}
    -2u_{k,1}^{z-} - u_{k,2}^{z-}\\
    -u_{k,1}^{z-} - 2u_{k,2}^{z-}\\
    2u_{k,3}^{z+} + u_{k,4}^{z+}\\
    u_{k,3}^{z+} + 2u_{k,4}^{z+}
    \end{bmatrix}.
\end{equation*}
The superscripts indicate that the value is evaluated in the boundary, e.g., 
$u^{x-}_{k,1}$ is the value in the left edge of support node $1$ in cell ${k}$. We apply the mass-matrix lumping, surface lumping and within-cell gradient term lumping techniques via the formulas: 
$$M^{lump}_{ij} = \delta_{ij} \sum_{j'=1}^{4}M_{ij'},$$ 
$$(L\vec{u}_k)^{\xi,\mathrm{surf,lump}}_{i,j} = \delta_{ij} \sum_{i=1}^{4}(L\vec{u}_k)^{\xi,\mathrm{surf}},$$and 
$$L_{\xi}^\mathrm{lump} = \delta_{ij} \sum_{i=1}^{4}L_{ij}.$$ 

The fully lumped BLD equation is 
\begin{multline}\label{eq: LO_lumped_BLD}
    \frac{d\vec{u}_k}{dt} + 2\frac{\vec{A_x}}{\Delta_x}\begin{bmatrix}
    -u_{k,1}^{x-}\\
    u_{k,2}^{x+}\\
    u_{k,3}^{x+}\\
    -u_{k,4}^{x-}
\end{bmatrix} + \frac{\vec{A_x}}{\Delta_x} \begin{bmatrix}     
    1&1&0&0\\
    -1&-1&0&0\\
    0&0&-1&-1\\
    0&0&1&1
    \end{bmatrix}\vec{u} \\
    + 2\frac{\vec{A_z}}{\Delta_z}\begin{bmatrix}
    -u_{k,1}^{z-}\\
    -u_{k,2}^{z-}\\
    u_{k,3}^{z+}\\
    u_{k,4}^{z+}
\end{bmatrix} + \frac{\vec{A_z}}{\Delta_z} \begin{bmatrix}     
    1&0&0&1\\
    0&1&1&0\\
    0&-1&-1&0\\
    -1&0&0&-1
    \end{bmatrix}\vec{u} = C\vec{u}_k + S.
\end{multline}

\subsection{High-order system}
We apply a similar BLD method to construct the spatial basis $X_i(x, z, t)$, that is, $$X_i(x, z, t) = \sum_k\sum_{q=1}^4 Z_{k, q}(x, z) U_{k, q, i}(t)$$
where $Z_{k, q}(x, z)$ is the basis function with the support nodes $q$ in cell $k$. Note that $X_i$ is orthogonal and $ U_{k, q, i} U_{k', q', i'} = \delta_{kk'} \delta_{qq'}\delta_{ii'}$ is achieved by SVD or QR decomposition during our low-rank calculations. To impose the constraints to basis functions $Z_{k, q}$  
\begin{equation}
    \int_{z_B}^{z^T}dz\int_{x_L}^{x^R}dx Z_{k, q} Z_{k', q'} = \delta_{kk'} \delta_{qq'},
\end{equation}
we can use the normalized basis functions $\frac{2}{\sqrt{\Delta x_k \Delta z_k}}B_{k,i}$ in each cell: 
\begin{equation} \label{eq: HO_basis_functions}
    \begin{split}
    Z_{k, 1}(x, z) = \frac{2}{\sqrt{\Delta x_k \Delta z_k}}\frac{x_R - x}{\Delta x_k}\frac{z_T - z}{\Delta z_k},\\
    Z_{k, 2}(x, z) = \frac{2}{\sqrt{\Delta x_k \Delta z_k}}\frac{x - x_L}{\Delta x_k}\frac{z_T - z}{\Delta z_k},\\
    Z_{k, 3}(x, z) = \frac{2}{\sqrt{\Delta x_k \Delta z_k}}\frac{x - x_L}{\Delta x_k}\frac{z - z_B}{\Delta z_k},\\
    Z_{k, 4}(x, z) = \frac{2}{\sqrt{\Delta x_k \Delta z_k}}\frac{x_R - x}{\Delta x_k}\frac{z - z_B}{\Delta z_k},
    \end{split}
\end{equation}
because $\int_{z_B}^{z^T}dz\int_{x_L}^{x^R}dx \, {B}_{k,i} {B}_{k,j} = 4\Delta x \Delta z \, \delta_{ij}$. 

Equation \eqref{eq: K_eq_2} can be written in a conservation form 
\begin{equation}\label{eq: K_eq_2_conservation}
\frac{\partial \vec{u}}{\partial t} + \vec{A_z}^{HO}\frac{\partial \vec{u}}{\partial z}+ \vec{A_x}^{HO}\frac{\partial \vec{u}}{\partial x} = S(\vec{u})
\end{equation}
where $\vec{u} = [K_1, K_2, ... K_r]^T$, $V = [V_1, V_2, ... V_r]$, $\vec{A_x}^{HO} = V^T \vec{A_x} V$, $\vec{A_z}^{HO} = V^T \vec{A_z} V$, $S(\vec{u}) = -\sigmat K + \frac{1}{4\pi}V^T \langle Y\rangle \langle Y \rangle^T V \sigmas K + \frac{1}{2 \sqrt{\pi}}Q$. 

There is no need to develop the weak form for Eqs.~\eqref{eq: S_eq_2} and \eqref{eq: L_eq_2} but we still need to calculate spatial integration terms like $\langle \partial_{\xi} X_p X_q \rangle_{\vec{r}}$ where $\xi = x$ or $z$:
\begin{equation}
\begin{split}
    \langle \partial_{\xi} X_p X_q \rangle_{\vec{r}} &= \int_{z_B}^{z^T}dz\int_{x_L}^{x^R}dx \, \partial_{\xi}\left ( \sum_k\sum_i^4 Z_{k,i}u_{k,i,p}\right)  \sum_{k'}\sum_{i'}^4 Z_{k',i'}u_{k',i',q}\\
    &=\sum_{k,k'} \sum_{i,i'} u_{k,i,p} u_{k',i',q} \int_{z_B}^{z^T}dz\int_{x_L}^{x^R}dx\, \partial_{\xi}Z_{k,i}Z_{k',i'}\\
    &= L_{\xi} \begin{bmatrix}
    \sum_{k,k'}u_{k,p,1}u_{k,q,1}\\
    \sum_{k,k'}u_{k,p,2}u_{k,q,2}\\
    \sum_{k,k'}u_{k,p,3}u_{k,q,3}\\
    \sum_{k,k'}u_{k,p,4}u_{k,q,4}
    \end{bmatrix},
\end{split}
\end{equation}
where $L_{\xi}$ is the lumped matrix from Eq.~\eqref{eq: LO_lumped_BLD}. 

\subsection{Slope limiter}
Here we briefly describe the slope limiter in our scheme to address artificial oscillations. First, we define $s_k^x$ as the slope in $x$ direction and $s_k^z$ in $z$ direction within each cell. Then we 
calculate the cell average 
$$\bar{u}_k = \frac{1}{4}(u_{k,1}, u_{k,2}, u_{k,3}, u_{k,4})$$
The neighbors of the cell $k$ in the left, right, top and bottom are defined as $\bar{u}_k^R$, $\bar{u}_k^L$, $\bar{u}_k^T$, $\bar{u}_k^B$, respectively. 
The double minmod limiter which preserves the diffusion limit \cite{mcclarren2008effects,smedley2015asymptotic} is
\begin{equation}
    s_k^x = \rm{minmod}\left(\frac{1}{2}(u_{k,2}+u_{k,3}-u_{k,1}-u_{k,4}), \rm{minmod}(\bar{u}_k^R - \bar{u}_k, \bar{u}_k - \bar{u}_k^L)\right)
\end{equation}

\begin{equation}
    s_k^z = \rm{minmod}\left(\frac{1}{2}(u_{k,3}+u_{k,4}-u_{k,1}-u_{k,2}), \rm{minmod}(\bar{u}_k^T - \bar{u}_k, \bar{u}_k - \bar{u}_k^B)\right)
\end{equation}
where the minmod operation is given by
\begin{equation}
\rm{minmod}(a, b)=
\begin{cases}
\begin{aligned}
a \,\,\,\,\, & \, |a|<|b| \,\,  \& \,\, ab>0,  \\
b \,\,\,\,\, & \, |a|>|b| \,\,  \& \,\, ab>0,   \\
0 \,\,\,\,\, & \, ab<0
\end{aligned}
\end{cases}.
\end{equation}
From the limiter, we change the value of the four support nodes to be 
\begin{equation} \label{eq: reconstructedslope}
    \begin{split}
    \Tilde{u}_{k,1} = \bar{u}_{k} - \frac{1}{2}s_k^x - \frac{1}{2}s_k^z,\\
    \Tilde{u}_{k,2} = \bar{u}_{k} + \frac{1}{2}s_k^x - \frac{1}{2}s_k^z,\\
    \Tilde{u}_{k,3} = \bar{u}_{k} + \frac{1}{2}s_k^x + \frac{1}{2}s_k^z,\\
    \Tilde{u}_{k,4} = \bar{u}_{k} - \frac{1}{2}s_k^x + \frac{1}{2}s_k^z.
    \end{split}
\end{equation}

\section{Numerical Results}
We demonstrate the accuracy and the computational efficiency of our HOLO algorithm with six benchmark problems. The plane source problem emphasizes the conservation fix and the modified Reed's problem highlights the diffusion limit in a heterogeneous problem \cite{Ryan_diff}. Another four 2D problems show the memory-saving feature and the benefits of high angular resolution. In all simulations the unit of length is $\mathrm{cm}$ and the particle speed is set to be $1 \ \mathrm{cm/s}$. We implement the double minmod limiter in the modified Reed's problem and the line source problem and the minmod limiter for other problems.
  
\subsection{Plane source problem}
The plane source problem \cite{ganapol1977generation,ganapol2001homogeneous,Ganapol2008} has been used to test a variety of radiation transport methods. It describes an initial pulse of particles emitted in an infinite medium with no source and absorption, which means the total number of particles is fixed during the evolution. The main purpose of this test is to show that our HOLO algorithm is conservative without loss of computational efficiency. In the  problem the initial condition is given by a Dirac-delta function placed in the center of a purely scattering media, where $\phi(x, 0) = \delta(x)$, $\sigmat = \sigmas = 1$. In all simulations, we fix the spatial resolution to $\Delta x = 0.02$, and the Courant–Friedrichs–Lewy (CFL) condition $\textrm{CFL} = \frac{c \Delta t}{\Delta x}$ to $0.2$, where $\Delta t$ is the time step and the particle speed $c$ is set to 1. We compare the numerical solutions with the analytical benchmark given by Ganapol. Note that in 1D problems a full rank solution has $r_\mathrm{F} = N + 1$ where $N$ is the order of the spherical harmonics. 

\begin{figure}[h!] 
\begin{minipage}{0.5\textwidth}
  \centering
\includegraphics[width=\textwidth]{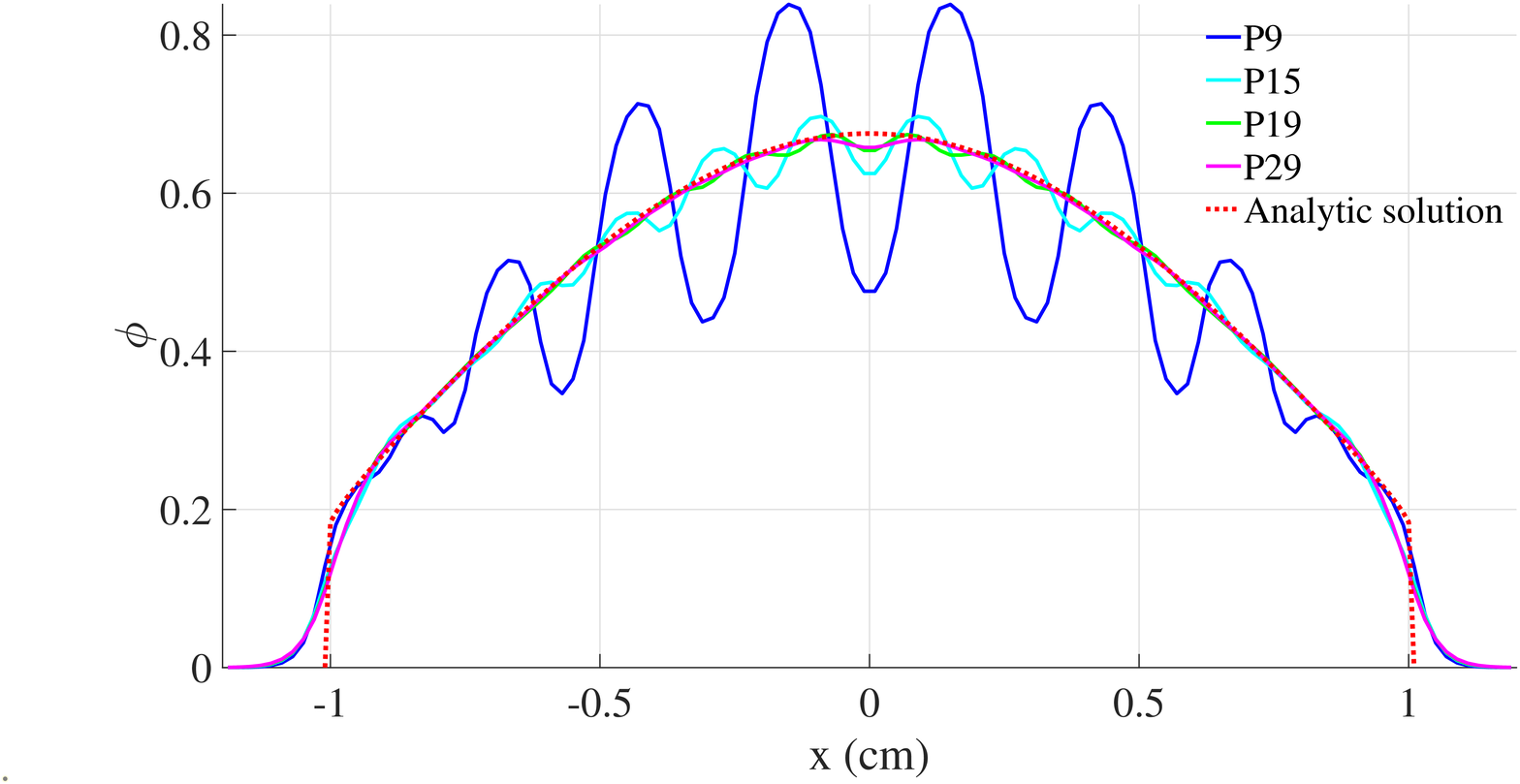}
\subcaption{$t = 1$, $N_x = 120$}
\label{fig: PP_classical_a}
\end{minipage}%
\begin{minipage}{0.5\textwidth}
  \centering
\includegraphics[width=\textwidth]{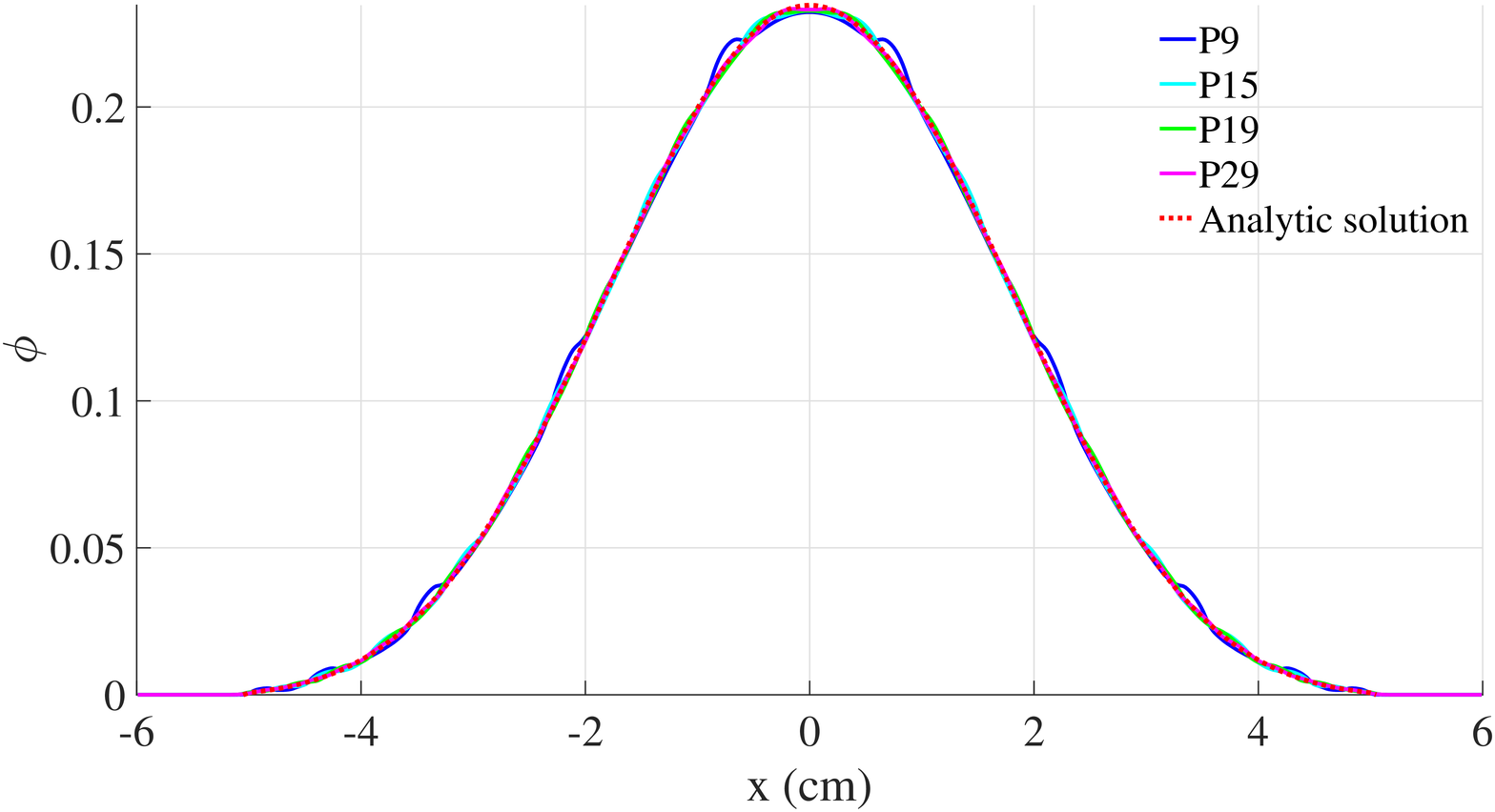}
\subcaption{$t = 5$, $N_x = 600$}\label{fig: PP_classical_b}
\end{minipage}%
\caption{The scalar flux $\phi$ of the plane source problem calculated with $P_7$, $P_{19}$ and $P_{29}$ expansions are compared to the benchmark solution. No rank reduction is performed here.}
\label{fig: PP_classical_results}
\end{figure}

The first set of simulations is designed to reveal what order of ~\PN~ is sufficient by comparing with results calculated by the classical full rank method. The plane source problem is considered as a difficult test 
because of the unavoidable oscillations. As shown in Figure \ref{fig: PP_classical_a}, the magnitude of spikes in the $P_9$ solution is much higher than $P_{29}$. These spikes contain uncollided particles moving at the characteristic speeds of the \PN ~equations. This figure also demonstrates the requirement of the high \PN~order in this case because even P$_{29}$ has extant oscillations at the early time. From Figure \ref{fig: PP_classical_b}, we can see that the order requirement is lower at a later time, and even $P_9$ is sufficient to capture the analytical solution. That is because there are few uncollided particles remaining at this late time. 

\begin{figure}[h!] 
\begin{minipage}{0.5\textwidth}
  \centering
\includegraphics[width=\textwidth]{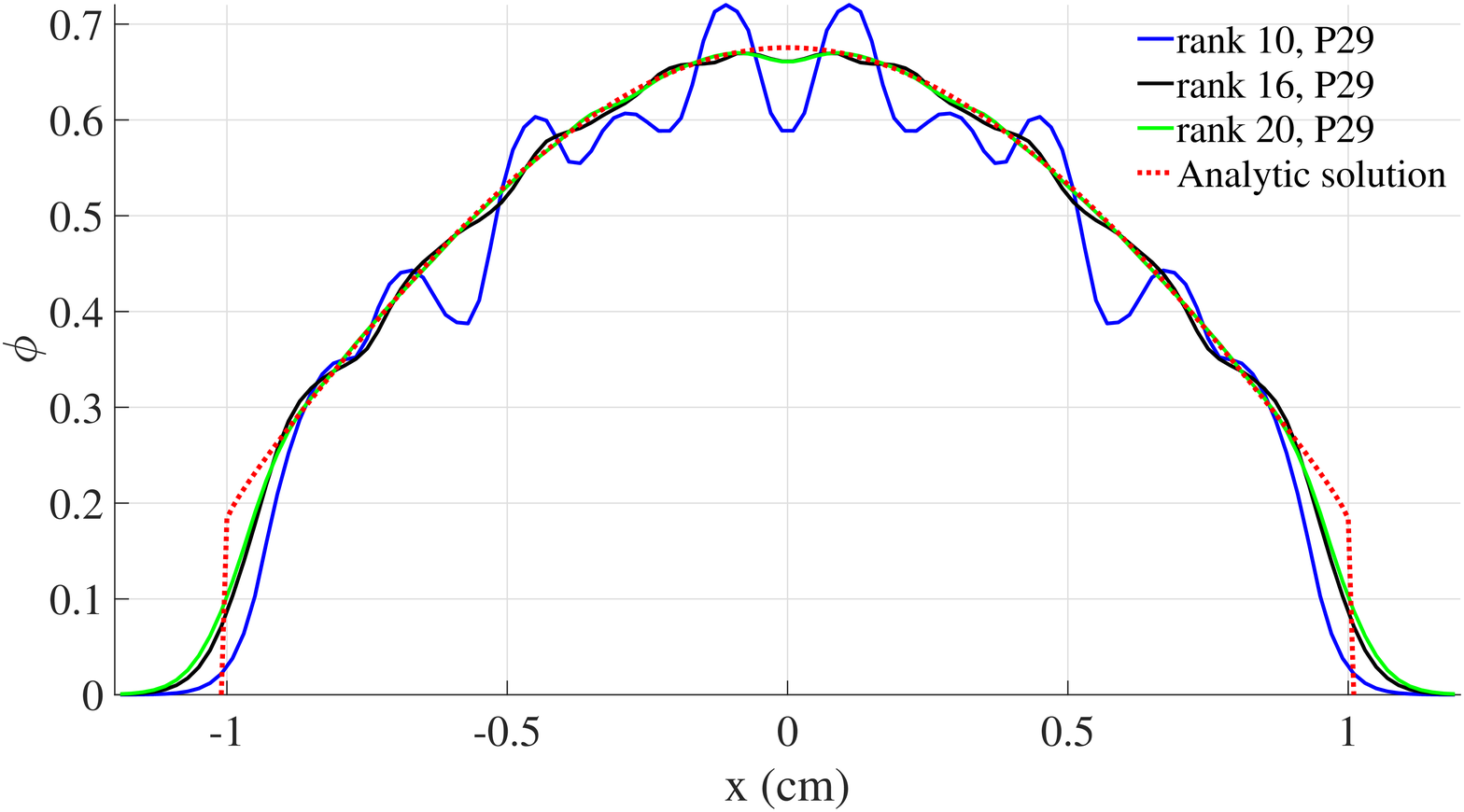}
\subcaption{$t = 1$, $N_x = 120$}\label{fig: PP_lowrank_a}
\end{minipage}%
\begin{minipage}{0.5\textwidth}
  \centering
\includegraphics[width=\textwidth]{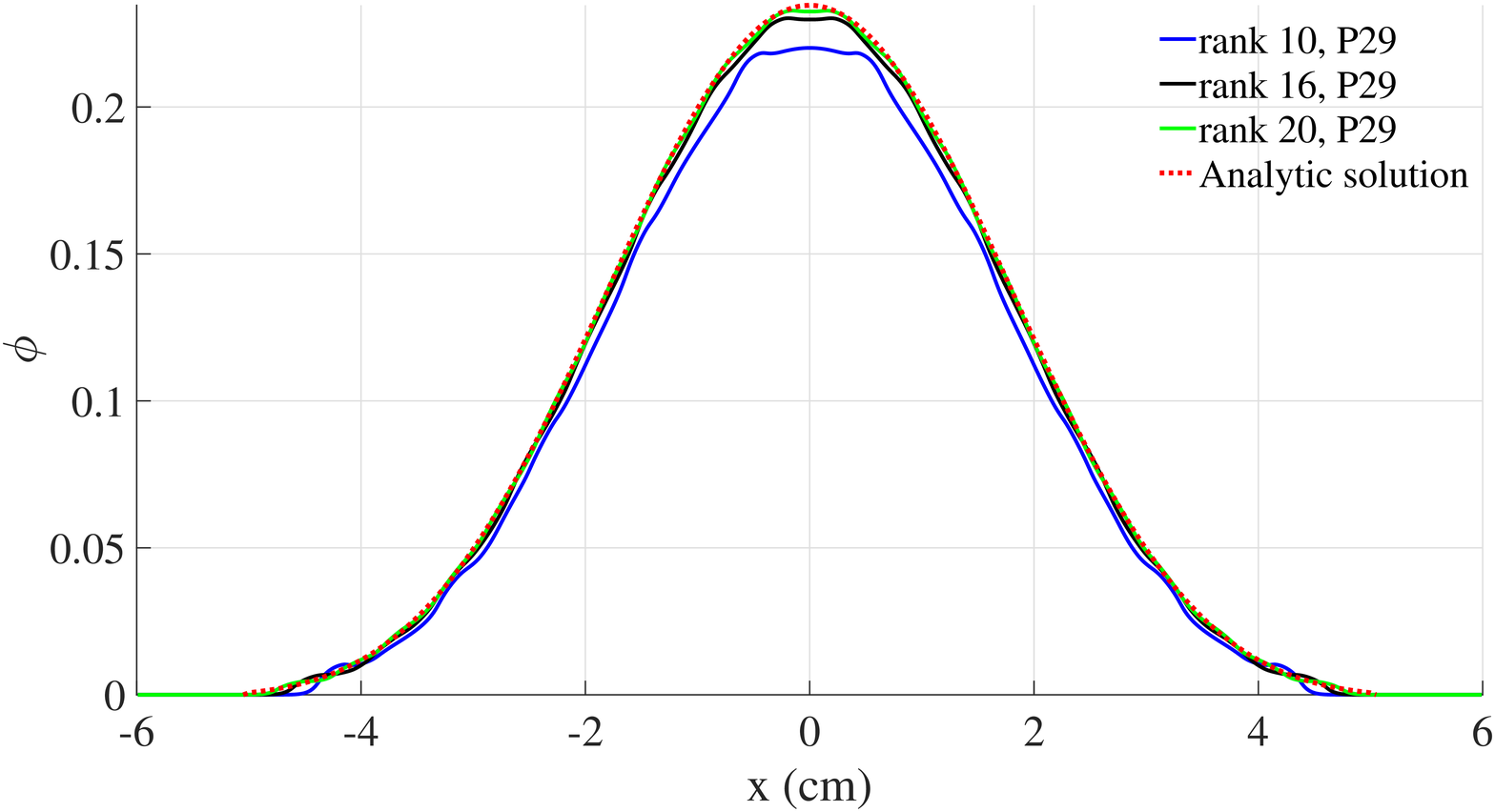}
\subcaption{$t = 5$, $N_x = 600$}\label{fig: PP_lowrank_b}
\end{minipage}%
\caption{The scalar flux $\phi$ of the plane source problem calculated with $P_{29}$ expansion and the standard, non-conservative DLR with ranks $16$, $20$ and $24$ are compared to the benchmark solution.}
\label{fig: PP_lowrank_results}
\end{figure}

Figure \ref{fig: PP_lowrank_results} presents the $P_{29}$ solutions with the low-rank method. 
Here the $P_{29}$ solution with rank 20, which is two-thirds of the full rank, is comparable in accuracy, as shown in Figure \ref{fig: PP_lowrank_a} and Figure \ref{fig: PP_lowrank_b}. We point out that this indicates a memory reduction of roughly one-third of full rank memory. However, the conservation loss of the low-rank method is exhibited in Figure \ref{fig: PP_lowrank_b}. Even though the solution with rank 10 approximates the shape of the true solution well, the area below the solution curve is lower than either the analytical or the numerical solutions with higher ranks due to the loss of conservation. 

\begin{figure}[h!] 
\begin{minipage}{0.5\textwidth}
  \centering
\includegraphics[width=\textwidth]{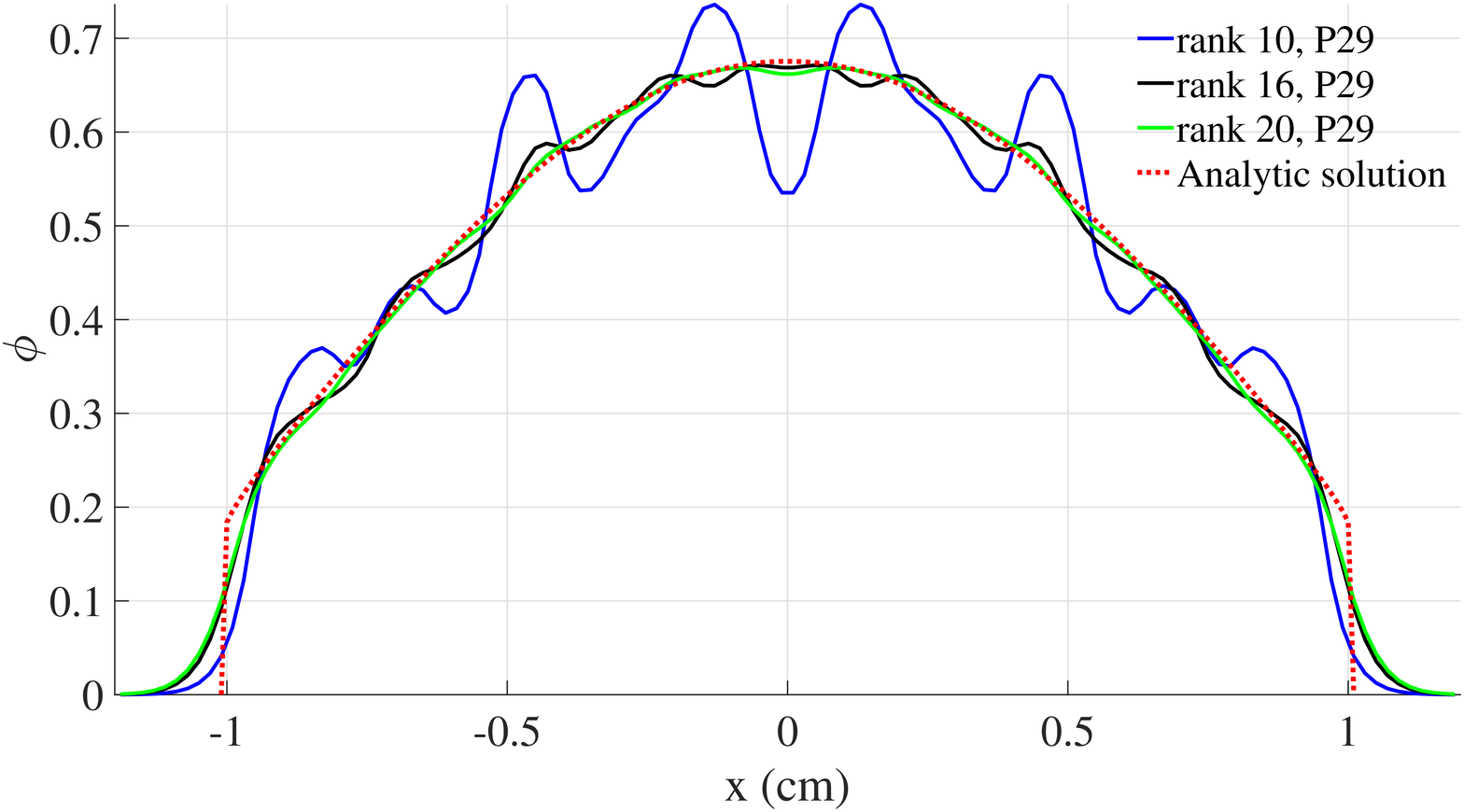}
\subcaption{$t = 1$, $N_x = 120$}\label{fig: PP_HOLO_a}
\end{minipage}%
\begin{minipage}{0.5\textwidth}
  \centering
\includegraphics[width=\textwidth]{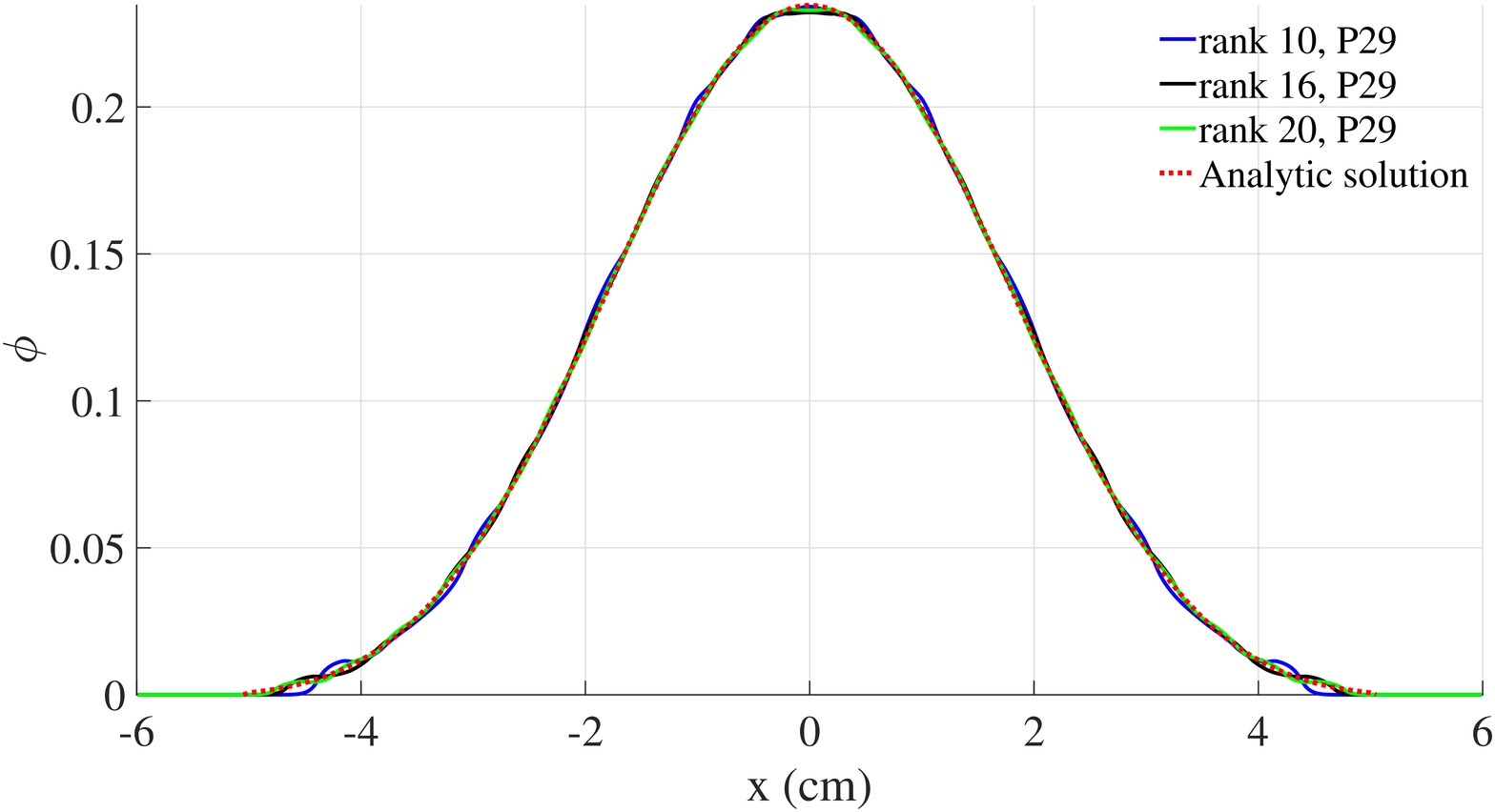}
\subcaption{$t = 5$, $N_x = 600$}\label{fig: PP_HOLO_b}
\end{minipage}%
\caption{The scalar flux $\phi$ of the plane source problem calculated by HOLO with a $P_{29}$ expansion and ranks $16$, $20$ and $24$  compared to the benchmark solution.}
\label{fig: PP_HOLO_results}
\end{figure}

This issue is solved with the HOLO algorithm. As we can see from Figure \ref{fig: PP_HOLO_b}, the $P_{29}$ solution with rank 16 is no longer lower than the analytical solution, and it is also a good approximation compared to the full rank $P_{29}$ solution. Figure \ref{fig: PP_HOLO_a} shows that a solution with rank 20 matches the analytical solution well. In this case the memory usage calculated by \eqref{eq: memory} is $0.185$ MB while the full rank memory is $0.230$ MB by \eqref{eq: memory_fullrank}, corresponding to a 20\% memory savings. Though this is a modest reduction in memory, as we will see, as the number of spatial dimensions increases, the memory reduction will increase.

\subsection{Modified Reed's problem}
\begin{figure}[h!]
	\centering
	\includegraphics[scale=0.22]{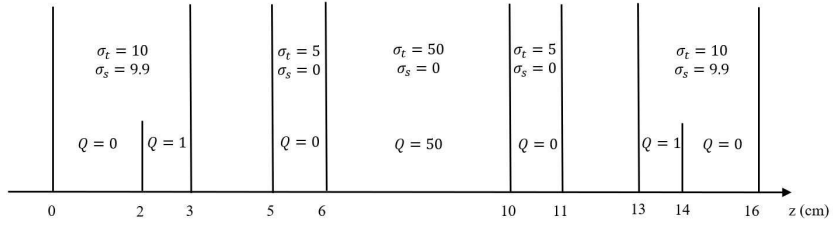}
	\caption{The material layout in Reed's problem where the blank zones are vacuum.}
	\label{fig: Reeds_layout}
\end{figure}

\begin{figure}[h!]
	\centering
	\includegraphics[scale=0.2]{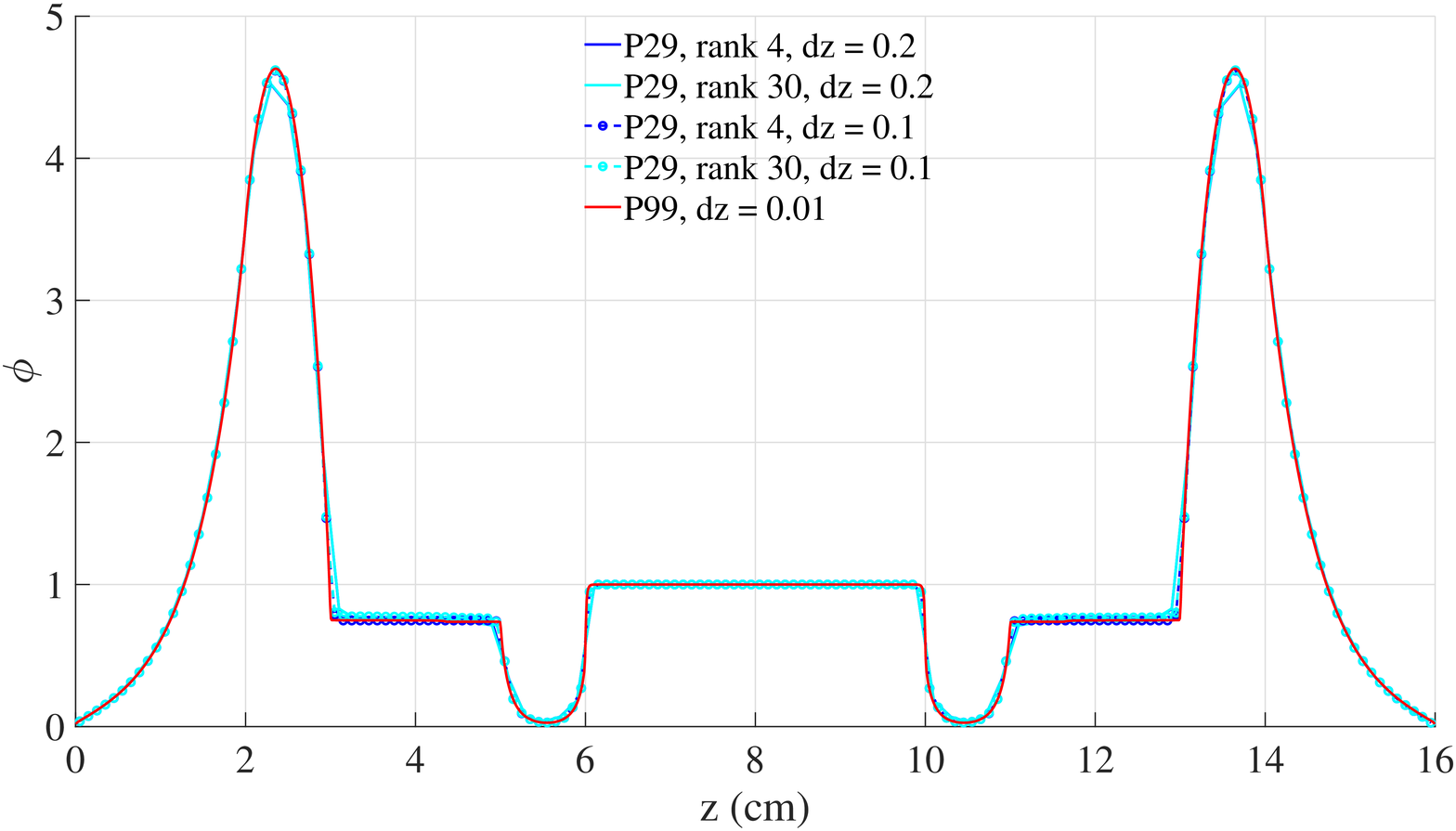}
	\caption{The scalar flux $\phi$ of the modified Reed's problem calculated by HOLO with different rank and spatial resolutions are compared to the high fidelity benchmark.}
	\label{fig: Reeds_results}
\end{figure}

The second test is a multi-material problem that aims to verify that our numerical scheme preserves the diffusion limit. The material layout is detailed in Figure \ref{fig: Reeds_layout}. Note that there are highly scattering regions, where the mean-free-path in these regions is 0.1 ($\sigmat = 10$, $\sigmas = 9.9$). A high-resolution and high-order full rank solution with $\Delta x = 0.01$, $P_{99}$, CFL$ = 0.05$ is used as a benchmark. We then compare our HOLO solutions with different spatial resolutions and rank to the benchmark at $t = 100$, which is very near steady-state. One important finding from Figure \ref{fig: Reeds_results} is that the solutions are not sensitive to the grid size. In strong scattering regions, all solutions match the benchmark well even with the gird size larger than the mean-free-path. We also notice that the full rank and low-rank solution with the same spatial resolution are identical on the scale of the figure. 

\subsection{Line source problem}
The line source problem is a two-dimensional pulsed source problem. Similar to its 1D version, the initial condition is given by $\psi(x, z, 0) = \delta(x)\delta(z)$, and the total and scattering cross-section are set to 1. We compute the HOLO solution to $t = 1$ in the computational domain $2.4 \times 2.4$ with spatial resolution $\Delta x = \Delta z = 0.02$ and CFL condition $0.2$.  Note that the total number of spatial degrees of freedom is fixed to be $m = 4\ N_x N_z =  57600$, which is far larger than the  number of angular bases. Thus the total memory usage is nearly independent of the order of the \PN~expansion and is a stronger function of the rank, e.g, Memory (bytes) $= 8 \times 2 \times (mr + nr + r^2 + 3m) \approx 16(r + 3)m$. 

\begin{figure}[h!] 
\begin{minipage}[b]{0.5\textwidth}
  \centering
\includegraphics[width=\textwidth]{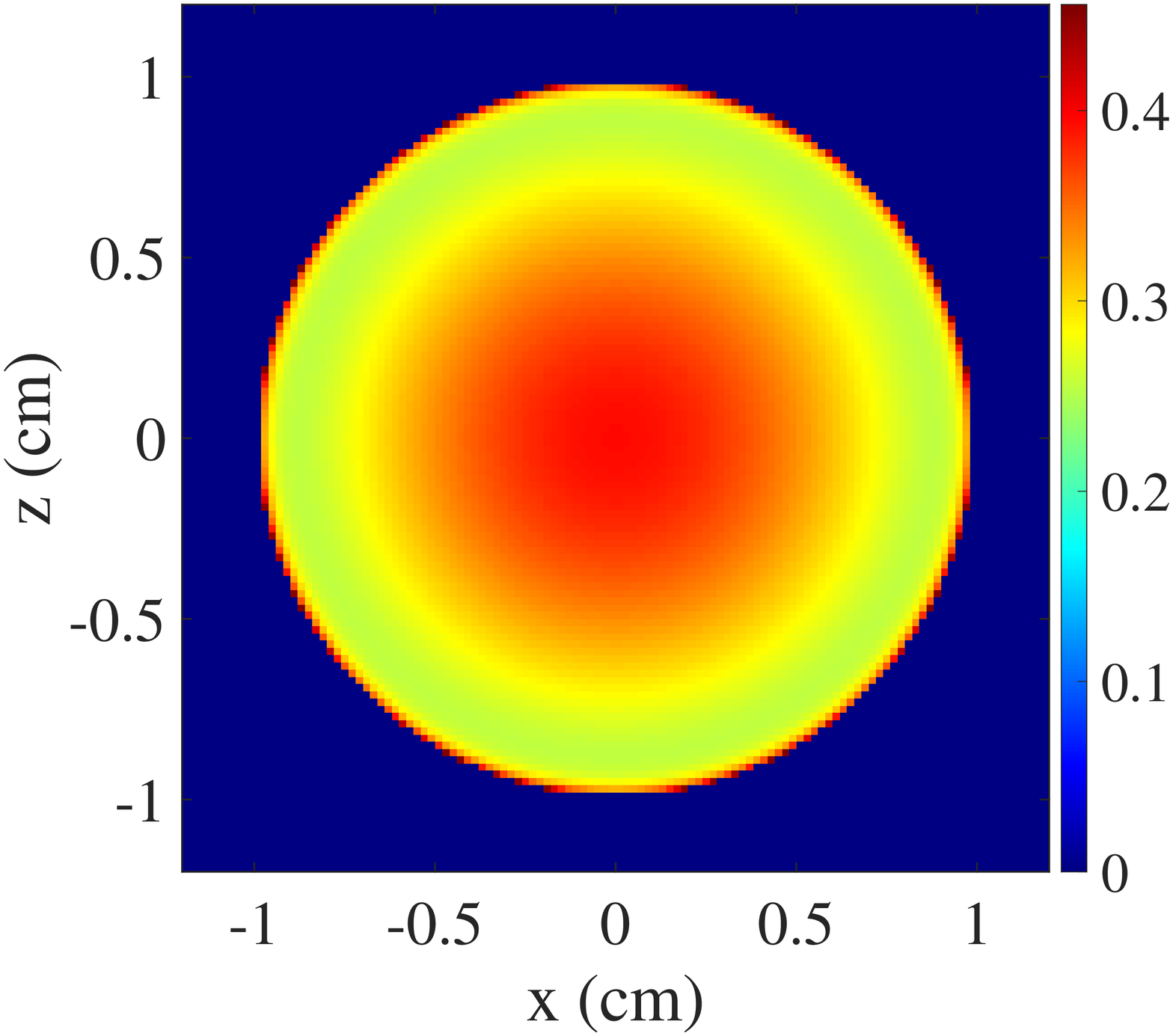}
\subcaption{Analytic solution}\label{fig: LS_HOLO_a}
\strut\end{minipage}%
\hfill\allowbreak%
\begin{minipage}[b]{0.5\textwidth}
  \centering
\includegraphics[width=\textwidth]{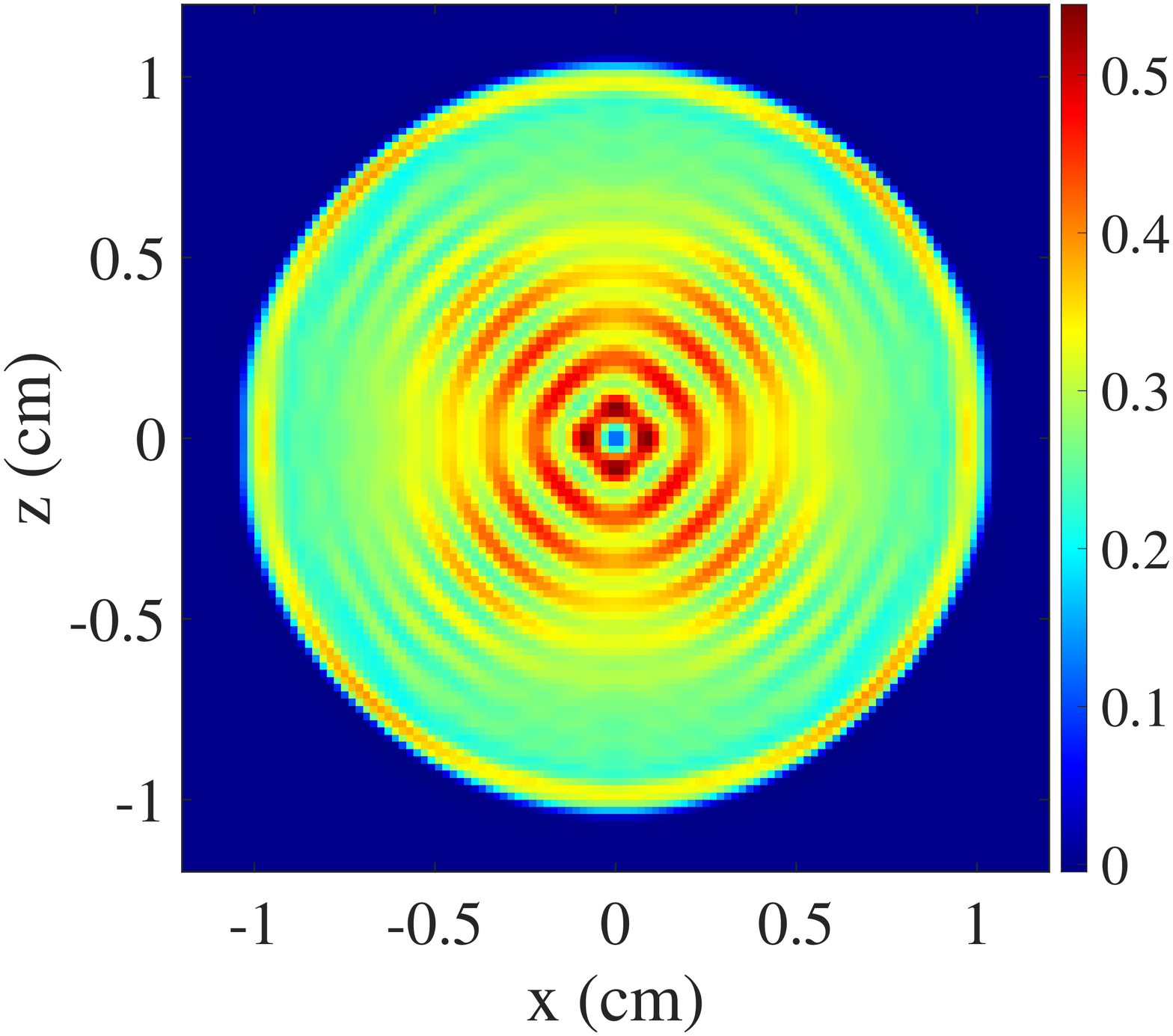}
\subcaption{$P_{23}$, rank 300}\label{fig: LS_HOLO_b}
\strut\end{minipage}%
\hfill\allowbreak%
\begin{minipage}[b]{0.5\textwidth}
  \centering
\includegraphics[width=\textwidth]{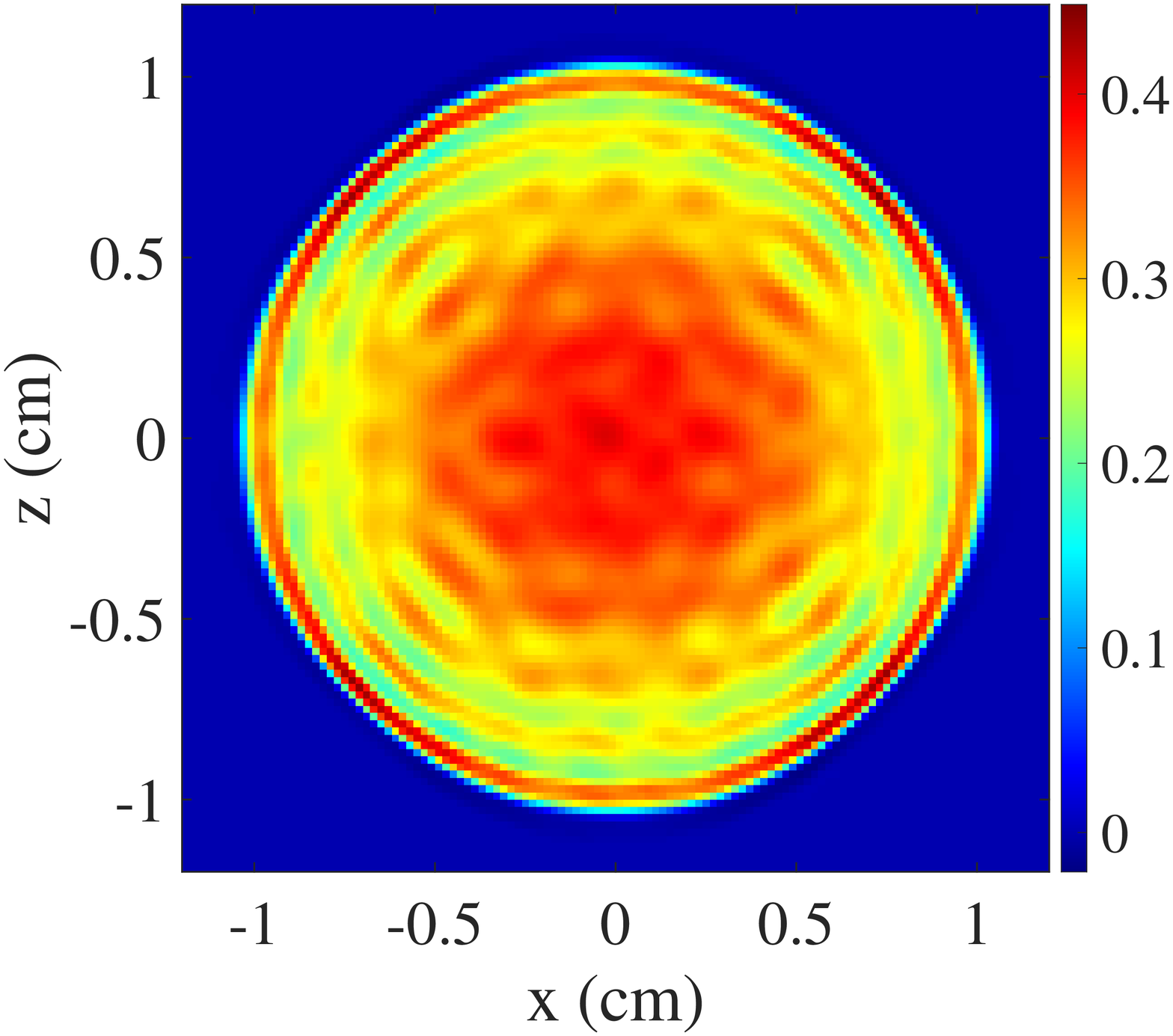}
\subcaption{$P_{39}$, rank 300}\label{fig: LS_HOLO_c}
\strut\end{minipage}%
\hfill\allowbreak%
\begin{minipage}[b]{0.5\textwidth}
  \centering
\includegraphics[width=\textwidth]{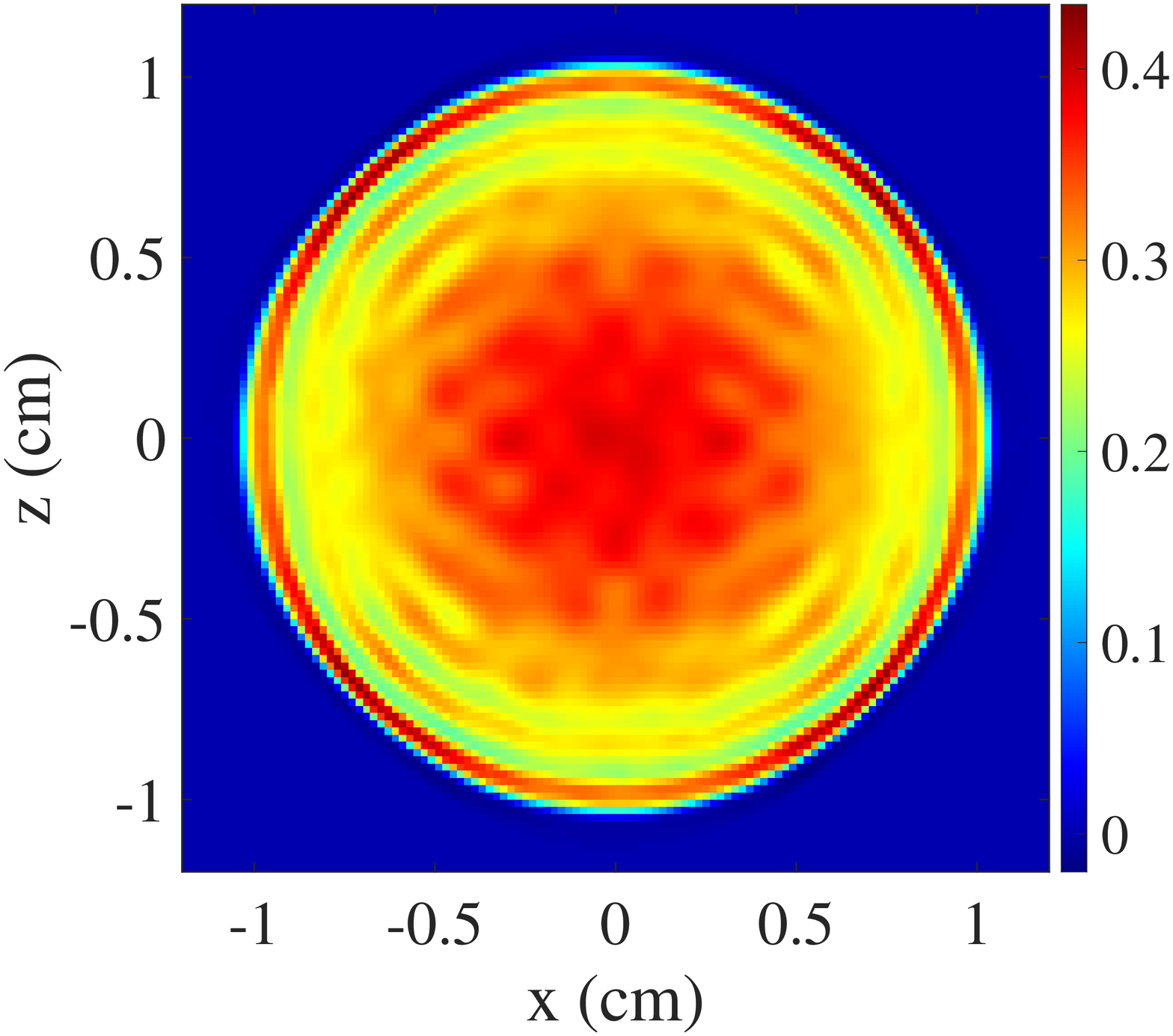}
\subcaption{$P_{59}$, rank 300}\label{fig: LS_HOLO_d}
\strut\end{minipage}%
\hfill\allowbreak%
\begin{minipage}[b]{0.5\textwidth}
  \centering
\includegraphics[width=\textwidth]{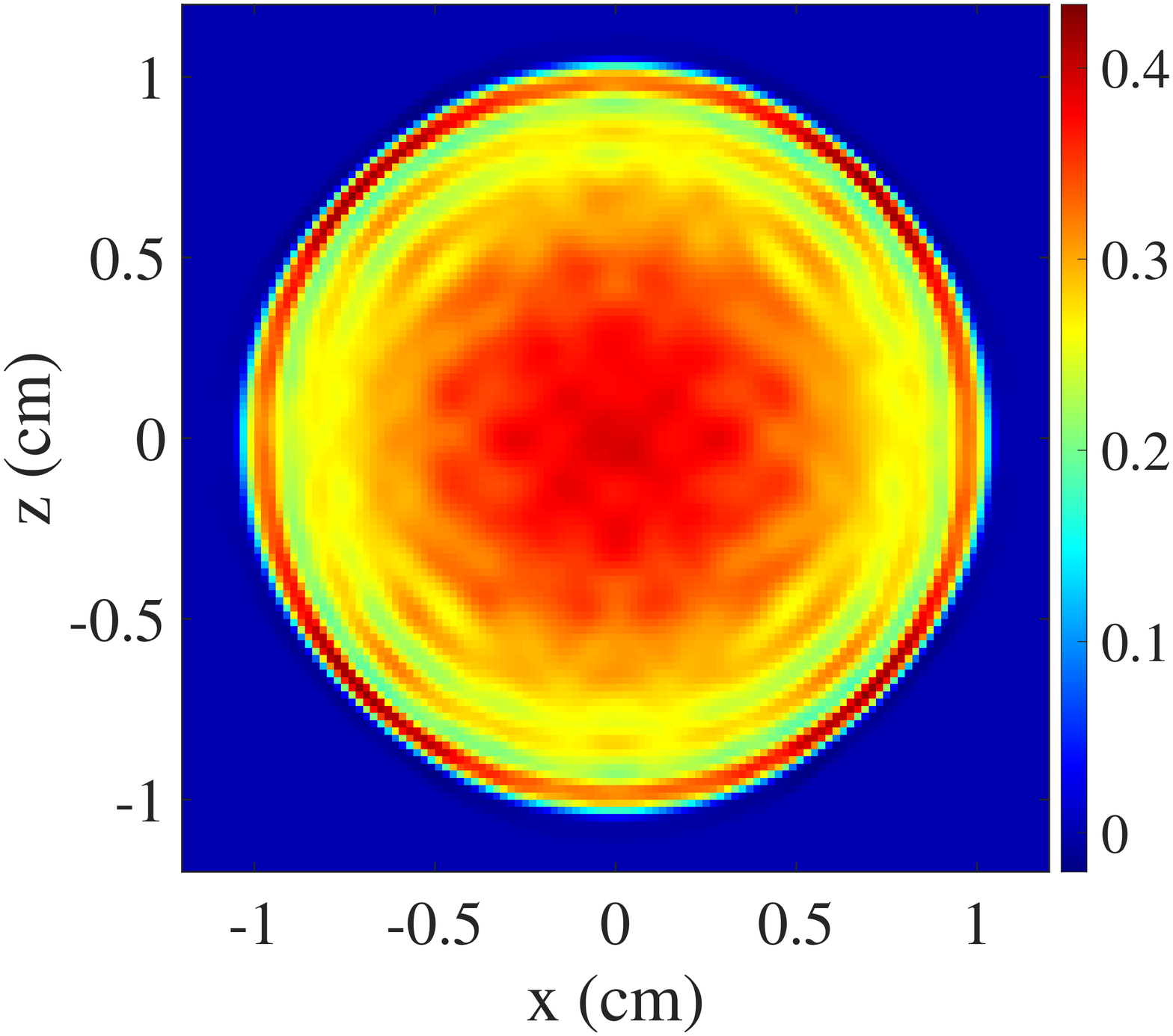}
\subcaption{$P_{99}$, rank 300}\label{fig: LS_HOLO_e}
\strut\end{minipage}%
\hfill\allowbreak%
\begin{minipage}[b]{0.5\textwidth}
  \centering
\includegraphics[width=\textwidth]{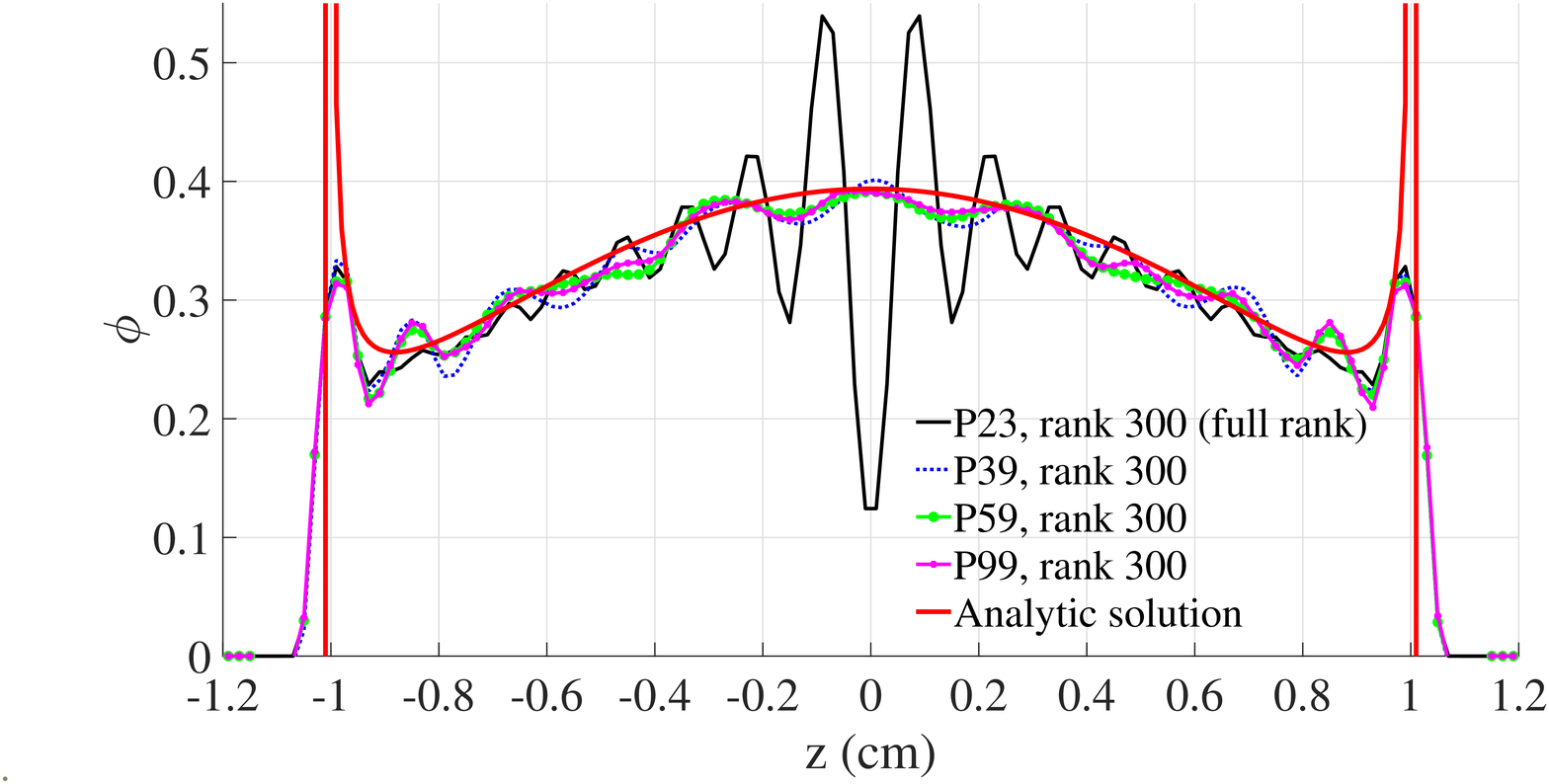}
\subcaption{The scalar flux on the cut along $x=0$.}\label{fig: LS_HOLO_cut}
\strut\end{minipage}%
\hfill\allowbreak%
\caption{The scalar flux to the line source problem calculated by HOLO with rank 300 are compared to the analytic benchmark.}
\label{fig: LS_HOLO_results}
\end{figure}

We compare the solutions with rank 300 (corresponding to the full rank $P_{23}$) and varying \PN~orders in the following simulations. We expect the solution can be refined by keeping the rank fixed and increasing the \PN~order. That is, we could have better results with a small amount of extra memory cost. Figure \ref{fig: LS_HOLO_b} shows the remarkable ring structure in the full rank $P_{23}$ solution. As we increase the \PN~order with the HOLO algorithm, the first noticeable change is that the solution range begins to match that of the analytical solution. Furthermore, the ring structure is no longer significant in Figures \ref{fig: LS_HOLO_c}, \ref{fig: LS_HOLO_d}, and \ref{fig: LS_HOLO_e}. Figure \ref{fig: LS_HOLO_cut} presents a more straightforward comparison, where the oscillations are reduced by using more angular basis functions. 

\begin{figure}[h!]
\centering
\includegraphics[scale=0.24]{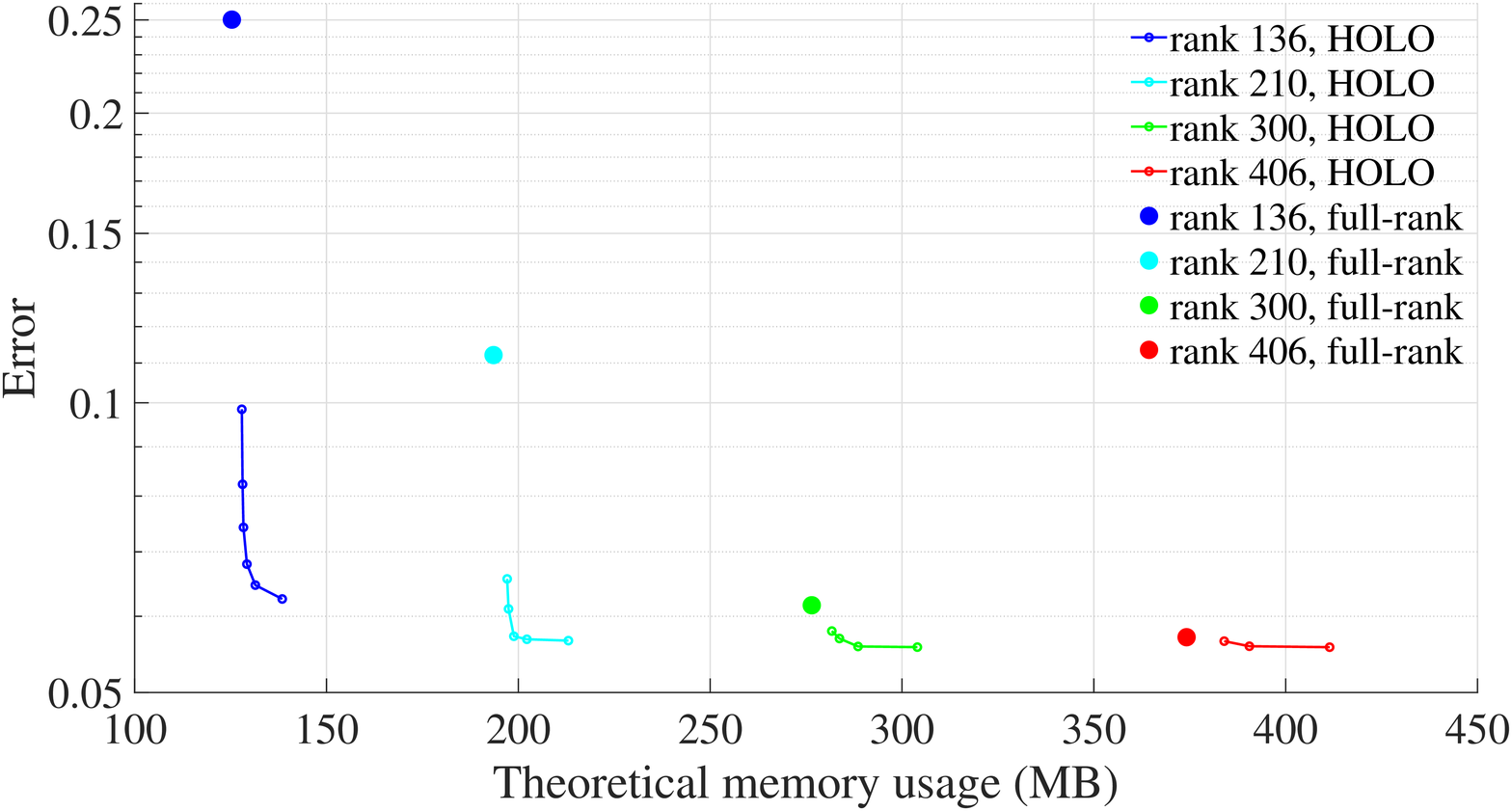}
\caption{The comparison of errors for the line source problem with different memory usage are shown. The solid dot represents the error of the full rank solution. Each dotted line denotes the error with a fixed rank that varies the number of angular basis functions $N$.}
\label{fig: LS_HOLO_error}
\end{figure}

\begin{figure}[h!]
	\centering
	\includegraphics[scale=0.24]{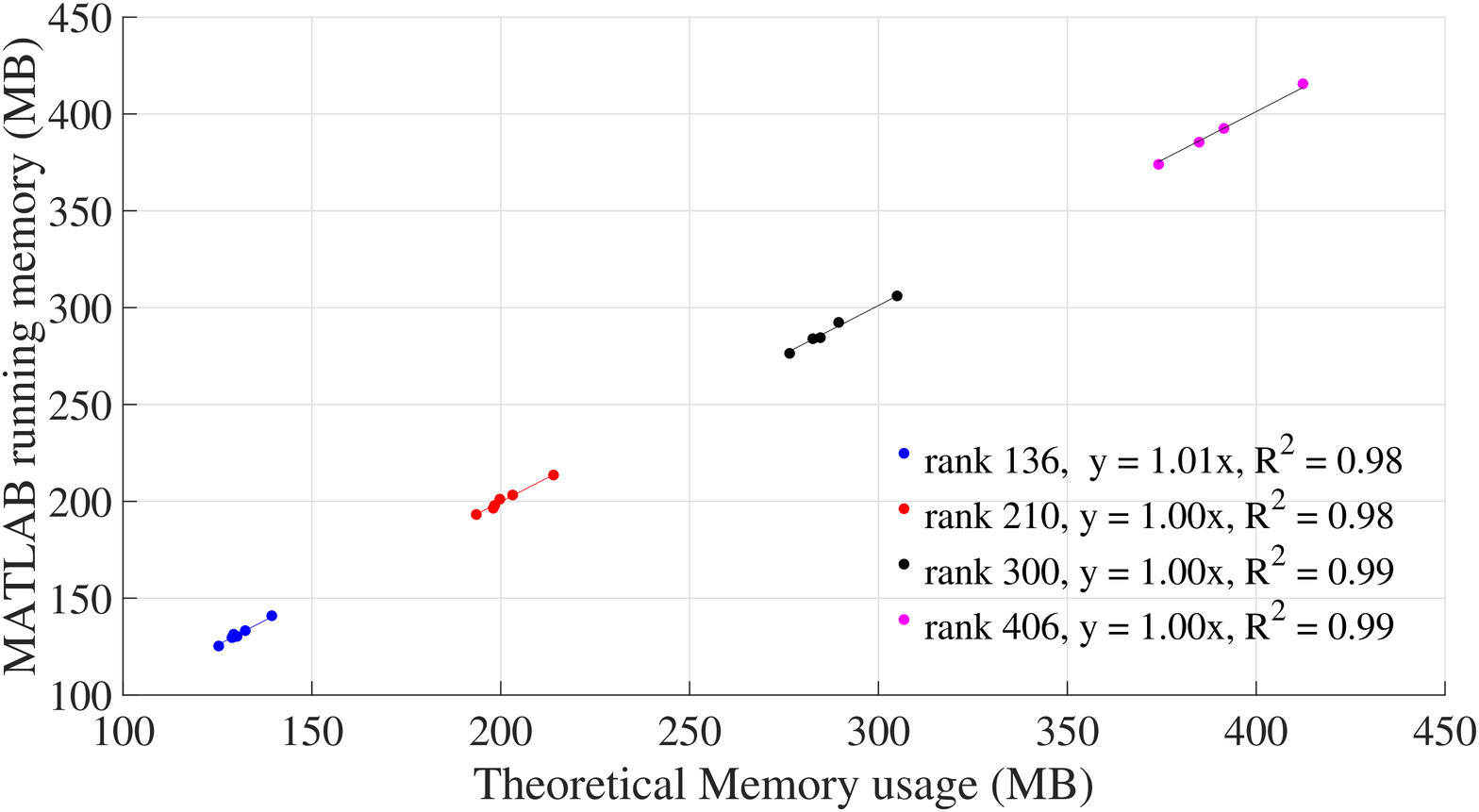}
	\caption{The relationship between the memory calculated by \eqref{eq: memory} or \eqref{eq: memory_fullrank} and the running memory in MATLAB for the simulations shown in Figure \ref{fig: LS_HOLO_error}.  The coefficients of determination($R^2$) for the linear are close to one, which indicates a strong linear relationship between the theoretical and actual memory.}
	\label{fig: LS_HOLO_memory}
\end{figure}

Figure \ref{fig: LS_HOLO_error} gives quantitative comparison, where we compare the root mean squares between the numerical solutions for the scalar flux and the analytic solution. It is apparent from the figure that the low-rank solution is more accurate than the full-rank solution with the same rank, which enables the choice to save memory or increase the accuracy. For example, the memory usage in full rank $P_{23}$ and $P_{39}$ with rank 300 are almost the same, as shown in the green line, but the accuracy is very different, as shown in Figure \ref{fig: LS_HOLO_cut}. Additionally, the rightmost dot of the dark blue line indicates that we can achieve an error of $0.06$ with memory $140$ MB; this accuracy cannot be obtained with less than $280$ MB in a full rank calculation, as shown in the large green dot. To demonstrate that the formulation in Eqs.~\eqref{eq: memory_fullrank} and \eqref{eq: memory} are correct representations of the required memory in practice, we measure the running memory in MATLAB with the ``memory" function, as shown in Figure \ref{fig: LS_HOLO_memory}. From this figure, we observe that our estimates are valid.

\begin{figure}[h!]
	\centering
	\includegraphics[scale=0.24]{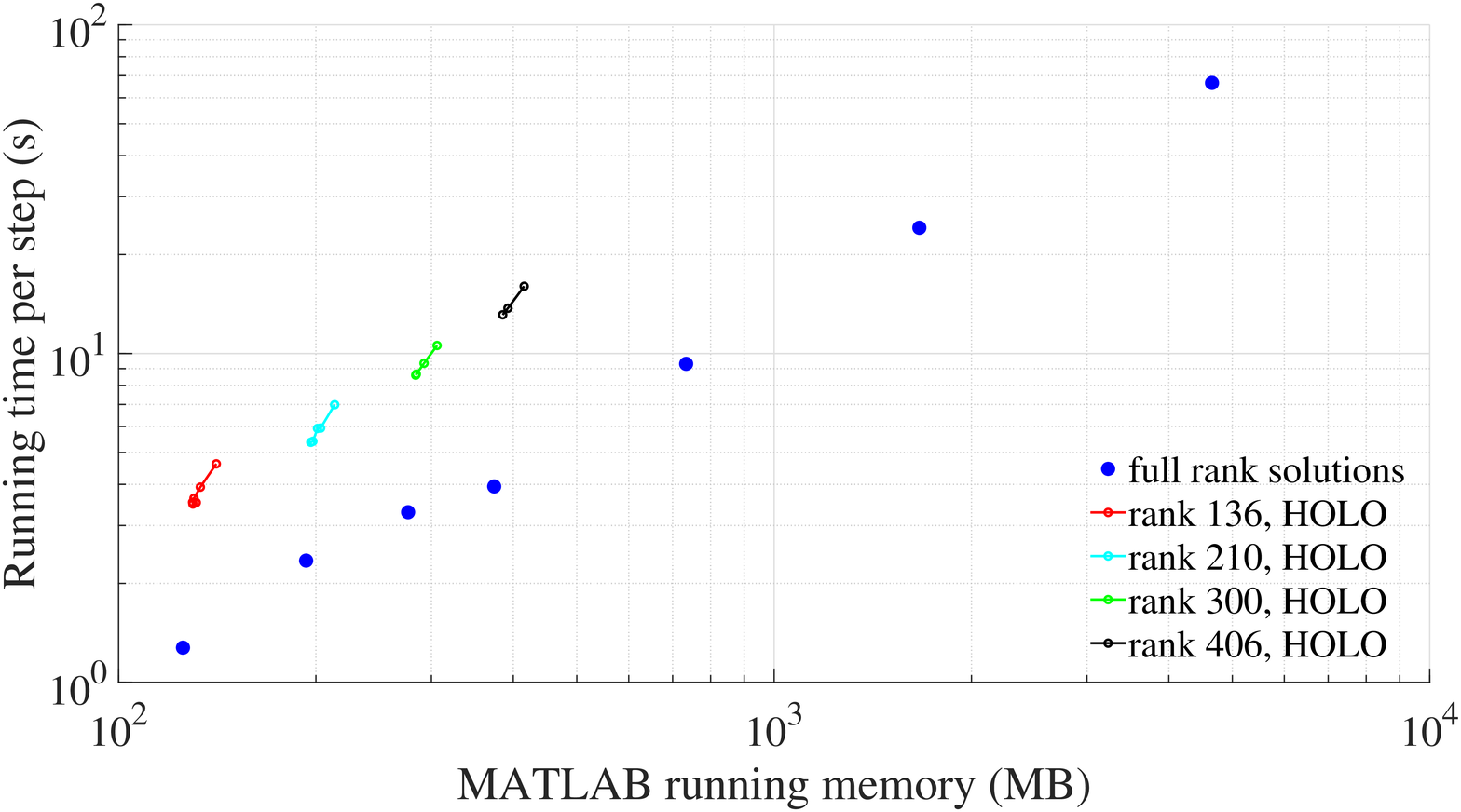}
	\caption{The comparison of memory and computational time for the line source problem with HOLO or full rank is shown. The blue solid dot represents the full rank solution with P$_{15}$, P$_{19}$, P$_{23}$, P$_{27}$, P$_{39}$, P$_{59}$ and P$_{99}$. Each dotted line denotes the error with a fixed rank that varies the number of angular basis functions $N$. Note that the full rank P$_{15}$ has rank 136, P$_{19}$ has rank 210, P$_{23}$ has rank 300, P$_{27}$ has rank 406. }
	\label{fig: LS_HOLO_memory_time}
\end{figure}

The computational cost of the HOLO algorithm is presented in Figure \ref{fig: LS_HOLO_memory_time}. We notice that the running time and the memory of the HOLO solutions are much lower than the full rank P$_{59}$ and P$_{99}$ solutions. For example, the HOLO solution with P$_{99}$ and rank 300 requires 200 MB memory and 7s running time each time step, while the full rank P$_{99}$ solution needs 5000 MB and 65s, respectively. Note that our results have the error decrease stagnate because we have reduced the angular error in the solution to be smaller than the spatial discretization error, as shown in Figure \ref{fig: LS_HOLO_error}. We also observe that the HOLO solution requires a longer running time than the classical full rank solution with the same rank due to the fact that the low-rank method has more arithmetic operations \cite{peng2019lowrank}. This indicates that to get the most benefit from a low-rank, HOLO approach, one should run the highest order in angle solution possible.

\subsection{Hohlraum problem}
\begin{figure}[h!] 
\begin{minipage}[b]{\textwidth}
\centering
\includegraphics[width=0.5\textwidth]{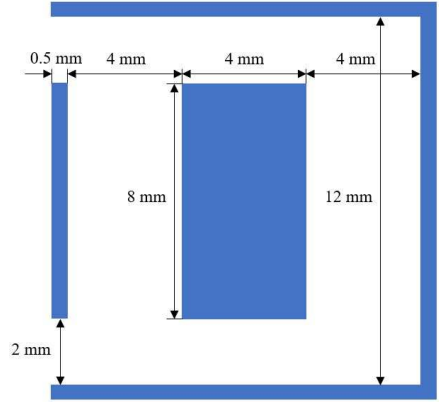}
\subcaption{The hohlraum}\label{fig: Hohlraum}
\strut\end{minipage}%
\hfill\allowbreak%
\begin{minipage}[b]{\textwidth}
\centering
\includegraphics[width=\textwidth]{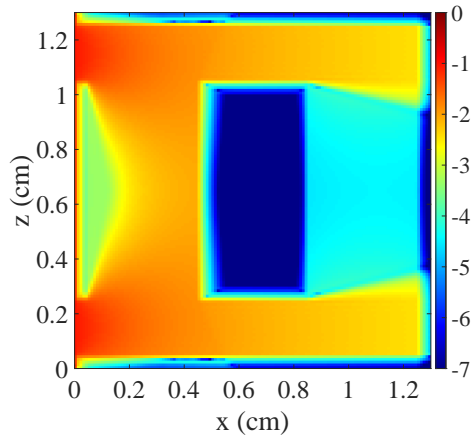}
\subcaption{P$_{141}$, full rank}\label{fig: Hohlraum_benchmark}
\strut\end{minipage}%
\hfill\allowbreak%
\caption{The layout of the Hohlraum test and the high-order benchmark solution. }
\label{fig: Hohlraum_example}
\end{figure}

We consider a modified Hohlraum problem \cite{Hauck2013} detailed in Figure \ref{fig: Hohlraum}. There is an isotopic source of  $Q = 1$ in the leftmost zone that is turned on at t = 0. The blank areas are dense materials with $\sigma_t = 100 \ \mathrm{cm^{-1}}$ and $\sigma_s = 1 \ \mathrm{cm^{-1}}$, the blank area are purely scattering materials with $\sigma_s = \sigma_t = 0.1 \ \mathrm{cm^{-1}}$. The high-fidelity $P_{141}$ solution is given in Figure \ref{fig: Hohlraum_benchmark}. The spatial grid is set to be $130 \times 130$ for the computational domain $[0, 1.3] \times [0, 1.3]$. The simulation time is 2.6s and the CFL number is chosen to be 0.2.

\begin{figure}[h!] 
\begin{minipage}[b]{0.5\textwidth}
  \centering
\includegraphics[width=\textwidth]{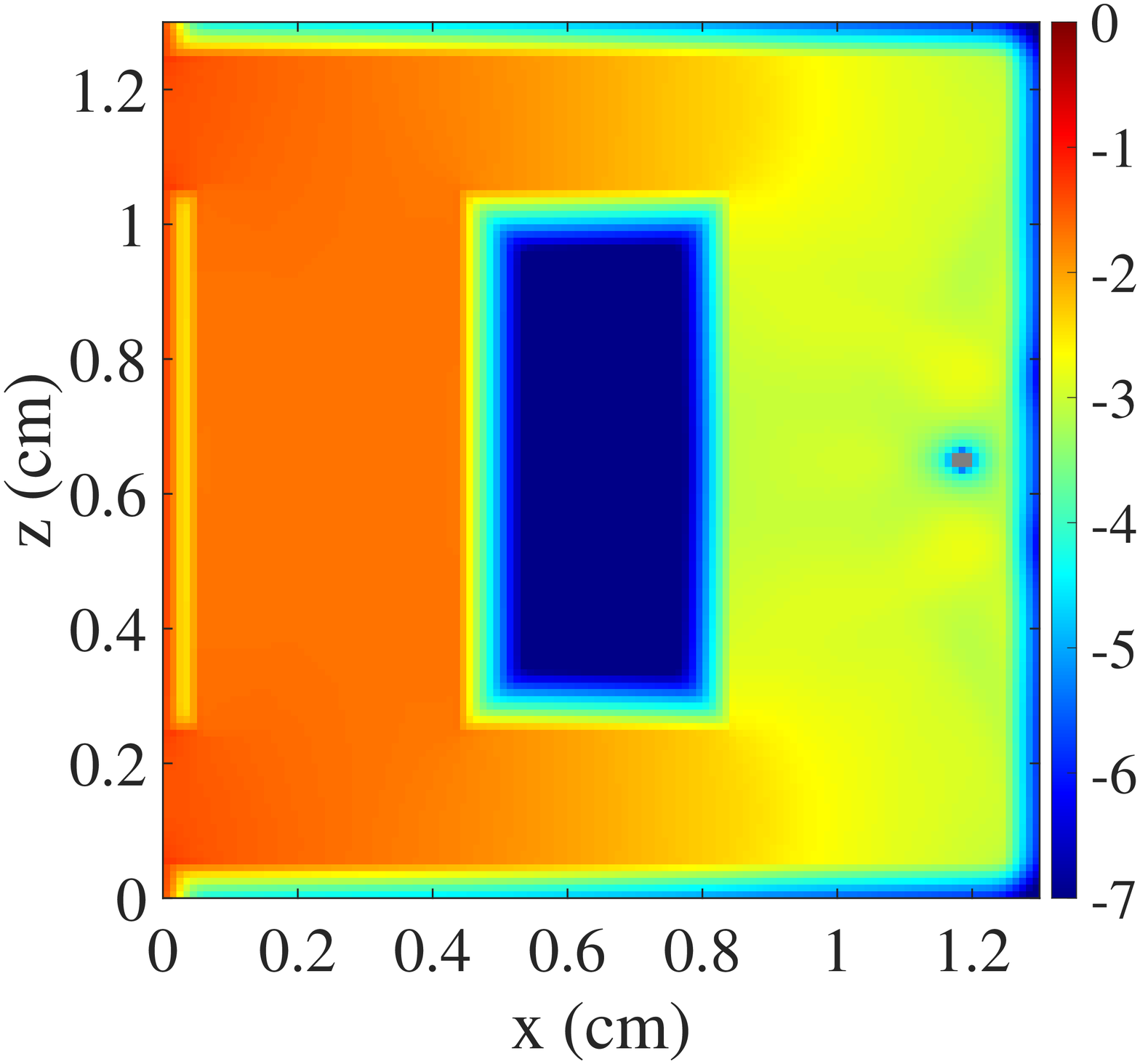}
\subcaption{P$_1$, rank 3, full rank}\label{fig: Hohlraum_HOLO_a}
\strut\end{minipage}%
\hfill\allowbreak%
\begin{minipage}[b]{0.5\textwidth}
  \centering
\includegraphics[width=\textwidth]{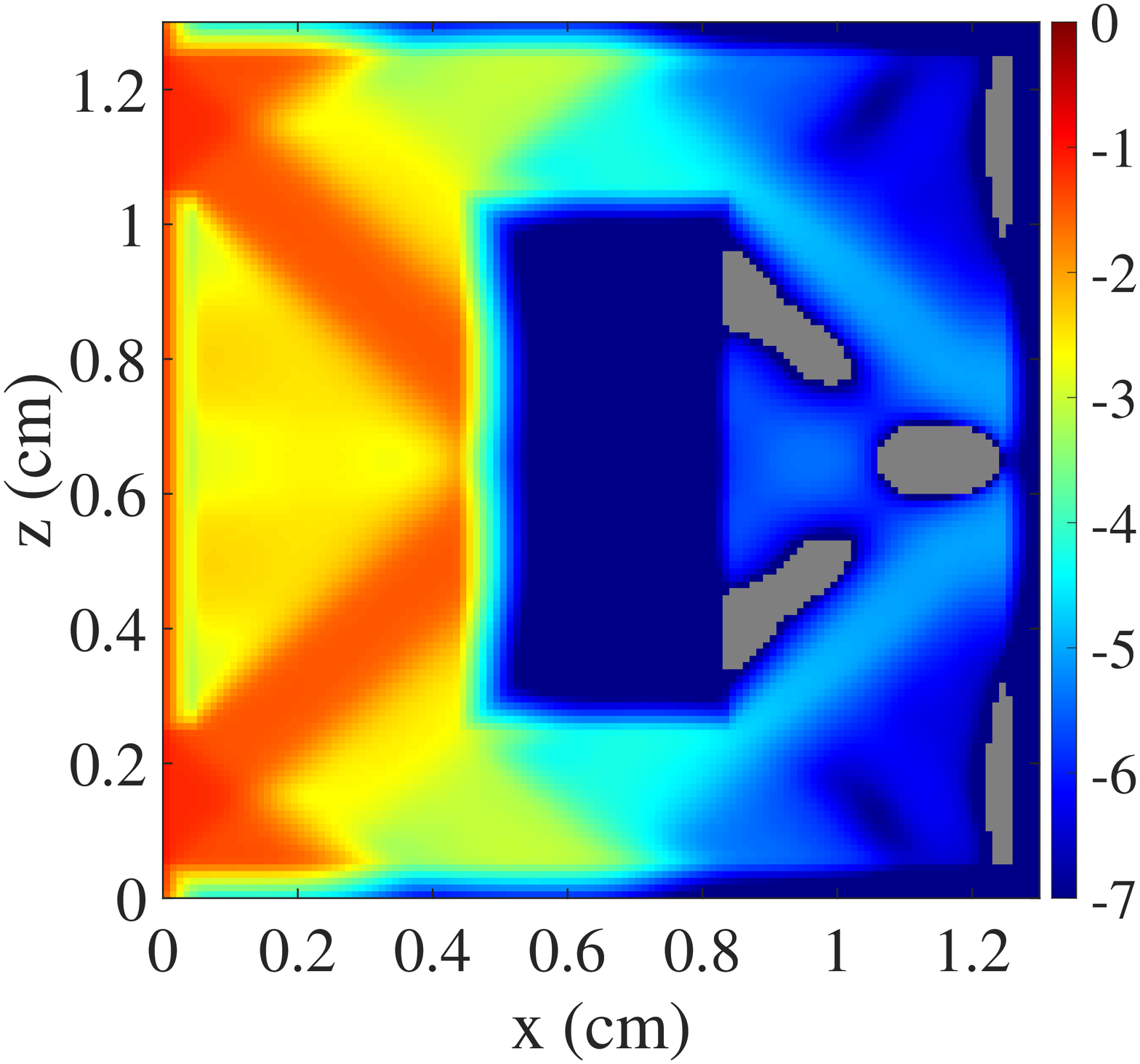}
\subcaption{P$_{39}$, rank 3, HOLO}\label{fig: Hohlraum_HOLO_b}
\strut\end{minipage}%
\hfill\allowbreak%
\begin{minipage}[b]{0.5\textwidth}
\centering
\includegraphics[width=\textwidth]{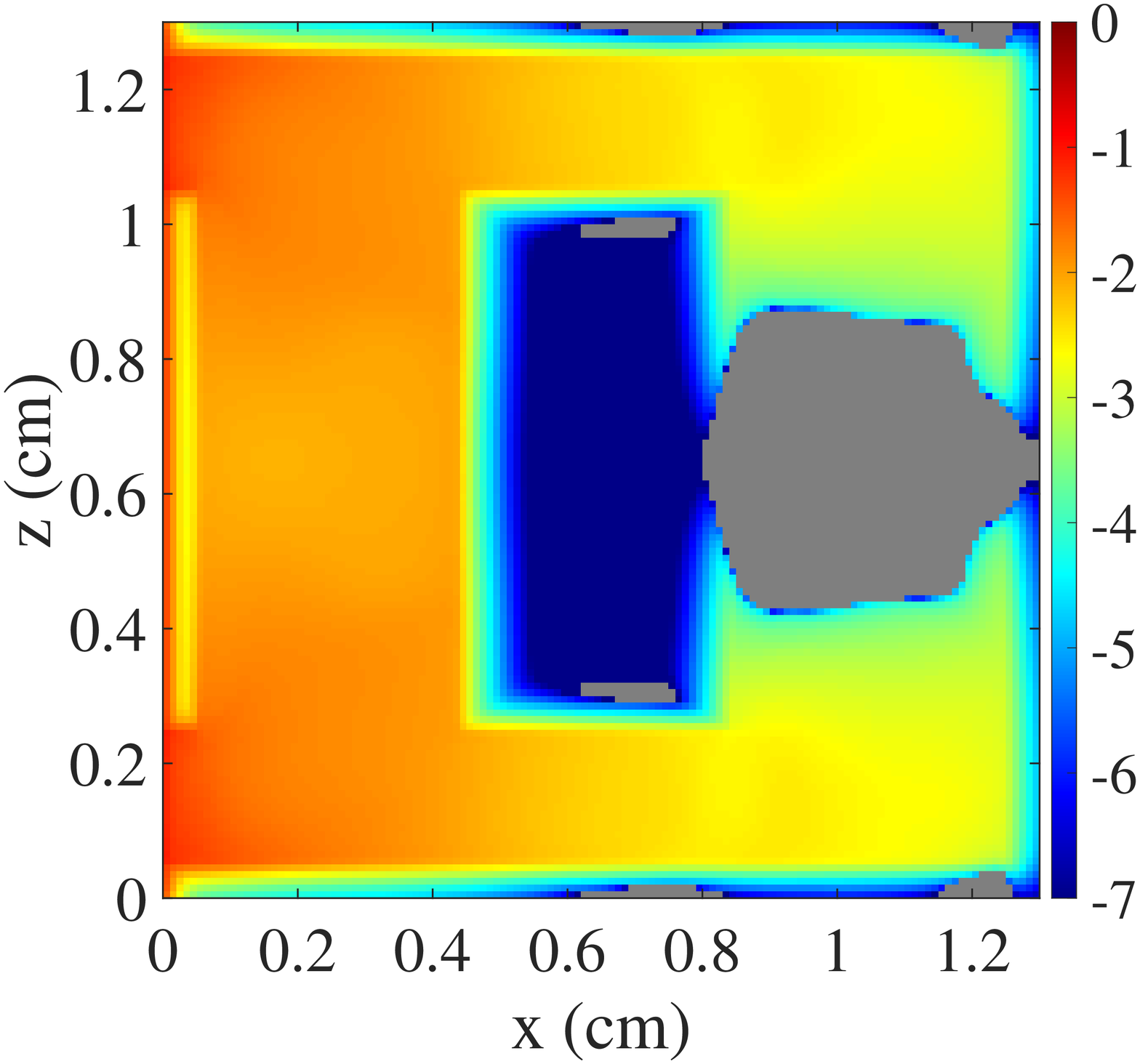}
\subcaption{P$_3$, rank 10, full rank}\label{fig: Hohlraum_HOLO_c}
\strut\end{minipage}%
\hfill\allowbreak%
\begin{minipage}[b]{0.5\textwidth}
\centering
\includegraphics[width=\textwidth]{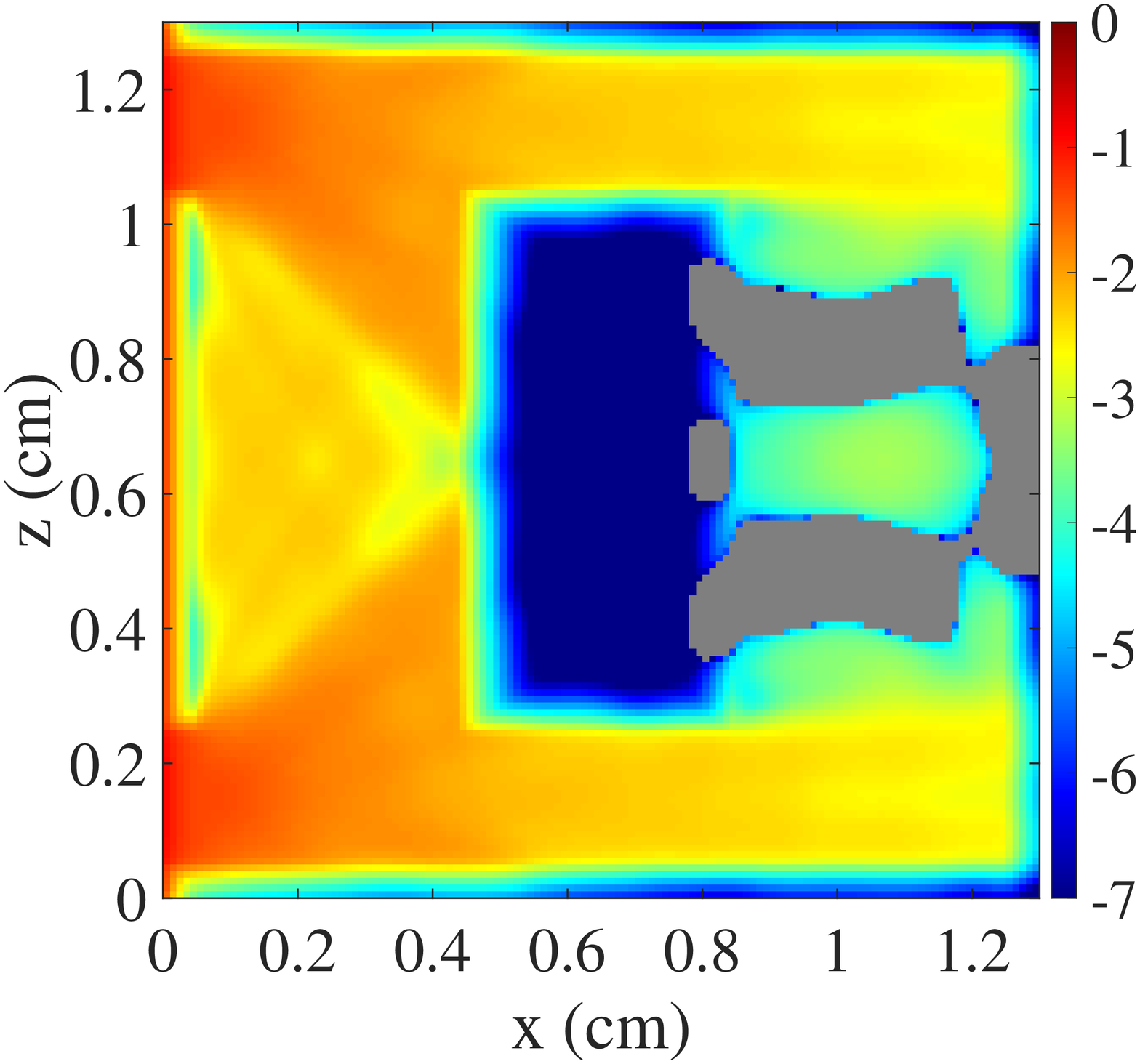}
\subcaption{P$_{39}$, rank 10, HOLO}\label{fig: Hohlraum_HOLO_d}
\strut\end{minipage}%
\hfill\allowbreak%
\begin{minipage}[b]{0.5\textwidth}
\centering
\includegraphics[width=\textwidth]{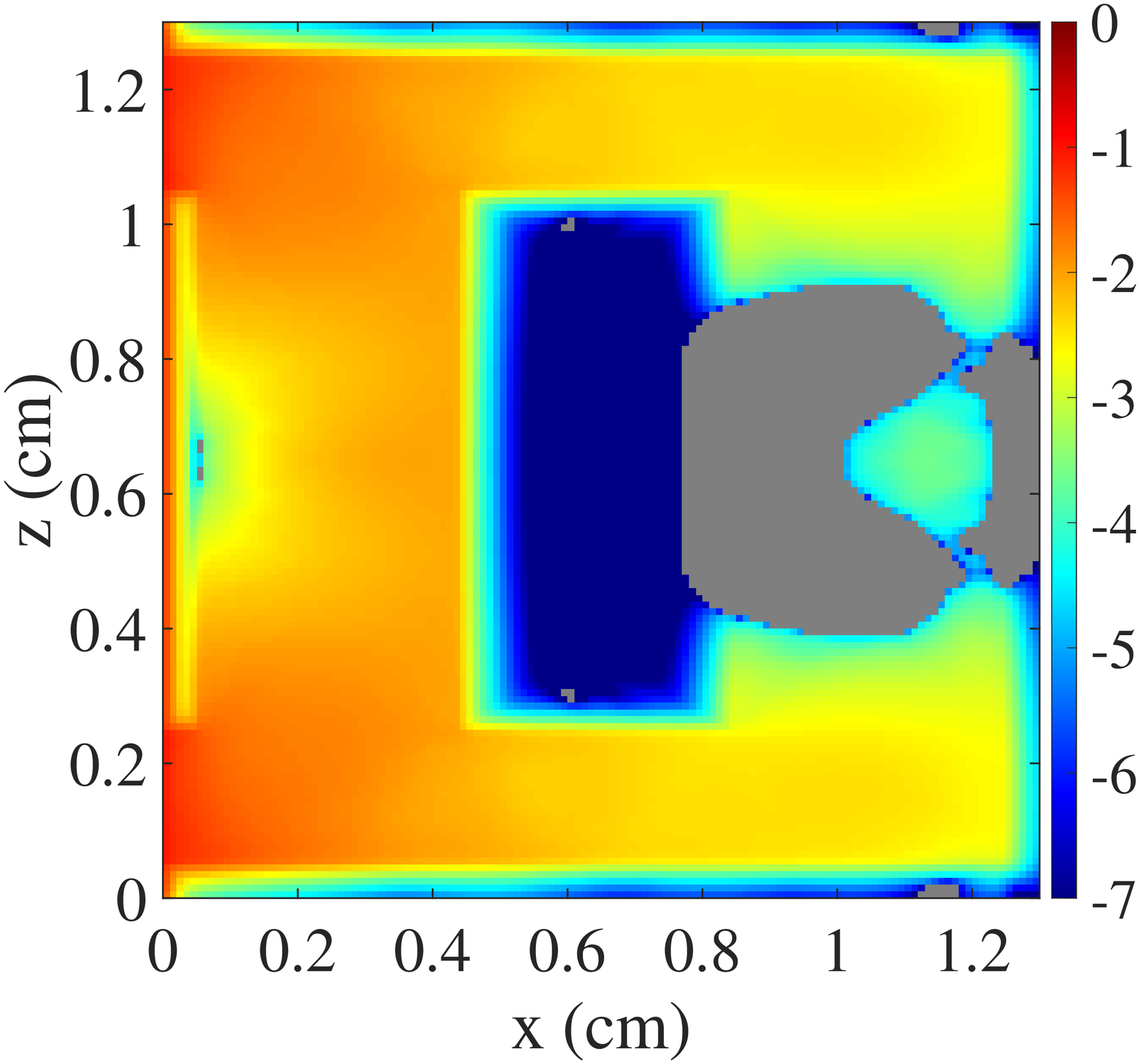}
\subcaption{P$_5$, rank 21, full rank}\label{fig: Hohlraum_HOLO_e}
\strut\end{minipage}%
\hfill\allowbreak%
\begin{minipage}[b]{0.5\textwidth}
\centering
\includegraphics[width=\textwidth]{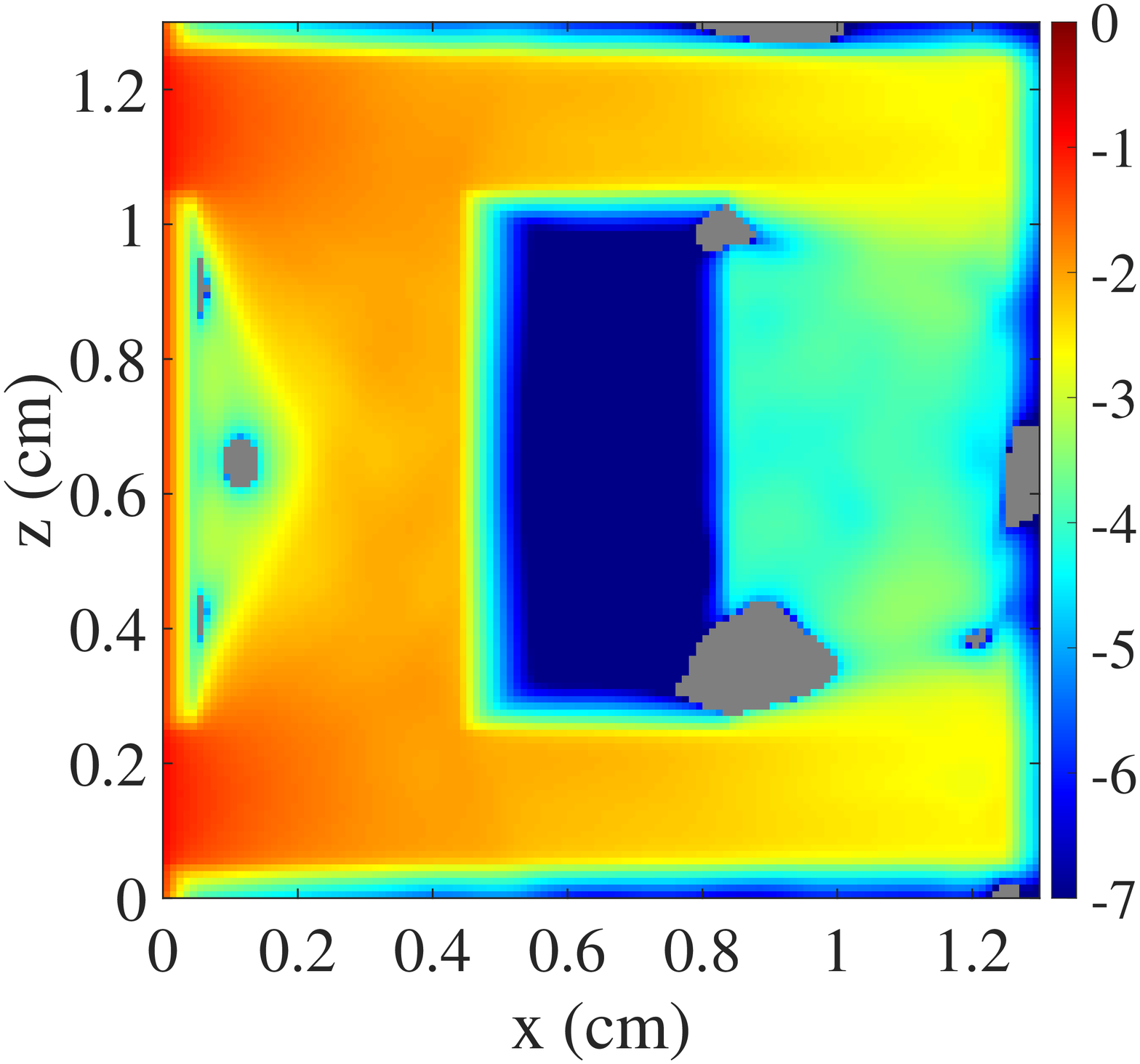}
\subcaption{P$_{39}$, rank 21, HOLO}\label{fig: Hohlraum_HOLO_f}
\strut\end{minipage}%
\hfill\allowbreak%
\begin{minipage}[b]{0.5\textwidth}
\centering
\includegraphics[width=\textwidth]{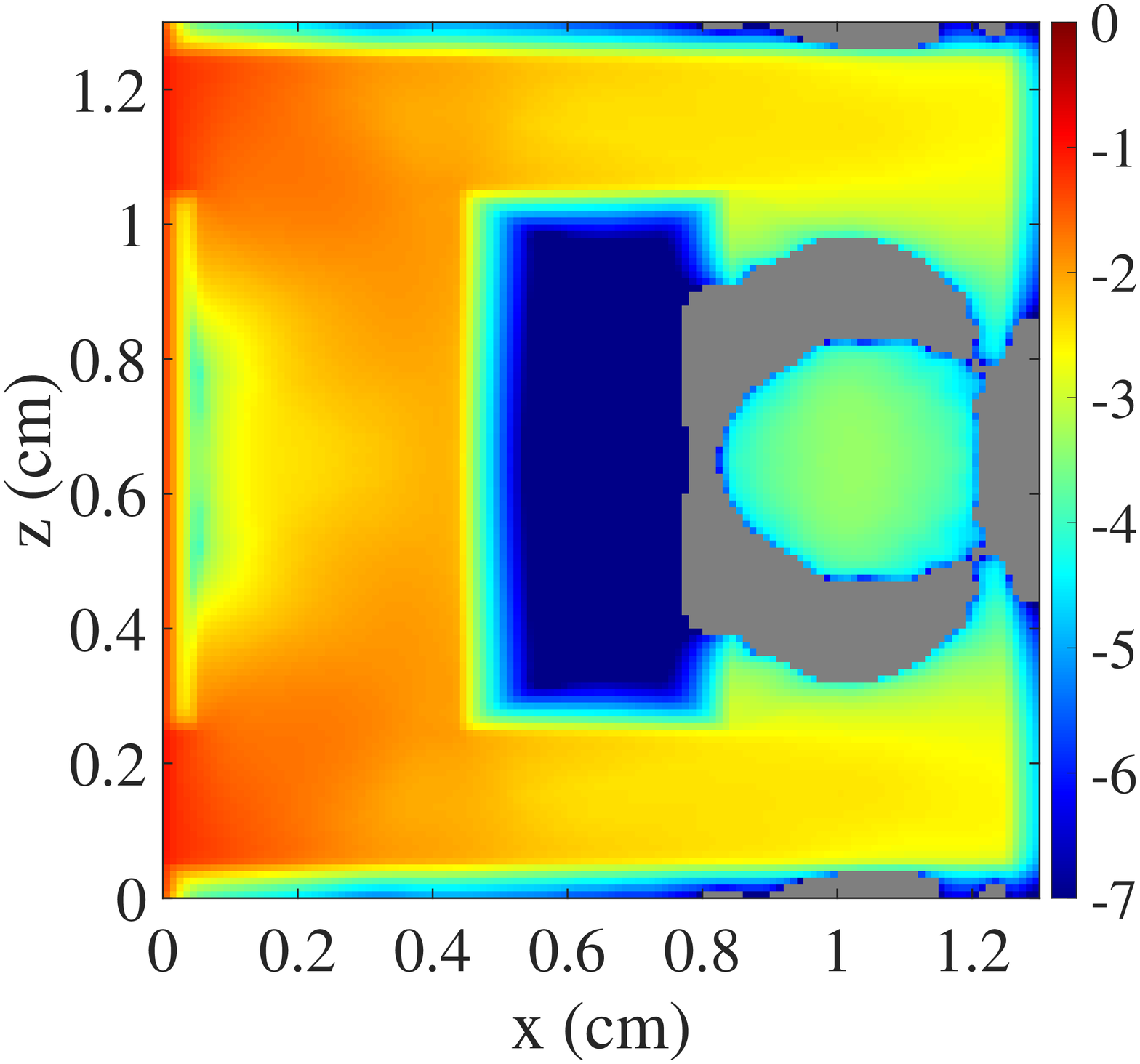}
\subcaption{P$_7$, rank 36, full rank}\label{fig: Hohlraum_HOLO_g}
\strut\end{minipage}%
\hfill\allowbreak%
\begin{minipage}[b]{0.5\textwidth}
\centering
\includegraphics[width=\textwidth]{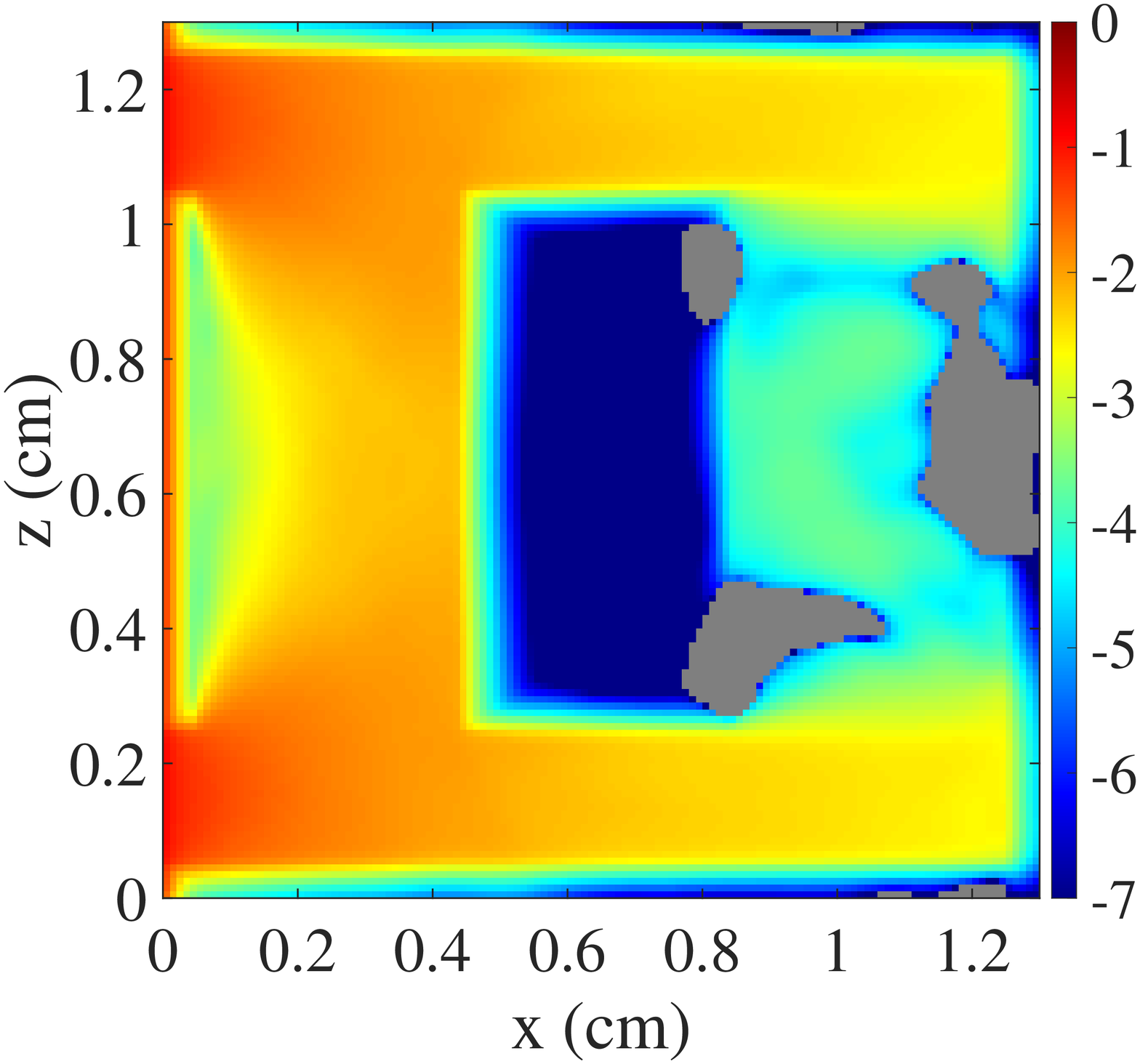}
\subcaption{P$_{39}$, rank 36, HOLO}\label{fig: Hohlraum_HOLO_h}
\strut\end{minipage}%
\hfill\allowbreak%
\caption{The scalar flux to the Hohlraum problem calculated by HOLO with P$_{39}$ are compared to solutions without  rank reduction. The color scale is logarithmic and negative regions are shaded gray.}
\label{fig: Hohlraum_HOLO_results}
\end{figure}

In this test, we compare the P$_{39}$ HOLO solutions to the full rank solutions with the same rank. From Figure \ref{fig: Hohlraum_benchmark}, we can see that the shape of particle distribution behind the first dense wall should be a triangle. But none of the full rank solutions can capture it, as shown in Figure \ref{fig: Hohlraum_HOLO_a}, \ref{fig: Hohlraum_HOLO_c}, \ref{fig: Hohlraum_HOLO_e} and \ref{fig: Hohlraum_HOLO_g}. In contrast, the HOLO solution with only rank 3 can preserve this feature, while the rank 36 is nearly identical to the benchmark except for the area behind the second obstacle. Note that rank 36 is considered a small rank for this problem: for the full rank P$_{39}$ corresponds to $n = r = 820$ and P$_{141}$ has $n = r = 10153$. In such a low-rank solution we cannot guarantee that the symmetries we expect in this problem (e.g., top/bottom symmetry) will be preserved by the projections. This is especially obvious on the logarithmic scale of Figure \ref{fig: Hohlraum_HOLO_results}.

\begin{figure}[h!] 
  \centering
\includegraphics[scale = 0.24]{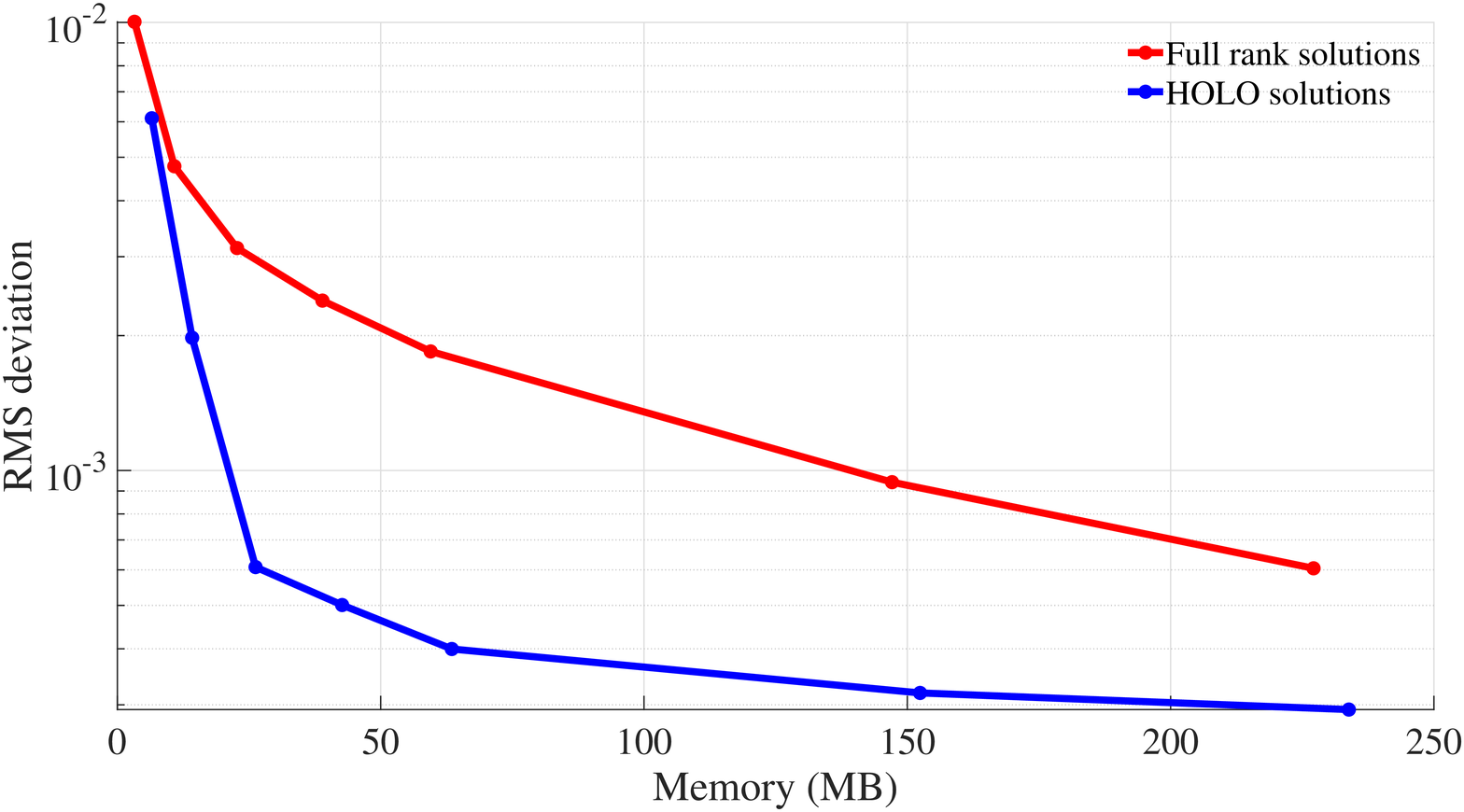}
\caption{The comparison of errors for the Hohlraum problem with different memory usage are shown. The red dot line represents the error of the full rank solution that varies the number of angular basis functions $N$. The blue dot line represents the error of the HOLO solutions with P$_{39}$ that varies the rank.}
\label{fig: Hohlraum_error}
\end{figure}

Figure \ref{fig: Hohlraum_error} shows the deviation of the HOLO solutions of P$_{39}$ and the full rank solutions ranging from P$_{1}$ to P$_{15}$ to the full rank P$_{39}$. We observe that the HOLO solution is more accurate than the full rank solution with the same memory usage and converges to the high-order full rank solution faster. 

\subsection{Lattice problem}
\begin{figure}[h!]
	\centering
	\includegraphics[width=0.6\textwidth]{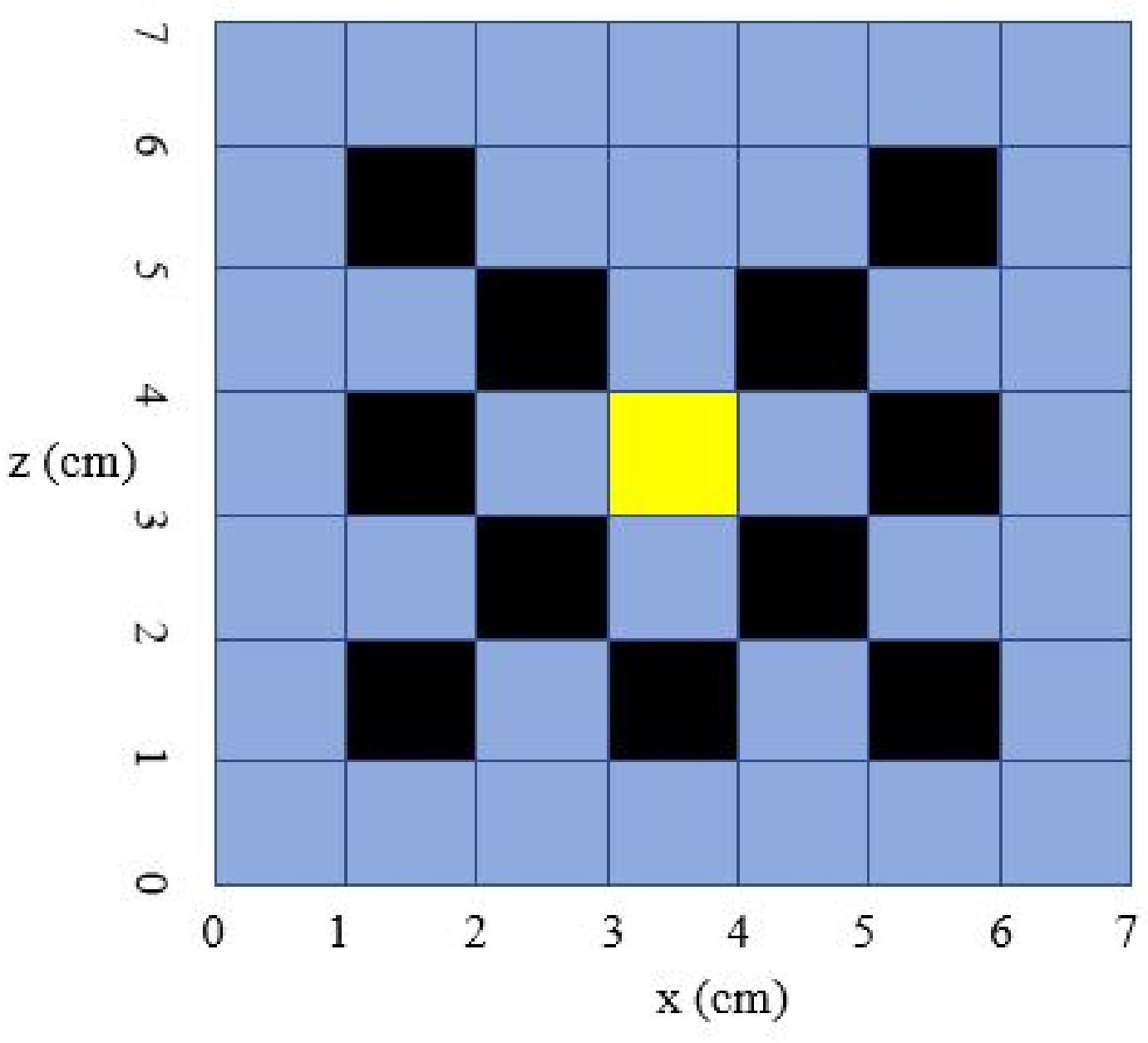}
	\caption{The material layout of the Lattice problem is shown. The blue zones are purely scattering region with $\sigma_s = \sigma_t = 1 \ \mathrm{cm^{-1}}$, the black are absorbing region with $\sigma_s = 0$, $\sigma_t = 10 \ \mathrm{cm^{-1}}$ and the yellow is the scattering region with an isotropic source $Q = 1$ which is turned on at $t = 0$. The checkerboard is surrounded by vacuum.}
	\label{fig: Lattice_layout}
\end{figure}

\begin{figure}[h!] 
\begin{minipage}[b]{0.5\textwidth}
  \centering
\includegraphics[width=\textwidth]{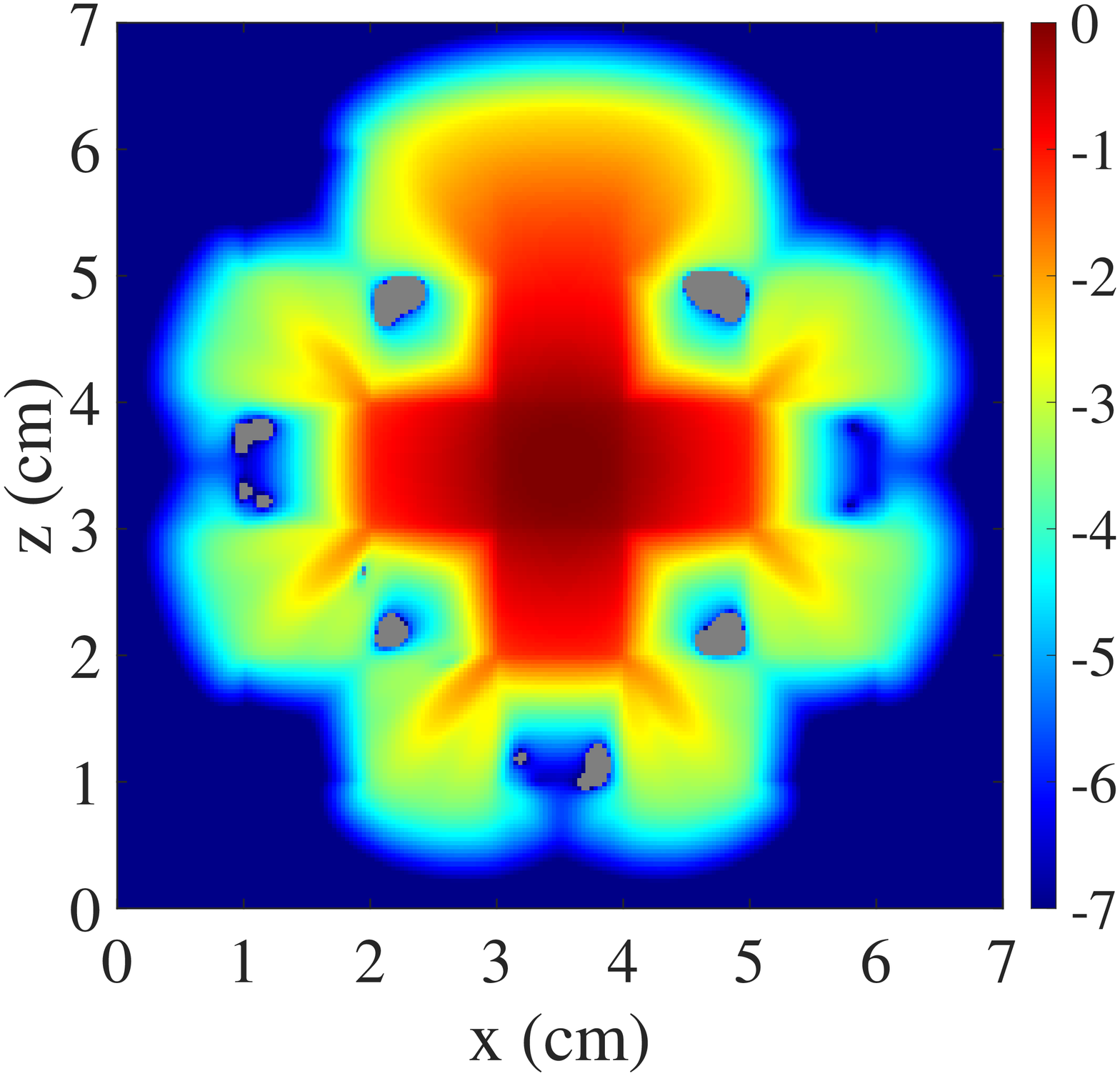}
\subcaption{P$_{39}$, rank 10}\label{fig: Lattice_HOLO_a}
\strut\end{minipage}%
\hfill\allowbreak%
\begin{minipage}[b]{0.5\textwidth}
  \centering
\includegraphics[width=\textwidth]{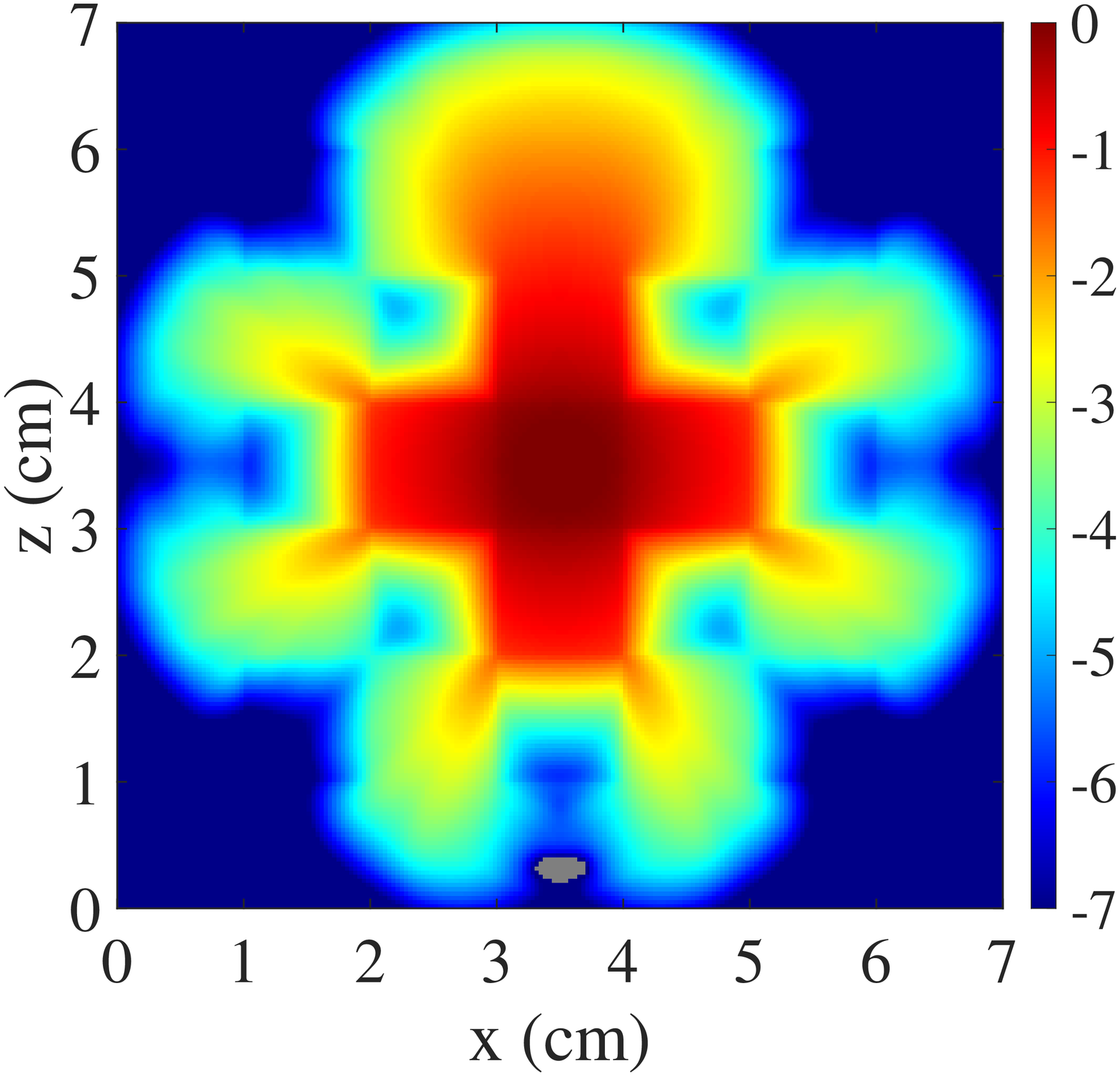}
\subcaption{P$_{39}$, rank 36}\label{fig: Lattice_HOLO_b}
\strut\end{minipage}%
\hfill\allowbreak%
\begin{minipage}[b]{0.5\textwidth}
  \centering
\includegraphics[width=\textwidth]{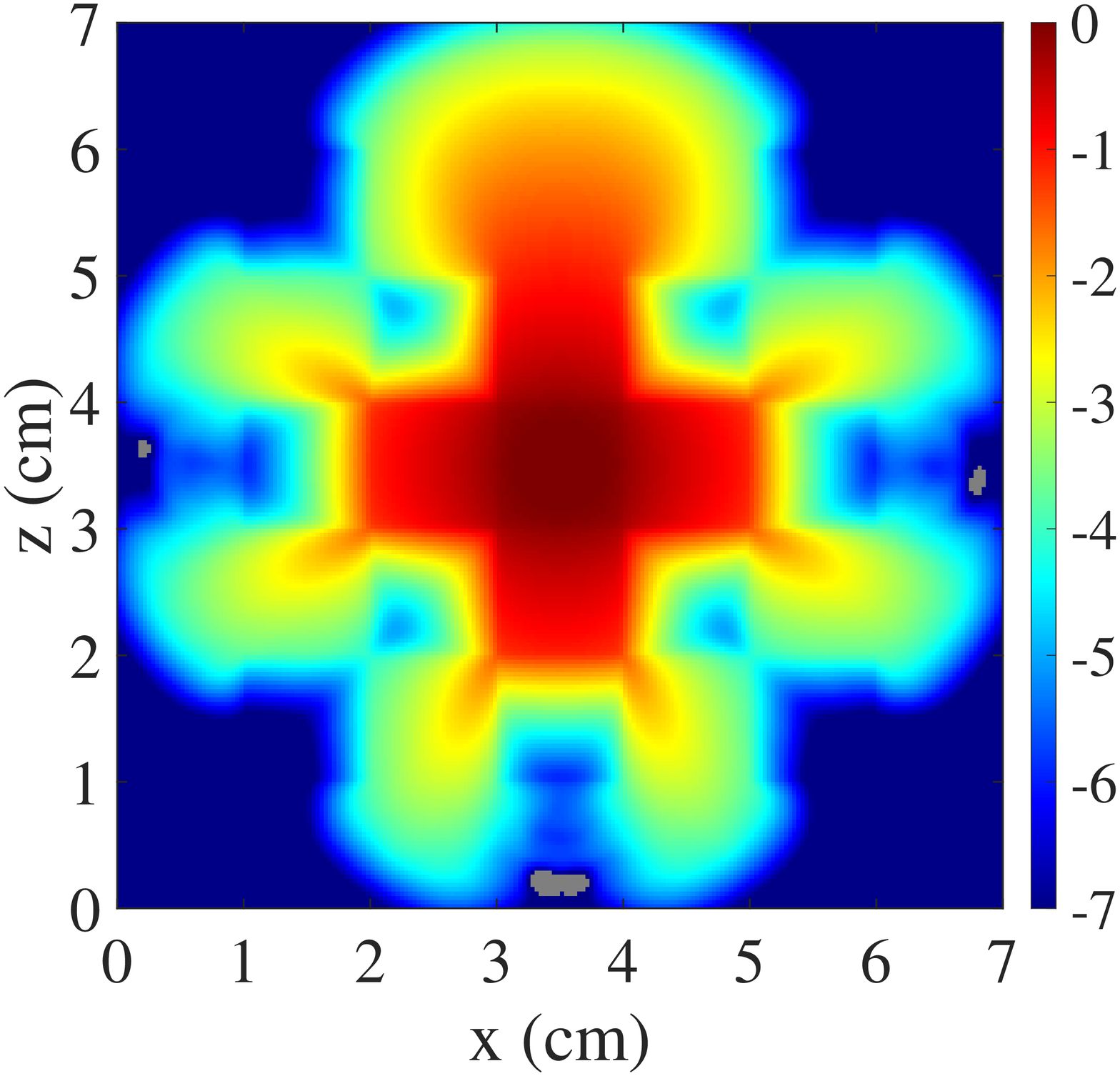}
\subcaption{P$_{39}$, rank 78}\label{fig: Lattice_HOLO_c}
\strut\end{minipage}%
\hfill\allowbreak%
\begin{minipage}[b]{0.5\textwidth}
  \centering
\includegraphics[width=\textwidth]{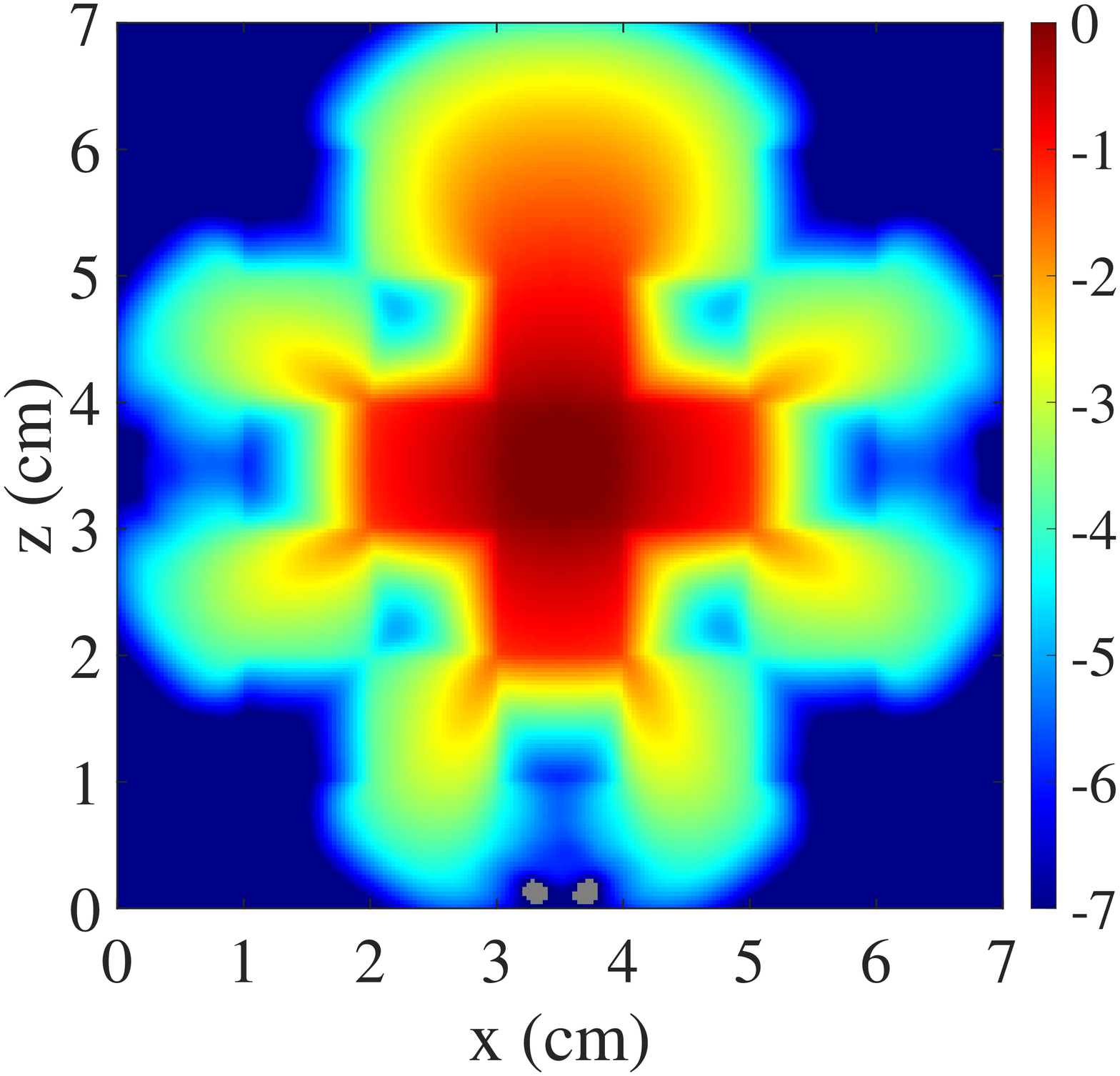}
\subcaption{P$_{39}$, rank 136}\label{fig: Lattice_HOLO_d}
\strut\end{minipage}%
\hfill\allowbreak%
\begin{minipage}[b]{0.5\textwidth}
  \centering
\includegraphics[width=\textwidth]{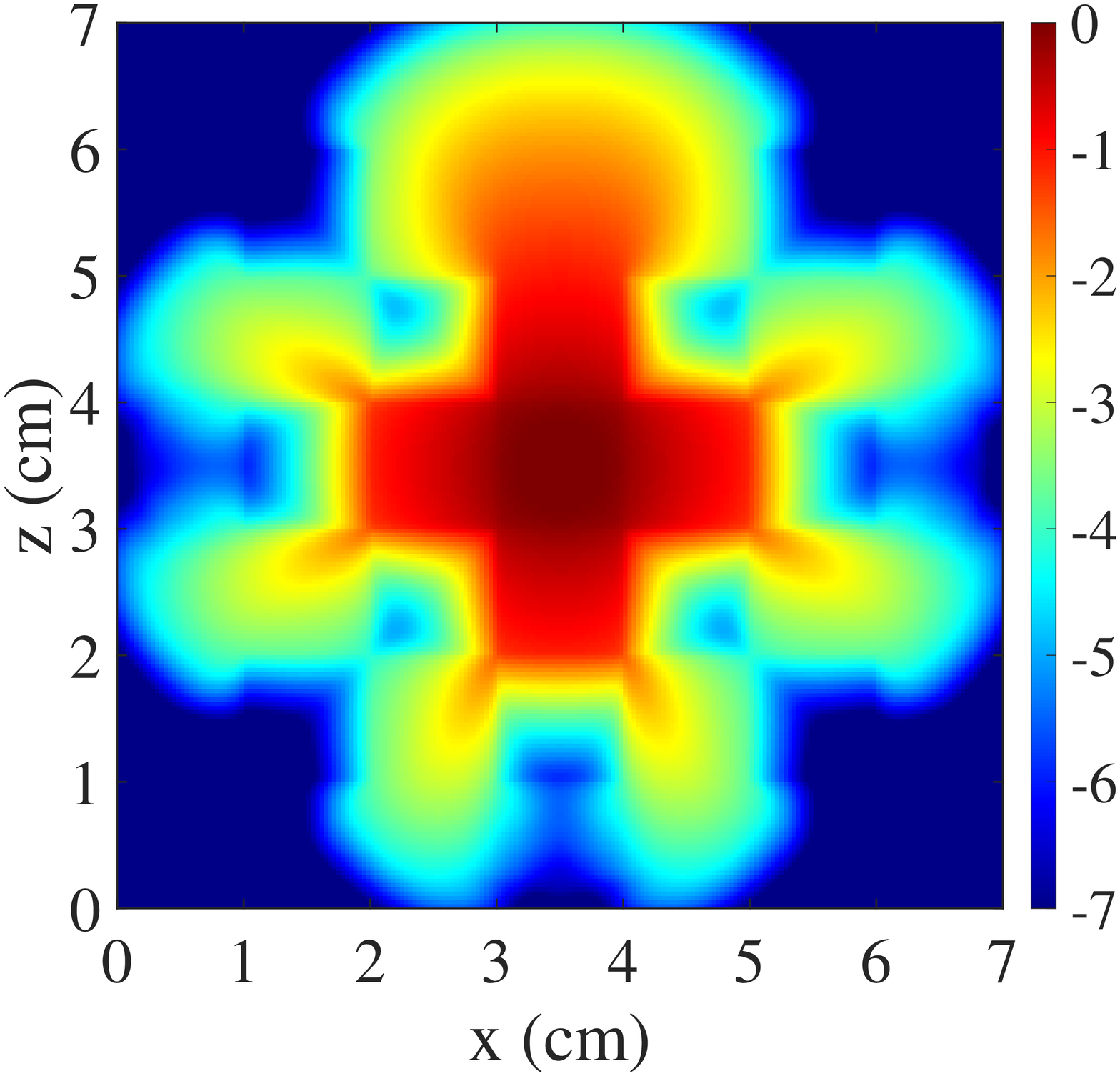}
\subcaption{P$_{39}$, rank 210}\label{fig: Lattice_HOLO_e}
\strut\end{minipage}%
\begin{minipage}[b]{0.5\textwidth}
  \centering
\includegraphics[width=\textwidth]{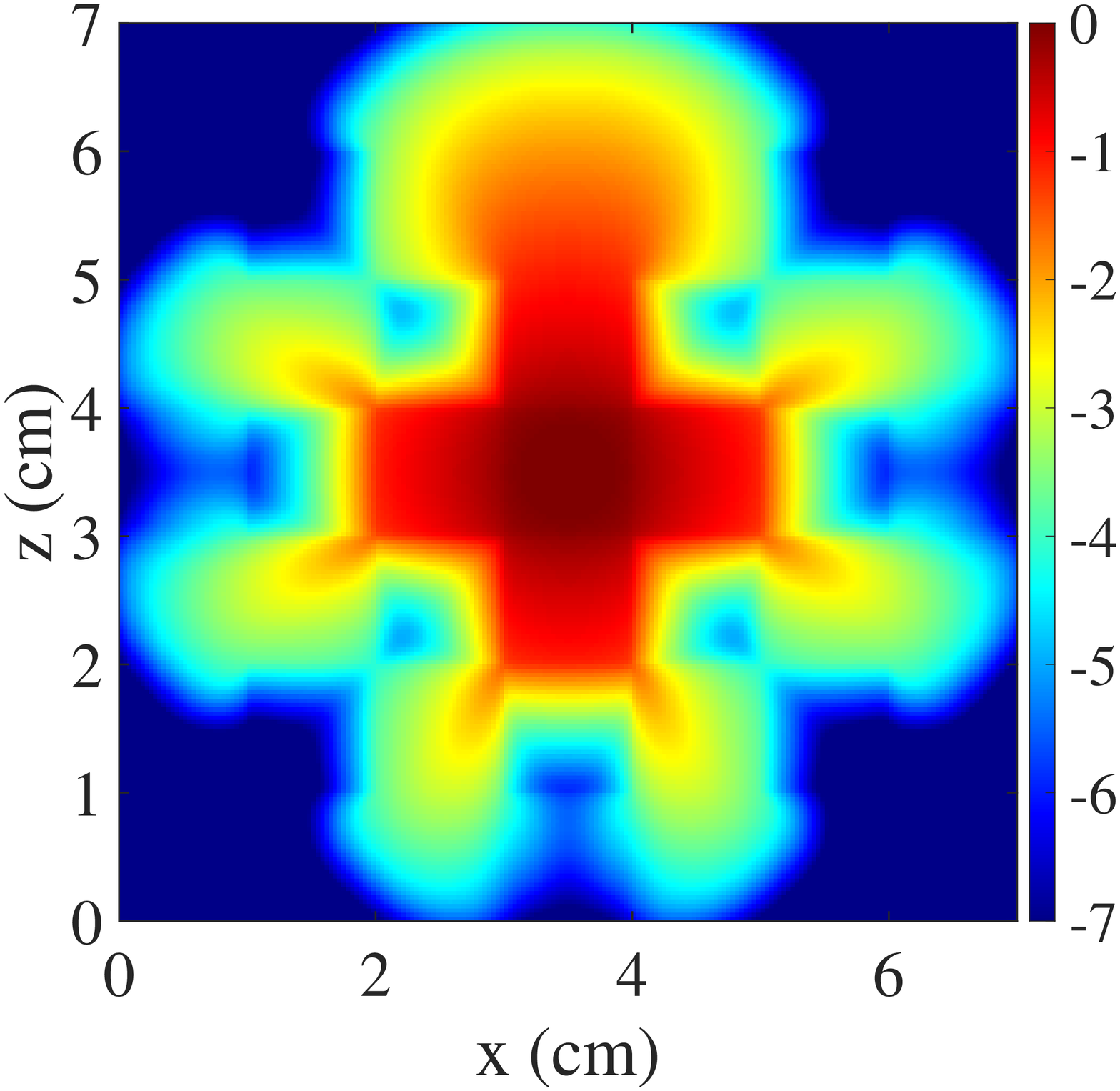}
\subcaption{P$_{39}$, full rank 820}\label{fig: Lattice_HOLO_f}
\strut\end{minipage}%
\caption{The scalar flux to the lattice problem calculated by HOLO with P$_{39}$ and different rank are compared to the full rank P$_{39}$ solution. The color scale is logarithmic and negative regions are shaded gray.}
\label{fig: Lattice_HOLO_results1}
\end{figure}

\begin{figure}[h!] 
\centering
\includegraphics[width=\textwidth]{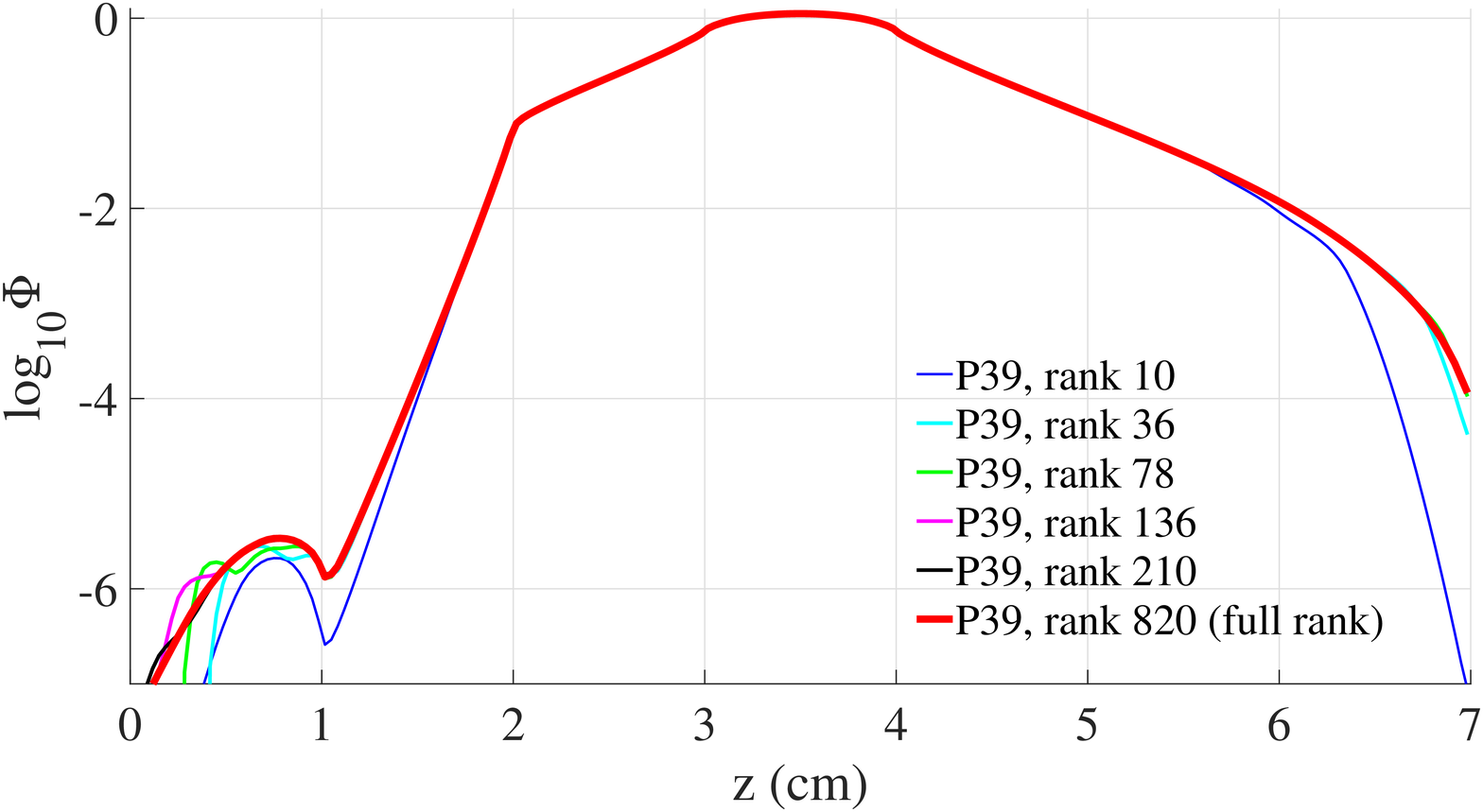}
\hfill\allowbreak%
\caption{The logarithm of the scalar flux along $x=3.5$.}
\label{fig: Lattice_HOLO_results2}
\end{figure}

Next, we solve a $7 \ cm \times 7 \ cm$ checkerboard problem as shown in Figure \ref{fig: Lattice_layout}. The spatial grid is $210 \times 210$. We run the simulations to $t = 3.2$s with $\mathrm{CFL} = 0.2$. Figure \ref{fig: Lattice_HOLO_results1} shows the $P_{39}$ solutions with different rank. From Figure \ref{fig: Lattice_HOLO_e}, we can see that the rank 210 solution is nearly identical to the full-rank solutions (a reduction of nearly a factor of 4). There are noticeable negative scalar flux regions in solutions with small rank, which is plotted in grey, as can be seen from Figure \ref{fig: Lattice_HOLO_a}, \ref{fig: Lattice_HOLO_b} and \ref{fig: Lattice_HOLO_c}. Specifically, the correct propagation speed is lost in the solution with rank 36, which is also shown in Figure \ref{fig: Lattice_HOLO_results2}. Therefore, we conclude that rank 36 is not sufficient for this problem.


\subsection{Double Chevron problem}
\begin{figure}[h!]
	\centering
	\includegraphics[width=0.6\textwidth]{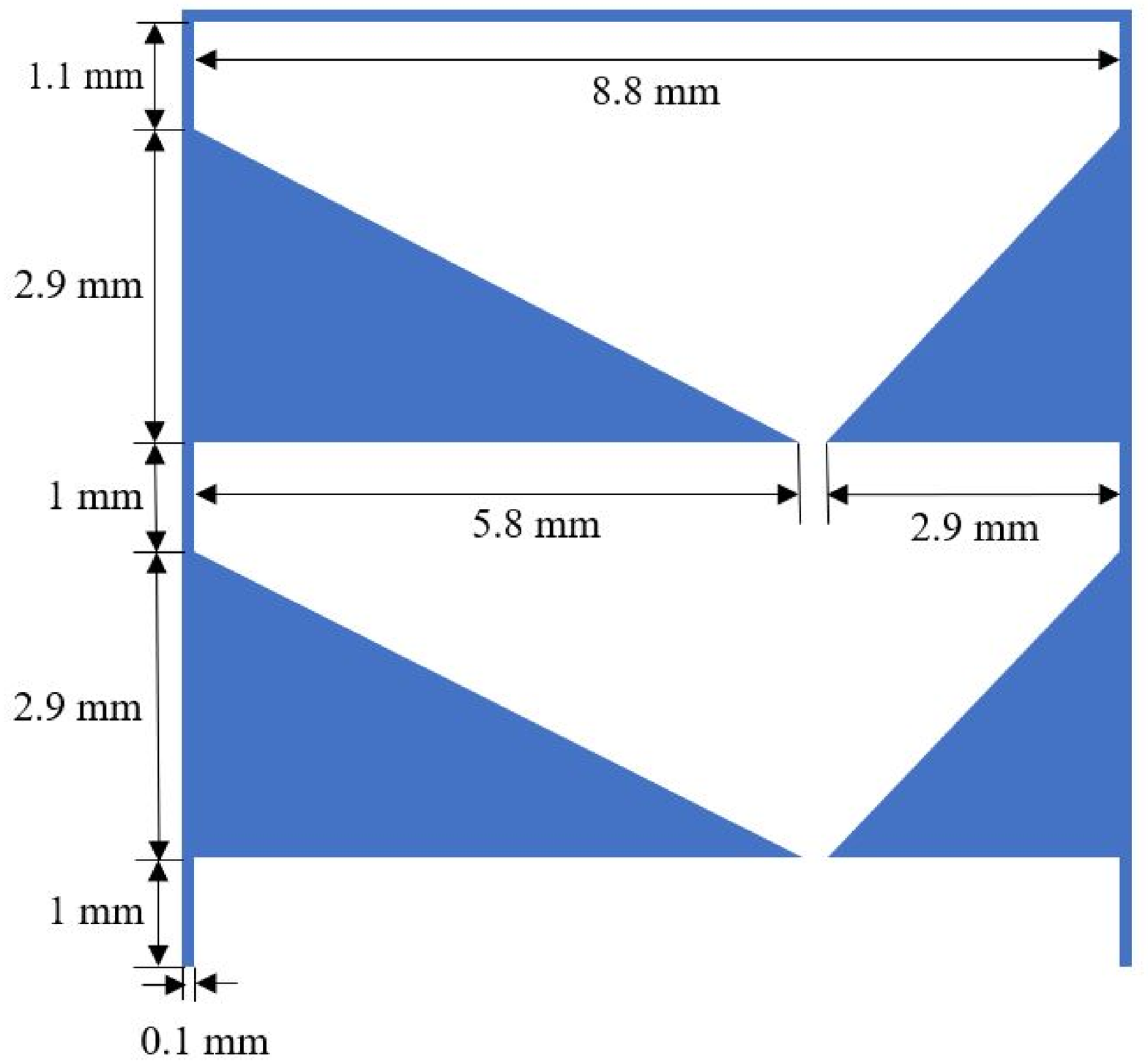}
	\caption{The material layout of the double chevron problem \cite{peng2019lowrank} is shown. The blue area are highly absorbing walls with $\sigma_t = 100 \ \mathrm{cm^{-1}}$ and $\sigma_s = 0.01 \ \mathrm{cm^{-1}}$, the blank is the scattering region with $\sigma_s = \sigma_t = 0.01 \ \mathrm{cm^{-1}}$. There is an incoming isotropic source $Q = 1$ at bottom which is turned on at $t = 0$ and other sides are surrounded by vacuum.}
	\label{fig: DC_layout}
\end{figure}

\begin{figure}[h!] 
\begin{minipage}[b]{0.5\textwidth}
\centering
\includegraphics[width=\textwidth]{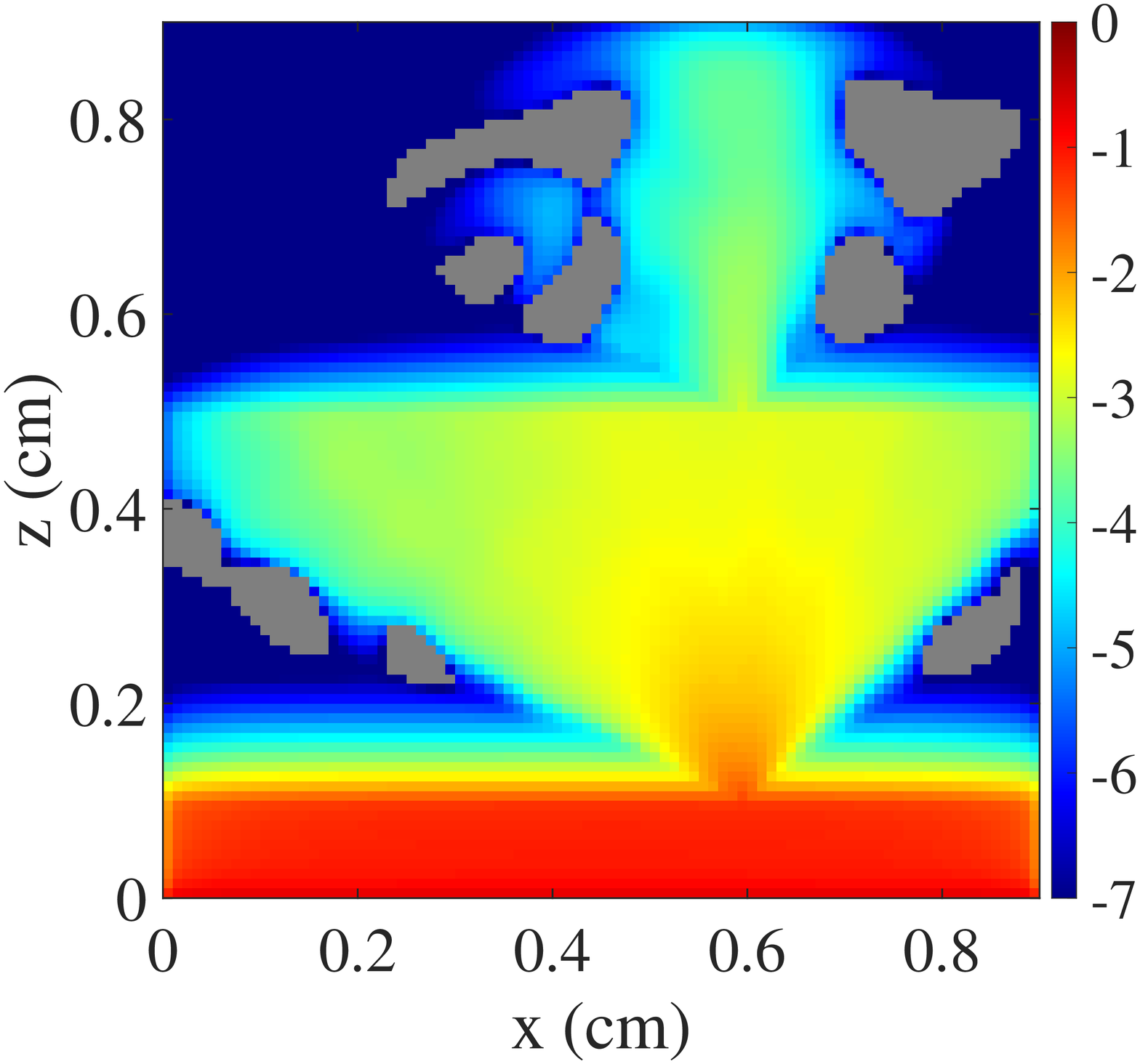}
\subcaption{P$_{99}$, rank 36}
\label{fig: DC_HOLO_a}
\strut\end{minipage}%
\hfill\allowbreak%
\begin{minipage}[b]{0.5\textwidth}
  \centering
\includegraphics[width=\textwidth]{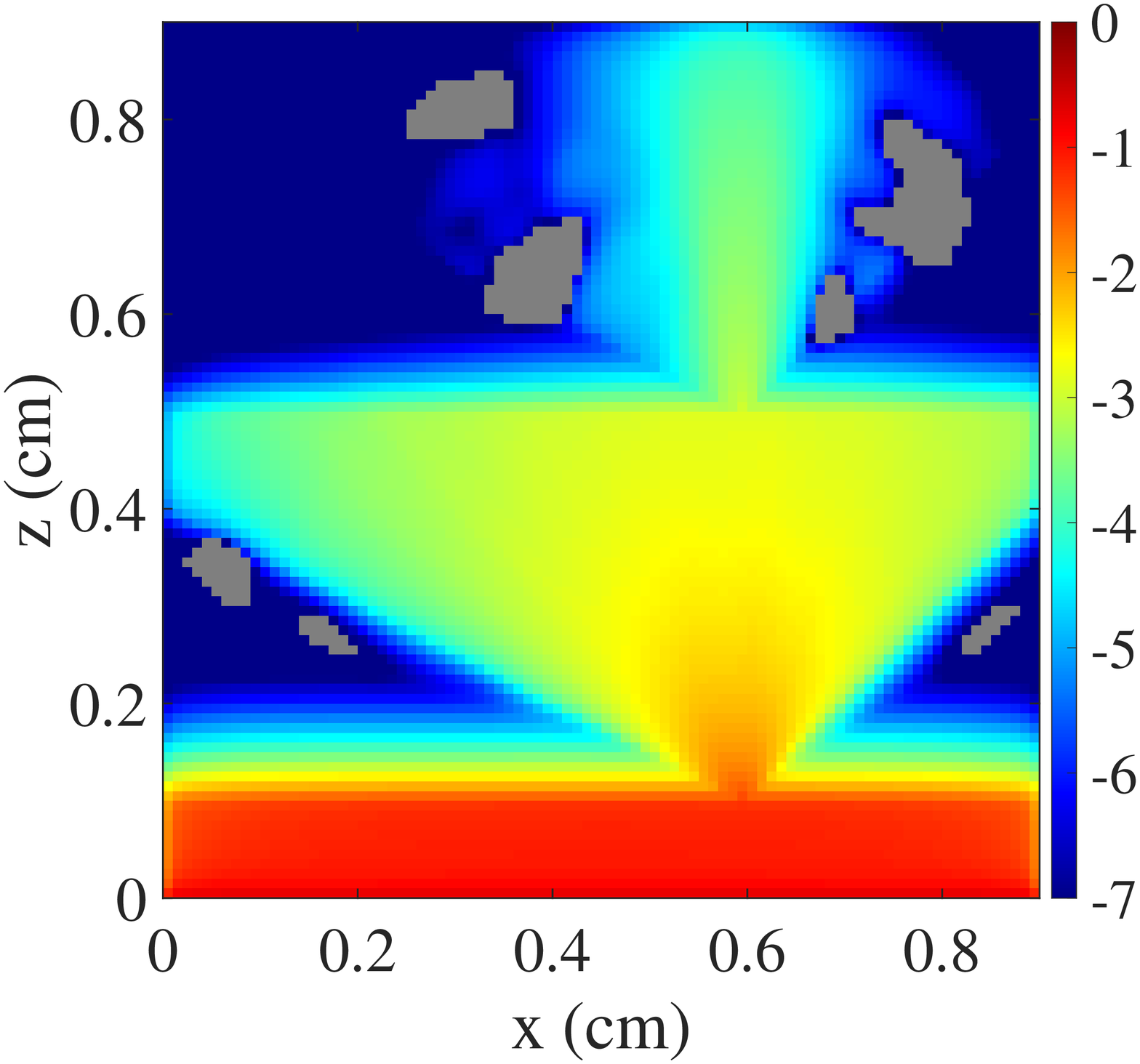}
\subcaption{P$_{99}$, rank 78}
\label{fig: DC_HOLO_b}
\strut\end{minipage}%
\hfill\allowbreak%
\begin{minipage}[b]{0.5\textwidth}
\centering
\includegraphics[width=\textwidth]{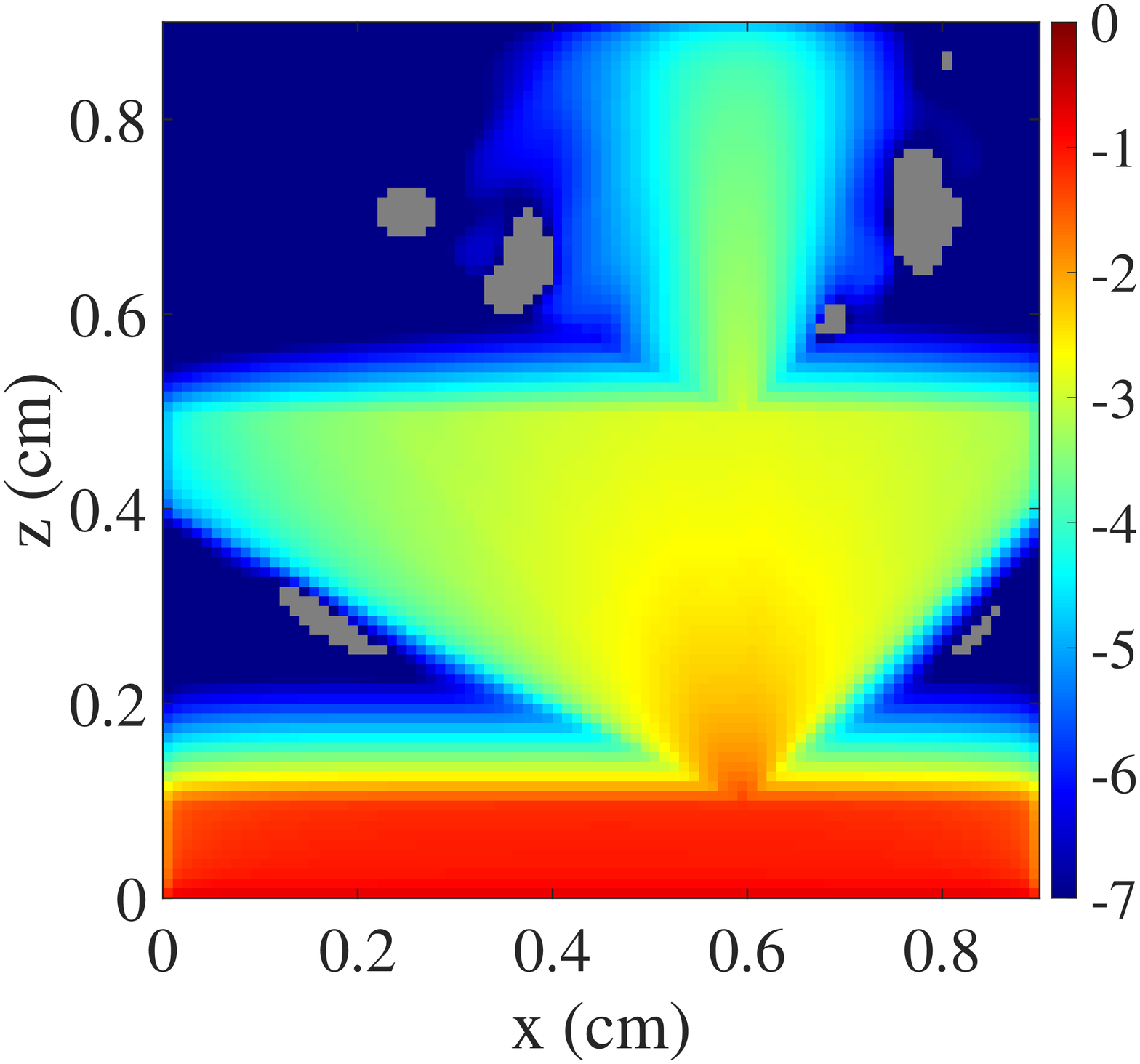}
\subcaption{P$_{99}$, rank 136}
\label{fig: DC_HOLO_c}
\strut\end{minipage}%
\hfill\allowbreak%
\begin{minipage}[b]{0.5\textwidth}
\centering
\includegraphics[width=\textwidth]{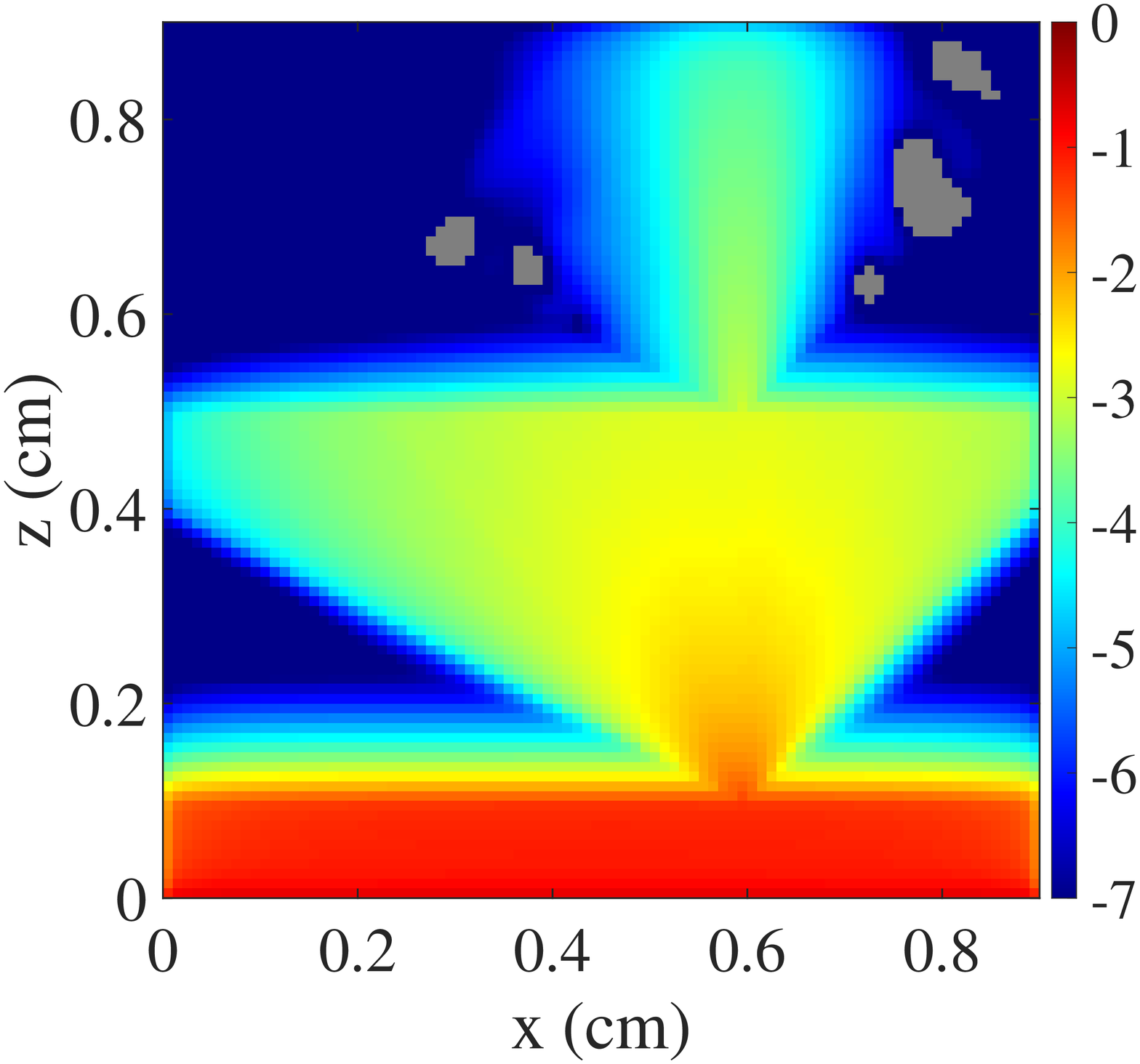}
\subcaption{P$_{99}$, rank 300}
\label{fig: DC_HOLO_d}
\strut\end{minipage}%
\hfill\allowbreak%
\begin{minipage}[b]{0.5\textwidth}
\centering
\includegraphics[width=\textwidth]{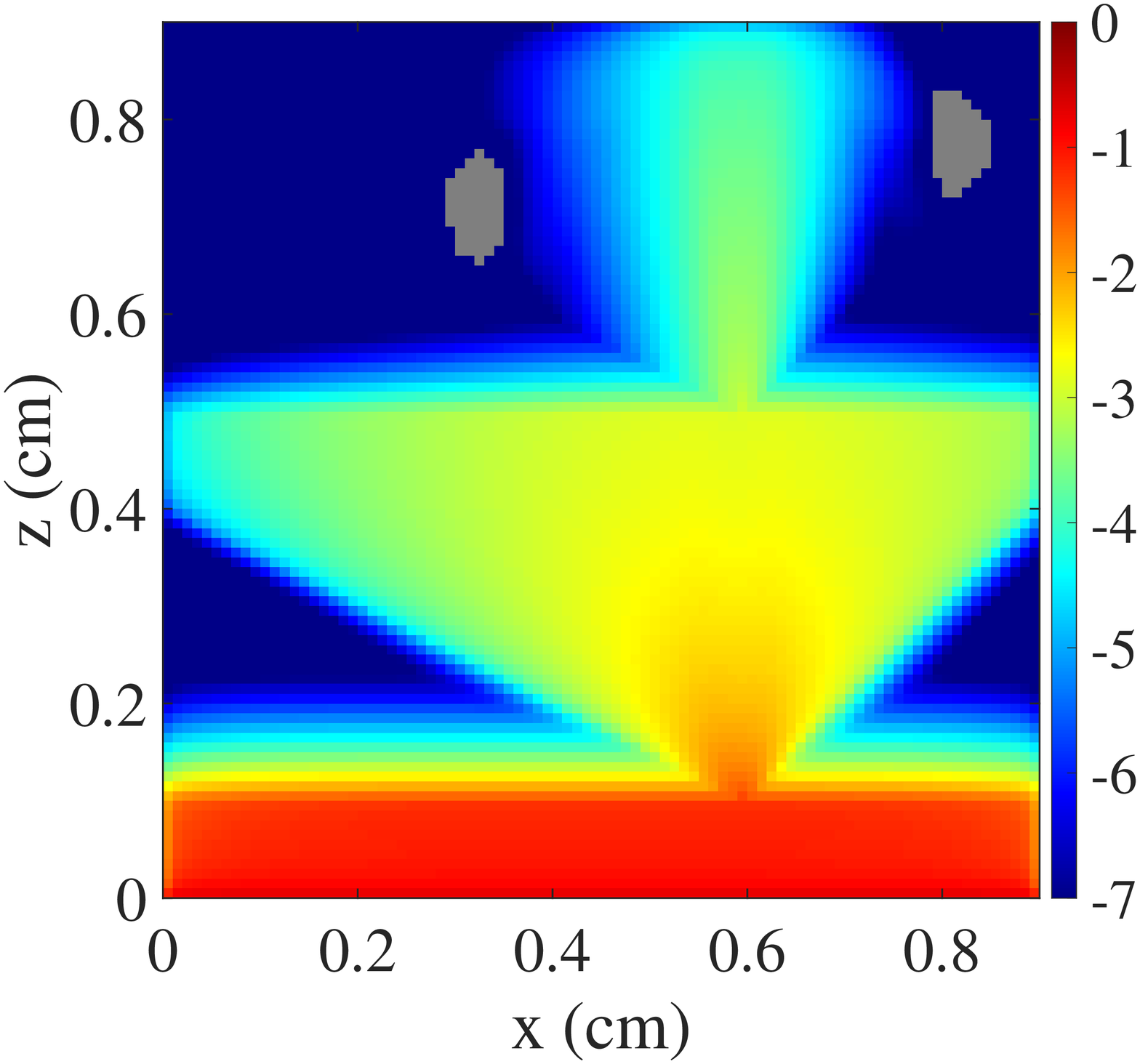}
\subcaption{P$_{99}$, full rank 5050}
\label{fig: DC_HOLO_e}
\strut\end{minipage}%
\hfill\allowbreak%
\begin{minipage}[b]{0.5\textwidth}
  \centering
\includegraphics[width=\textwidth]{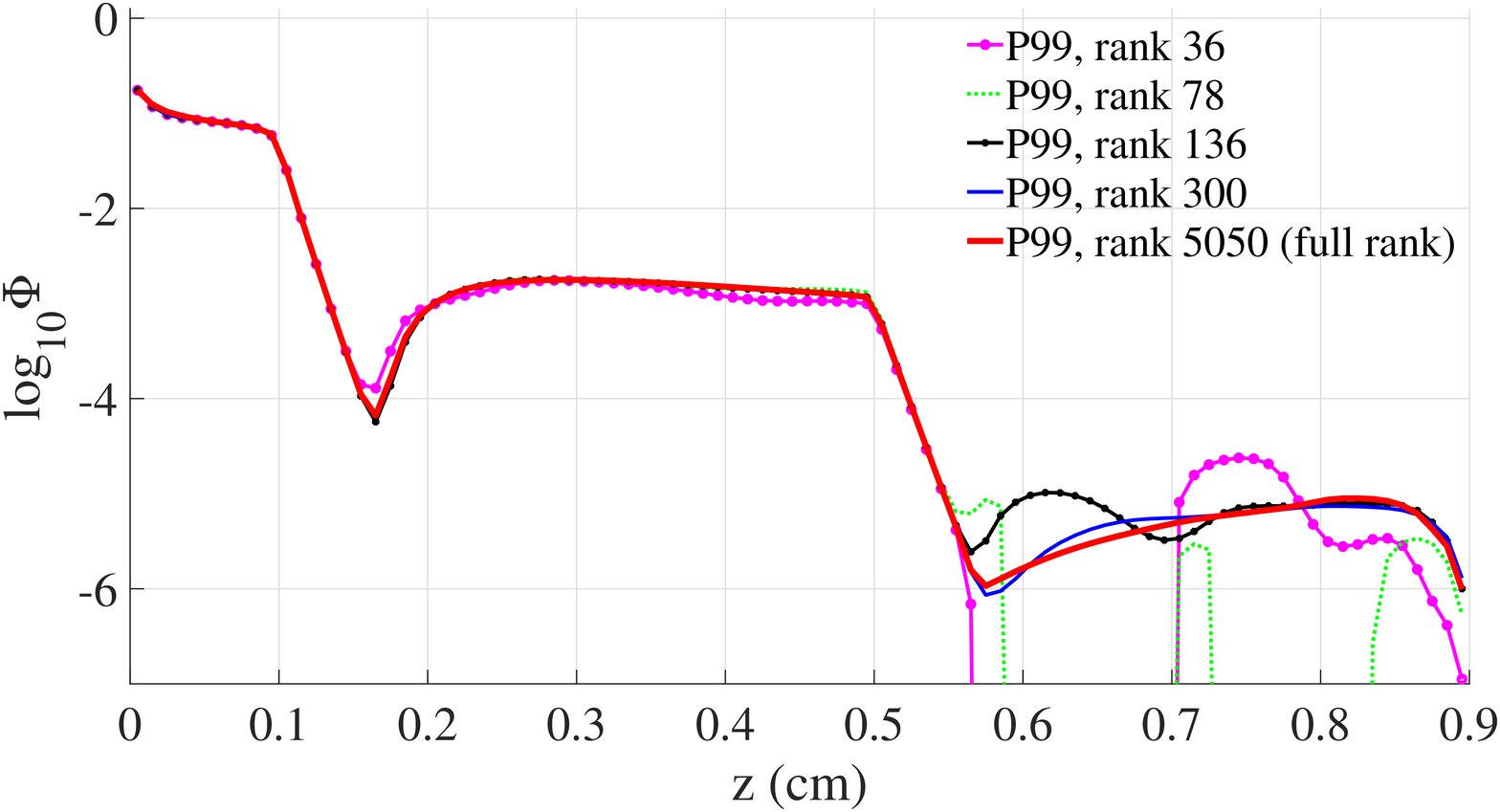}
\subcaption{The scalar flux on the cut along $x=0.45$.}
\label{fig: DC_HOLO_f}
\strut\end{minipage}%
\hfill\allowbreak%
\caption{Solutions to the double chevron problem at t = 0.9s with P$_{99}$ and different rank. The color scale is logarithmic and negative regions are shaded gray.}
\label{fig: DC_HOLO_results}
\end{figure}

We use an asymmetric double chevron problem detailed in Figure \ref{fig: DC_layout} as our final benchmark. Numerical solutions at t = 0.9 s are computed using a $90 \times 90$ spatial grid with a CFL number of 0.2. This problem was originally designed so that $m \approx n$ to get the largest possible benefit of the dynamic low-rank method. In this test, we compare the HOLO solutions of P$_{99}$ to the rank 5050, full rank P$_{99}$ solution. From Figure \ref{fig: DC_HOLO_results}, we can see that the solution with rank 300 is close to the full rank solution over a range of 6 orders of magnitude in this problem, while the rank 36 and 78 solutions cannot capture the particle distribution behind the second chevron. By calculating the memory using \eqref{eq: memory} and \eqref{eq: memory_fullrank}, we find that $93 \%$ of the memory can be saved by applying the HOLO algorithm with rank 300.

\section{Conclusions}
We have presented a HOLO algorithm to overcome the conservation issues in the dynamical low-rank method for radiative transfer. The key idea is to use the low-rank results to calculate the closure term of a two-moment system. When combined with a discontinuous Galerkin scheme we obtain a method that preserves the diffusion limit. These two improvements go a long way to making the method robust enough for a variety of physics applications. 

Our methods use explicit time integration techniques to advance the solution in time.  Future work should incorporate implicit time discretization techniques, perhaps similar to those recently developed to remove the backwards-in-time substep of the DLR method \cite{ceruti2020unconventional}. Additionally, other transport models (e.g., discrete ordinates) and energy-dependent problems should be fruitful areas of future research.

\newpage


\clearpage
\bibliography{mybibfile}

\end{document}